\documentclass{aa} 
\usepackage[varg]{txfonts}
\usepackage{natbib}
\usepackage{graphicx}
\usepackage{subcaption}
\usepackage{sidecap}
\usepackage{mwe}
\usepackage{color}
\usepackage{mathtools}
\usepackage{booktabs} 
\usepackage{multicol} 
\usepackage{multirow} 
\usepackage{array} 
\usepackage{mathabx} 
\usepackage{arydshln}
\usepackage{verbatim} 
\usepackage{lscape} 
\usepackage{pdflscape} 
\usepackage[export]{adjustbox} 
\usepackage{color,epstopdf} 
\DeclareGraphicsExtensions{.ps}
\usepackage{amsmath}
\newcommand{\RN}[1]{\textup{\uppercase\expandafter{\romannumeral#1}}}
\newcolumntype{L}[1]{>{\raggedright\let\newline\\\arraybackslash\hspace{0pt}}m{#1}}
\begin{document}

\title{Molecular complexity on disk-scales uncovered by ALMA} 


\subtitle{The chemical composition of the high-mass protostar AFGL~4176} 

\author{Eva G. B\o gelund\inst{\ref{inst1}}
	\and Andrew G. Barr\inst{\ref{inst1}}
	\and Vianney Taquet\inst{\ref{inst1},\ref{inst2}}
	\and Niels F. W. Ligterink\inst{\ref{inst3}}
	\and Magnus V. Persson\inst{\ref{inst4}}
	\and Michiel R. Hogerheijde\inst{\ref{inst1},\ref{inst5}}
	\and Ewine F. van Dishoeck\inst{\ref{inst1},\ref{inst6}}}  


\institute{Leiden Observatory, Leiden University, PO Box 9513, 2300
	RA Leiden, The Netherlands\label{inst1} \newline \email{bogelund@strw.leidenuniv.nl}
	\and INAF, Osservatorio Astrofisico di Arcetri, Largo E. Fermi 5, 50125 Firenze, Italy \label{inst2}
	\and Center for Space and Habitability (CSH), University of Bern, Sidlerstrasse 5, 3012 Bern, Switzerland\label{inst3}
	\and  Department of Space, Earth and Environment, Chalmers University of Technology, 
	Onsala Space Observatory, 439 92, Onsala, Sweden\label{inst4}
	\and Anton Pannekoek Institute for Astronomy, University of Amsterdam, Science Park 904, 1098 XH Amsterdam, The Netherlands\label{inst5}
	\and Max-Planck Institut für Extraterrestrische Physik, Giessenbachstr. 1, 85748 Garching, Germany\label{inst6}
}

\date{Submitted 29/10/2018} 

\abstract
{The chemical composition of high-mass protostars reflects the physical evolution associated with different stages of star formation. In addition, the spatial distribution and velocity structure of different molecular species provide valuable information of the physical structure of these embedded objects. Despite an increasing number of interferometric studies, there is still a high demand for high-angular resolution data to study chemical compositions and velocity structures for these objects.}
{The molecular inventory of the forming high-mass star AFGL~4176, located at a distance of $\sim$3.7~kpc, is studied in detail at high angular resolution of $\sim$0.35\arcsec, equivalent to $\sim$1285 au at the distance of AFGL~4176. This high resolution makes it possible to separate the emission associated with the inner hot envelope and disk around the forming star from that of its cool outer envelope. The composition of AFGL~4176 is compared with other, both high- and low-mass sources, and placed in the broader context of star-formation.}
{Using the Atacama Large Millimeter/submillimeter Array (ALMA) the chemical inventory of AFGL~4176 is characterised. The high sensitivity of ALMA makes it possible to identify weak and optically thin lines and allows for many isotopologues to be detected, providing a more complete and accurate inventory of the source. For the detected species, excitation temperatures in the range 120 -- 320 K are determined and column densities are derived assuming LTE and using optically thin lines. The spatial distribution of a number of species is studied.}
{A total of 23 different molecular species and their isotopologues are detected in the spectrum towards AFGL~4176. The most abundant species is methanol (CH$_3$OH) with a column density of 5.5$\times$10$^{18}$ cm$^{-2}$ in a beam of $\sim$0.3\arcsec, derived from its $^{13}$C-isotopologue. The remaining species are present at levels between 0.003 \% and 15 \% with respect to methanol. Hints that N-bearing species peak slightly closer to the location of the peak continuum emission than the O-bearing species are seen. A single species, propyne (CH$_3$C$_2$H), displays a double-peaked distribution.}  
{AFGL~4176 comprises a rich chemical inventory including many complex species present on disk-scales. On average, the derived column density ratios with respect to methanol of O-bearing species are higher than those derived for N-bearing species by a factor of three. This may indicate that AFGL~4176 is a relatively young source since nitrogen chemistry generally takes longer to evolve in the gas-phase. Taking methanol as a reference, the composition of AFGL~4176 more closely resembles that of the low-mass protostar IRAS~16293--2422B than that of high-mass star-forming regions located near the Galactic centre. This similarity hints that the chemical composition of complex species is already set in the cold cloud stage and implies that AFGL~4176 is a young source whose chemical composition has not yet been strongly processed by the central protostar.}

 
\maketitle
\section{Introduction} \label{sec:introduction}
The molecular composition of a star-forming region can be used to probe the physical conditions of its environment, define its evolutionary stage, identify chemical processes and in addition, sets the stage and starting conditions for chemistry in disks and eventually planetary systems. A large number of molecular species, ranging from simple to complex, that is molecules consisting of six or more atoms, have been identified in various interstellar environments, from giant molecular clouds to dense cores, protostars and protoplanetary disks \citep[see reviews by][]{Herbst2009, Caselli2012, Tielens2013, Sakai2013}. In context to the formation of high- and low-mass stars, the hot core or hot corino stage displays a particular rich chemistry. At this stage, the young protostar heats its surroundings and creates a bubble of warm ($\sim$200~K) gas, enriched in complex molecules. This complexity is a result of chemistry in the warm gas combined with thermal desorption of the icy mantles of dust grains \citep[e.g.][]{Charnley1992}. 

Over the last decades, many surveys, mostly using single-dish telescopes, have been undertaken to investigate the chemical complexity of high-mass hot cores \citep[e.g.][]{Blake1987, Hatchell1998, vanderTak2000, Ikeda2001, Bisschop2007, Kalenskii2010, Isokoski2013, Rivilla2017, McGuire2017, Suzuki2018, McGuire2018b}. Much focus has been on the hot cores associated with Orion and Sagittarius B2 (hereafter Sgr B2), famous for their high abundances of complex molecules \citep[see e.g.][and references therein]{Neill2014, Crockett2014}, although recently, the low-mass counterparts of these sources have also been under investigations \citep[e.g.][]{Schoier2002, Cazaux2003, Bottinelli2004}. A wealth of information on the chemistry associated with hot cores has been provided by these observations, although most are limited by the, in general, large beam sizes of single-dish telescopes. The consequence hereof is that observations do not only sample the hot core, but also the surrounding environments associated with the protostar, such as the large-scale envelope or outflows \citep[e.g.][]{Fayolle2015}. Generally, this results in multi-component molecular emission, where each component may be characterised by a different line width, velocity, excitation temperature and column density. Furthermore, beam dilution effects may result in large uncertainties on derived molecular column densities if not accounted for correctly. 

The emergence of interferometers such as the Submillimeter Array (SMA), the NOrthern Extended Millimeter Array (NOEMA) and, in particular, the Atacama Large Millimeter/submillimeter Array (ALMA), which provide much higher spatial resolutions than single-dish telescopes, has made it possible to study the molecular emission associated with hot cores on much smaller scales than were previously accessible. This means, that for the first time, an opportunity to "look into" the hot cores themselves is provided whereby the challenges of many single-dish studies can be overcome. In addition, the unprecedented sensitivity of ALMA has made possible the detection of a wealth of weak lines ensuring a more accurate characterization of the chemistry associated with the cores. 

To date, the chemical inventory of only a handful of sources has been extensively studied with interferometers. These include the low-mass protobinary system IRAS~16293--2422 \citep[hereafter IRAS~16293,][]{Jorgensen2016} and the high-mass star-forming regions associated with Sgr B2(N) \citep{Belloche2016} and Orion KL \citep{Brouillet2013, Pagani2017, Favre2017, Tercero2018, Peng2019}. Therefore, there is a substantial need for the continued investigation of additional hot cores in order to build up a database of the molecular inventories and temperatures characterising these sources. Such a database will provide the statistics needed for new insights into the physical and chemical processes at play during the formation of hot cores and will help the classification of sources according to evolutionary stage. 

To this end, the high-mass hot core of AFGL~4176 has been investigated with ALMA and for the first time, a comprehensive study of the chemical inventory of the source is presented. The results of this work are compared with other high- and low-mass sources, in addition to the predictions of hot core chemical models.

\subsection{AFGL~4176}
AFGL~4176, located in the southern hemisphere at 13$^{\rm{h}}$43$^{\rm{m}}$01.704$^{\rm{s}}$, $-62^{\circ}$08\arcmin51.23\arcsec (ICRS J2000), was first identified by \citet{Henning1984} through its bright infrared spectrum as a young and massive star embedded in a thick dusty envelope. The source has been further characterised by \citet{Beltran2006}, who used large-scale millimetre continuum observations carried out with the Swedish-ESO Submillimetre Telescope (SEST), to identify a compact core of approximately 1120 M$_{\odot}$ with a diameter of 1 pc and luminosity of 2$\times$10$^5$ L$_{\odot}$ (assuming a distance of 5.3 kpc). It should be noted however, that the distance to AFGL~4176 is not well-constrained and cited values range from 3.5 to 5.3 kpc \citep[see,][and references therein]{Boley2012}, with the most frequently cited distance being 4.2 kpc, based on observations of CH$_3$OH masers \citep{Green2011}. However, in this work we will assume a distance of 3.7 kpc, based on the recent second release of \textit{Gaia} data, which places the source at a distance of 3.7$\substack{+2.6 \\ -1.6}$ kpc \citep{Bailer-Jones2018}.

In addition to the large-scale envelope, strong evidence that the system contains a Keplerian-like disk is presented by \citet{Johnston2015} who use observations of CH$_3$CN obtained with ALMA to trace the disk kinematics on scales of $\sim$1200 au. A disk-like structure is consistent with the models reported by \citet{Boley2012} who combine interferometric and photometric observations of AFGL~4176 and use radiative transfer and geometric models to characterise the source. Although the observations are generally well described by one-dimensional models, \citet{Boley2012} find substantial deviations from spherical symmetry at scales of tens to hundreds of astronomical units. On these scales, the observations are better described by a multi-component model consisting of a Gaussian halo and an inclined circumstellar disk. Knots of shocked H$_2$ emission have also been identified around AFGL~4176, potentially indicating an outflow, though no preferred spatial direction was revealed \citep{DeBuizer2003}. 

A limited number of detections of molecular species have been reported towards AFGL~4176. CO$_2$ and H$_2$O have been identified in observations carried out with the \textit{Infrared Space Observatory} \citep{vanDishoek1996, vanDishoek1996_water, Boonman2003} and detections of CO, NH$_3$ and CH$_3$CN by the Atacama Pathfinder Experiment (APEX) and ALMA have been reported by \citet{Johnston2014, Johnston2015}. In addition, a number of CH$_3$OH masers are reported in the vicinity of the source \citep{Phillips1998, Green2011}. However, so far no reports of a more comprehensive chemical inventory of the source exist. 

This paper presents an extensive study of the molecular species detected towards AFGL~4176, in addition to those previously reported. The work is based on the same set of high sensitivity, high resolution ALMA observations as analysed by \citet{Johnston2015}, but focuses on identifying and characterising all molecular species with transitions in the observed frequency range associated with the source, rather than the disk kinematics. The high sensitivity of ALMA makes it possible to identify weak and optically thin lines while the unique spatial resolving power ensures that the analysed emission stems from the disk around the central forming star rather than the large-scale surrounding envelope. 

The structure of the paper is as follows: Sect. \ref{sec:obs_and_method} introduces the observations, calibration process and methods used for identifying molecular species. Section \ref{sec:results} lists all detected species, our derived molecular column densities and excitation temperatures and discusses the spatial distribution of selected molecules. Section \ref{sec:discussion} compares the results for AFGL~4176 with observations of other objects and with model predictions. Finally, sect. \ref{sec:summary} summaries the results and conclusions.

\section{Observations and methods} \label{sec:obs_and_method}
\subsection{Observations} \label{subsec:obs}
Observations of AFGL~4176 were carried out with ALMA during Cycle 1 \citep[program 2012.1.00469.S, see][for first results]{Johnston2015} with 39 antennas in the array, between August 16, 2014 and August 17, 2014 using the Band 6 receivers, covering the frequency range of 211 -- 275 GHz. Four spectral windows were obtained covering a total bandwidth of $\sim$4.7 GHz. These consist of two wide windows, of 1875 MHz, centred at 240.5 and 254.0 GHz and two narrow windows, of 468.75 MHz, centred at 239.0 and 256.3 GHz. The spectral resolution of the observations is 976.6 kHz ($\sim$1.2 km s$^{-1}$) and 244 kHz ($\sim$0.3 km s$^{-1}$) for the wide and narrow windows, respectively. The angular resolution is $\sim$0.35$\arcsec$, equivalent to $\sim$1285 au at the distance of AFGL~4176. 

The data were downloaded from the ALMA archive and reduced via the delivered pipeline script using the Common Astronomy Software Applications (CASA) version 4.2.1. Bandpass and absolute flux calibration was carried out, respectively, using J1617-5848 and Titan, on August 16 and J1427-4206 and Ceres on August 17. The phase and gain calibration was carried out, respectively, using J1308-6707 and J1329-5608 on both days. A conservative flux calibration accuracy of 20\% has been adopted. This uncertainty only contributes moderately to the total uncertainty of the presented results. The data were continuum subtracted using the most line-free channels and corrected for primary beam attenuation. 

The continuum and line data were imaged separately in CASA version 5.1.1-5 using a pixel size of 0.04$\arcsec$, a velocity resolution for the spectral cubes of 1.2 km s$^{-1}$ and Briggs weighting with a robust parameter of 1.5. The peak continuum emission is 29 mJy beam$^{-1}$ (5.7 K at 247 GHz) with an rms noise of approximately 0.5 mJy beam$^{-1}$ (0.1 K at 247 GHz) in a beam of 0$\overset{\second}{.}$33$\times$0$\overset{\second}{.}$31. The coordinates of the continuum peak were determined by 2D Gaussian fitting in the image plane to be 13$^{\rm{h}}$43$^{\rm{m}}$01.699$^{\rm{s}} \pm$ 0.003$^{\rm{s}}$, -62$^{\circ}$08\arcmin51.25\arcsec $\pm$ 0.02\arcsec (IRCS J2000). Table \ref{tab:spw_overview} lists the frequencies covered and the rms noise per 1.2 km s$^{-1}$ channel derived for each of the spectral windows.  

\begin{table}[]
\begin{small}
	\centering
	\caption{Overview of spectral cubes.}
	\label{tab:spw_overview}
	\begin{tabular}{cccc}
		\toprule
		Frequency & Beam & \multicolumn{2}{c}{rms noise} \\
		\cmidrule{3-4}
		$[$GHz$]$ & [$\arcsec \times \arcsec$ (P.A.$^{\circ}$)] & [mJy beam$^{-1}$] & [K] \\
		\midrule
		238.838 -- 239.306 & 0$\overset{\second}{.}$35$\times$0$\overset{\second}{.}$29 (-31.6) & 1.9 & 0.40 \\
		239.604 -- 241.478 & 0$\overset{\second}{.}$34$\times$0$\overset{\second}{.}$30 (-31.2) & 1.5 & 0.31 \\
		253.107 -- 254.980 & 0$\overset{\second}{.}$33$\times$0$\overset{\second}{.}$28 (-30.8) & 1.7 & 0.35 \\
		256.115 -- 256.583 & 0$\overset{\second}{.}$32$\times$0$\overset{\second}{.}$28 (-32.3) & 2.2 & 0.46 \\	
		\bottomrule
	\end{tabular}
	\tablefoot{The listed rms noise is determined over the line-free channels. The channel width is 1.2 km s$^{-1}$.}
\end{small}
\end{table}

\begin{figure}
	\centering
	\includegraphics[width=0.48\textwidth]{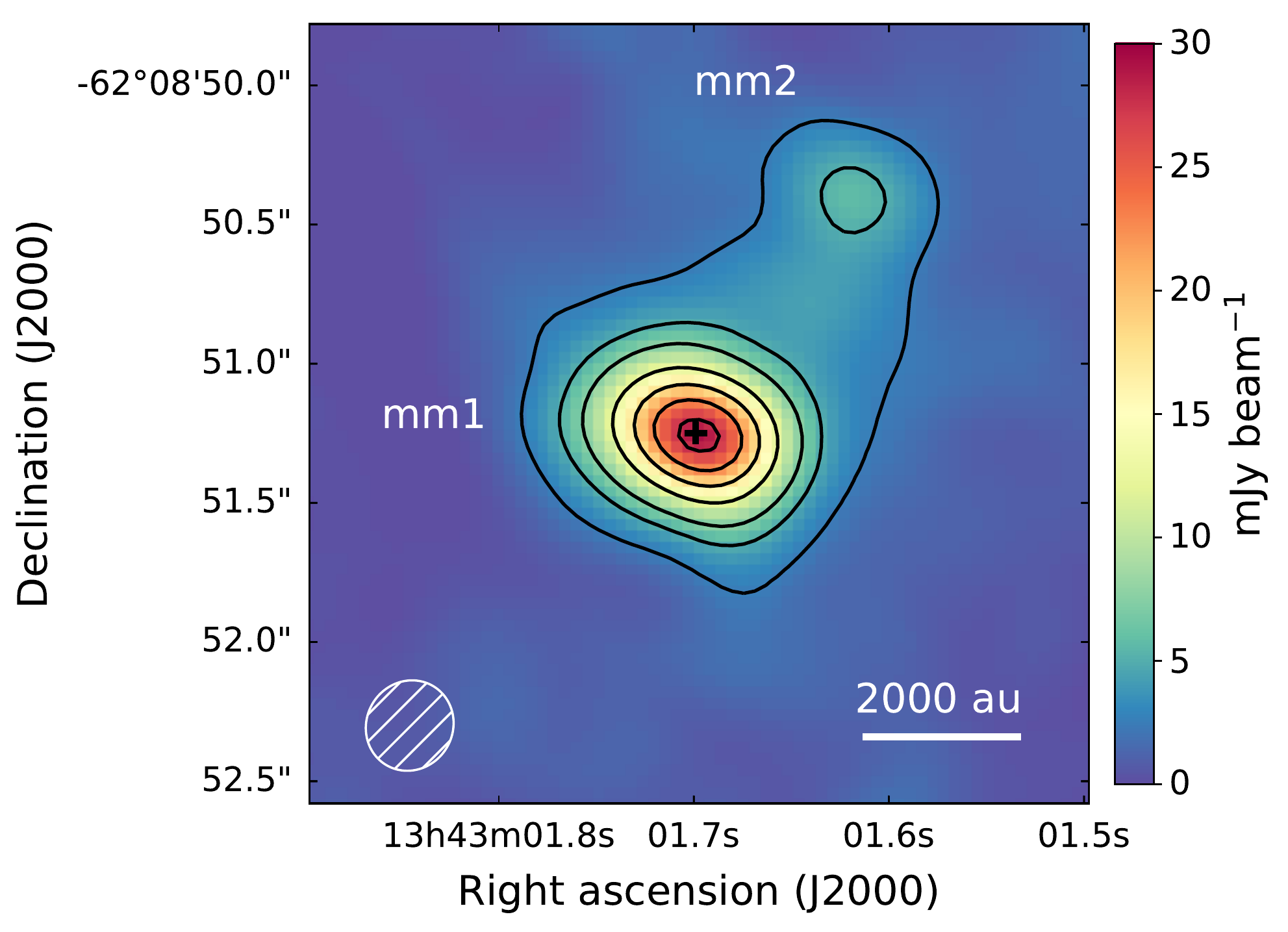}
	\caption[]{Continuum image of AFGL~4176 at 1.2 mm. Contours are [5, 10, 15, 25, 35, 45, 55]$\sigma$, with $\sigma$ = 0.5 mJy beam$^{-1}$. The peak continuum location at which the spectra have been extracted is marked by the black cross. The synthesised beam (0$\overset{\second}{.}$33$\times$0$\overset{\second}{.}$31 $\sim$1210$\times$1140 au) is shown in the bottom left corner.}
	\label{fig:Map}
\end{figure}

For each of the imaged cubes, a spectrum is extracted at the location of the peak continuum emission. Each spectrum represents the average over an area equivalent to the size of the synthesised beam ($\sim$0$\overset{\second}{.}$3). As a consequence, all derived molecular column densities are thus synthesised beam-averaged and probe the warmest, inner regions of the hot core. Figure \ref{fig:Map} shows the continuum and the location at which the spectra were extracted. In addition to the main continuum peak, labelled mm1 by \citet{Johnston2015}, a secondary peak is observed $\sim$1$\arcsec$ north-west of the primary peak; this peak is labelled mm2. A counterpart to this secondary peak is observed to the south-east of mm1 (outside the plotted region). These two peaks are located perpendicular to the major axis of mm1 and may indicate a large-scale outflow, consistent with the CO observations presented by \citet{Johnston2015}. For this work however, the focus is on the main continuum peak and all subsequent discussion refers to this source only.   

\subsection{Methods for line identification and modelling} \label{subsec:method}
For the identification of spectral lines, catalogued transition frequencies from the JPL \citep[Jet Propulsion Laboratory\footnote{http://spec.jpl.nasa.gov},][]{Pickett1998} and CDMS \citep[Cologne Database for Molecular Spectroscopy\footnote{https://cdms.astro.uni-koeln.de/cdms/portal/},][]{Muller2001,Muller2005,Endres2016} molecular databases are compared with the extracted spectra. Observed lines are considered detected if the peak signal-to-noise ratio is three or higher. Species with fewer than five detected lines are considered tentative detections. This criterion may be too strict for some of the simpler molecules with sparse, but strong rotational spectra (e.g. SO) and these can likely be considered detections. Using the CASSIS\footnote{Centre d'Analyse Scientifique de Spectres Instrumentaux et Synthétiques: http://cassis.irap.omp.eu} line analysis software and assuming local thermodynamic equilibrium (LTE) and optically thin lines, a synthetic spectrum is produced for each identified species. This is done by providing CASSIS with the following parameters: excitation temperature, $T_{\textrm{ex}}$ [K], column density of the species, $N_{\textrm{s}}$ [cm$^{-2}$], source velocity, $v_{\textrm{LSR}}$ [km s$^{-1}$], line width at half maximum [km s$^{-1}$], and angular size of the emitting region (assumed to be equal to the area of the synthesised beam), $\theta_{\textrm{s}}$ [$^{\second}$]. Note that $N_{\textrm{s}}$ is a synthesised beam-averaged column density, not a source-averaged column density.

For two species, CH$_3$CN and HC$_3$N, vibrationally excited transitions are detected (see sect. \ref{sec:results}). For vibrationally excited CH$_3$CN the JPL database is used. This entry utilises a partition function in which vibrational contributions are taken into account. In contrast, the CDMS entries for vibrationally excited HC$_3$N, and isotopologues thereof, do not include vibrational contributions to the partition function but list these separately. Therefore, vibrational correction factors have been applied to all listed values of vibrationally excited HC$_3$N. These vibrational correction factors are retrieved from the CDMS site\footnote{https://cdms.astro.uni-koeln.de/cdms/portal/catalog/379/}. At 225 K, the vibrational correction factor to the partition function of HC$_3$N and its isotopologues are 1.17 and 1.48, respectively.

Excitation temperatures and column densities are determined for species which have three or more unblended lines detected, that is lines with a signal-to-noise ratio of three or higher, whose emission can mainly be attributed to one molecule, and these span upper state energies of at least 100 K. This is done by creating grids of models varying $T_{\textrm{ex}}$ and $N_{\textrm{s}}$ and identifying the model with the minimal $\chi^2$ as the best fit. The CASSIS software computes the $\chi^2$ value for each synthetic spectrum in the model grid taking into account the channels within a range of $\pm$~10~km~s$^{-1}$ of the catalogue frequency of all unblended lines detected for each species. Table \ref{tab:model_grid} lists the model grids for each of the fitted species. The uncertainty on  $T_{\textrm{ex}}$ and $N_{\textrm{s}}$ is listed as the standard deviation of models within the 95$\%$ confidence level. For most species these are about 20$\%$, though for CH$_3$CHO and (CH$_2$OH)$_2$ the uncertainty on $T_{\textrm{ex}}$ is up to 85$\%$. In both cases, the larger uncertainty on $T_{\textrm{ex}}$ is likely due to the relatively low signal-to-noise (S/N $\sim$4) of a number of the unblended lines detected for these species. The uncertainty on column density ratios with respect to methanol is calculated through the propagation of errors. For species where less than three unblended lines are detected, or species where upper state energies of the detected lines do not span more than 100 K, the column density is derived assuming a fixed excitation temperature. In the case of CH$_3$OH and the isotopologues of SO$_2$, the excitation temperature is assumed to be 120 K and 150 K, based on the $^{13}$C-methanol isotopologue and main isotopologue of SO$_2$, respectively. For all other species, the excitation temperature is assumed to be 200 K. This value is the average of the best-fit excitation temperatures derived for the 12 species listed at the top of Tables \ref{tab:line_summary} and \ref{tab:line_models}. However, since the spread in best-fit excitation temperatures in fairly large, standard deviation of $\sim$70 K, column densities are also derived assuming excitation temperatures of 130 K and 270 K. For most species, the column densities derived at these temperatures are within 50$\%$ of the value derived assuming $T_{\textrm{ex}}$ = 200 K. For CH$_3$COCH$_3$, column densities at 130 K and 270 K are within a factor of three of the value derived assuming $T_{\textrm{ex}}$ = 200 K while for vibrationally excited HC$_3$N, the column density derived at 130 K is a factor five higher than the value derived at $T_{\textrm{ex}}$ = 200 K.

To ensure that no lines are incorrectly assigned, three checks are conducted. First, that the best-fit model for each species does not predict lines at frequencies, covered by the observations, where no emission is detected. Second, that no other species, for which the spectroscopy is known and listed in either of the databases mentioned above, can reproduce the line without predicting lines at frequencies where no emission is detected. Finally, that isotopically rare species do not predict lines of the main isotopologue where no emission is detected.

The excitation temperatures and column densities for $^{13}$CH$_3$OH and CH$_3$CN are derived first because their lines are very numerous, bright and span a large range of upper state energies. Based on fits to these species, a source velocity of $-$53.5~km s$^{-1}$ and FWHM line widths of 6 km s$^{-1}$ are found. These values are kept fixed for the subsequent fitting of other species to minimise the number of free parameters. However, it should be noted that different molecules may trace different gas components and that the fixed source velocity and line width represent the average conditions of the sources. For example, a slight velocity shift is observed for the transitions of SO$_2$. Leaving the source velocity as a free parameter for this species results in a best-fit value of $-$52.0 ~km~s$^{-1}$. The best-fit column density and excitation temperature at this source velocity are within the uncertainty of the values derived assuming the source velocity to be $-$53.5~km~s$^{-1}$.

Finally, in order to compare the derived molecular column densities across different objects, CH$_3$OH and H$_{2}$ are used as references. Methanol is used as a reference because it is usually one of the most abundant species in hot cores and is thought to be the parent molecule for most complex organics. However, because this species is very abundant, many of its lines are optically thick and therefore its column density cannot be derived directly. The column density of CH$_3$OH is instead estimated based on the best-fit value for its $^{13}$C-isotopologue, adopting a $^{12}$C/$^{13}$C values of 60, derived assuming a galactocentric distance ($d_{\textrm{GC}}$) of 6.64~kpc and the relation for $^{12}$C/$^{13}$C reported by \cite{Milam2005}. One has to keep in mind that $^{12}$C/$^{13}$C may still deviate from the galactrocentric trend (for example, HC$_{3}$N / HCC$^{13}$CN is tentatively found to be $\sim$16 for AFGL~4176, see sect. \ref{subsec:isotope_ratios}) and therefore can cause an uncertainty in the CH$_{3}$OH column density.

The H$_{2}$ column density is determined from the dust continuum according to equation \ref{eq:H2}:

\begin{equation}
	N_{\rm H_{2}} = \frac{100 \cdot I_{\nu}}{\Omega_{beam} \cdot \mu_{\textrm{H}_2} \cdot m_{\textrm{H}} \cdot \kappa_{\nu} \cdot B_{\nu}(T)}, 
	\label{eq:H2}
\end{equation}

\noindent where $I_{\nu}$ is the continuum intensity, $\Omega_{beam}$ is the solid angle covered by the beam, $\mu_{\rm H_{2}}$ = 2.33 is the mean molecular mass per H$_{2}$ molecule, $m_{\rm H}$ is the mass of the hydrogen atom, $\kappa_{\rm \nu}$ is the dust opacity, $B_{\rm \nu}$ is the Planck function at T = 200~K, and the factor 100 accounts for the gas-to-dust ratio. For our data, $I_{\rm \nu}$ = 29 mJy beam$^{-1}$ and $\kappa_{\rm \nu}$ = 1.0 cm$^{2}$ g$^{-1}$ \citep{Ossenkopf1994}. The resulting H$_{2}$ column density is found to be $N_{\rm H_{2}}$ = 4$\times$10$^{23}$ cm$^{-2}$. This equation does not assume the Rayleigh-Jeans limit, but includes the Planck correction.

\section{Results} \label{sec:results}
A total of 354 lines are identified towards AFGL~4176 with a signal-to-noise ratio of three or above. Of these, 324 lines can be assigned to a total of 23 different molecular species or their isotopologues. Fifteen species have five or more detected transitions, while eight species have fewer than five detected transitions and are therefore considered tentative detections. For the remaining 30 lines no match to known transitions was found. A list of frequencies and peak intensities for these unidentified lines is given in appendix \ref{app:unidentified_lines}. With a total covered bandwidth of $\sim$4.7~GHz, the line density is roughly 75 lines per GHz or one line per 13.3 MHz (ALMA Band 6). For comparison, the ALMA-PILS survey towards the low-mass protobinary system IRAS~16293B found one line per 3.4 MHz \citep[ALMA Band 7,][]{Jorgensen2016}. The high line density in AFGL~4176 means that detected lines are often blended with emission from other species. Therefore, as noted above, great caution is exercised when lines are assigned to species.

Table \ref{tab:line_summary} list all identified species, isotopologues and isomers. The Table also summarises the number of identified lines, both unblended and blended, the range of upper state energies and Einstein A coefficients covered by these lines as well as the derived excitation temperature and column density for each species. It should be noted that hyperfine transitions with the same catalogued frequency are counted as a single line since these are indistinguishable in the data. After the identification and modelling of individual species the synthetic spectra are summed to obtain a full model for AFGL~4176. Figure \ref{fig:spw0} shows the full model for the spectral window centred at 239.1 GHz (see appendix \ref{app:full_models} for full model of other spectral windows).

\begin{table*}[]
	\centering
	\caption{Summary of detected lines.}
	\label{tab:line_summary}
	\begin{tabular}{llccccccc}
		\toprule
		Species & Name & \multicolumn{4}{c}{N$_{\textrm{lines}}$} & $E_{\textrm{up}}$ & $A_{\textrm{ij}}$ & Catalogue \\
		\cline{3-6}
		& & U & B & OT & Total & [K] & $\times$10$^{-5}$ [s$^{-1}$] & \\
		\midrule
		$^{13}$CH$_3$OH & Methanol & 7 & 5 & 0 & 12 & 60 -- 644 & 2.31 -- 8.81 & CDMS \\		
		CH$_3$C$_2$H & Propyne & 9 & 1 & 0 & 10 & 86 -- 346 & 3.85 -- 5.83 & CDMS \\
		CH$_3$CN & Methyl cyanide (Acetonitrile) & 9 & 0 & 1 & 9 & 80 -- 537 & \phantom{0}73 -- 118 & JPL \\		
		CH$_3$CN, v$_8$=1 & & 14 & 2 & 0 & 16 & 600 -- 955\phantom{0}& \phantom{0}94 -- 111 & JPL \\		
		CH$_3$CHO & Acetaldehyde & 4 & 4 & 0 & 8 & 93 -- 490 & 3.85 -- 56.3 & JPL \\	
		NH$_2$CHO & Formamide & 10 & 3 & 0 & 13 & 68 -- 439 & 2.48 -- 115\phantom{.} & CDMS \\
		H$_2$CS & & 7 & 1 & 0 & 8 & 46 -- 519 & 5.43 -- 20.5 & CDMS \\				
		CH$_3$OCH$_3$ & Methyl ether & 17 & 5 & 0 & 22 & 26 -- 502 & 2.40 -- 8.72 & CDMS \\			
		C$_2$H$_5$OH & Ethanol & 7 & 4 & 0 & 11 & 35 -- 450 & 1.93 -- 41\phantom{.0} & CDMS \\
		C$_2$H$_3$CN & Vinylcyanide & 8 & 4 & 0 & 12 & 153 -- 388\phantom{0} & 114 -- 136& CDMS \\
		CH$_3$OCHO & Methyl formate & 14 & 5 & 0 & 19 & 101 -- 473\phantom{0} & 0.66 -- 24.8 & JPL \\		
		\textit{aGg'}(CH$_2$OH)$_2$ & Ethylene glycol & 14 & 16 & 0 & 30 & 144 -- 327\phantom{0} & 3.76 -- 40.9 & CDMS \\
		\textit{gGg'}(CH$_2$OH)$_2$ &  & 12 & 8 & 0 & 20 & 62 -- 229 & 2.93 -- 14.3 & CDMS \\		
		SO$_2$ & Sulphur dioxide & 6 & 0 & 4 & 6 & 36 -- 333 & 2.67 -- 13.3 & JPL \\
		\midrule
		CH$_3$OH & & 31 & 10 & 15 & 41 & 20 -- 950 & 1.56 -- 8.80 & JPL \\		
		H$_2$CCO & Ketene & 1 & 0 & 0 & 1 & 88 & 15.5 & CDMS \\	
		HNCO & Isocyanic acid & 1 & 0 & 0 & 1 & 113 & 19.0 & CDMS \\			
		NS & Nitrogen sulphide & 4 & 2 & 0 & 6 & 39 & 0.93 -- 28.4 & JPL \\		
		C$^{34}$S & Carbon sulphide & 1 & 0 & 0 & 1 & 35 & 28.6 & JPL \\		
		t-HCOOH & Formic acid & 1 & 1 & 0 & 2 & 70 & 15.7 & CDMS \\			
		SO & Sulphur monoxide & 1 & 0 & 0 & 1 & 100 & 0.43 & JPL \\			
		$^{34}$SO & & 1 & 0 & 0 & 1 & 56 & 20.4 & JPL \\	
		HC$_3$N, v=0 & Cyanoacetylene & 1 & 0 & 1 & 1 & 177 & 132 & CDMS \\
		HC$_3$N, v$_7$=2 & & 1 & 2 & 0 & 3 & 820 -- 823 & 132 -- 133 & CDMS \\
		HCC$^{13}$CN & & 1 & 0 & 0 & 1 & 177 & 130 & CDMS \\
		HCCC$^{15}$N & & 1 & 0 & 0 & 1 & 184 & 134 & CDMS \\
		HCC$^{13}$CN, v$_7$=1 & & 2 & 0 & 0 & 2 & 495 -- 496 & 130 -- 131 & CDMS \\
		C$_2$H$_5$CN & Ethyl cyanide (Propionitrile) & 9 & 4 & 0 & 13 & \phantom{0}79 -- 189 & 6.22 -- 142\phantom{.} & CDMS \\
		CH$_3$COCH$_3$ & Acetone & 13 & 2 & 0 & 15 & \phantom{0}74 -- 235 & 2.43 -- 659\phantom{.} & JPL \\	
		CH$_2$(OH)CHO & Glycolaldehyde & 2 & 1 & 0 & 3 & 111 -- 143 & 12.0 -- 27.9 & CDMS \\
		O$^{13}$CS & Carbonyl sulphide & 1 & 0 & 0 & 1 & 134 & 4.80 & CDMS \\	
		$^{33}$SO$_2$ & & 10 & 8 & 0 & 18 & \phantom{0}72 -- 471 & 0.13 -- 19.6 & JPL \\	
		$^{34}$SO$_2$ &  & 1 & 1 & 0 & 2 & \phantom{0}82 -- 182 & 2.66 -- 12.8 & JPL \\			
		SO$^{18}$O &  & 1 & 2 & 0 & 3 & 69 -- 89 & 0.11 -- 18.3 & JPL \\								
		\bottomrule
	\end{tabular}
	\tablefoot{U = Unblended lines, B = Blended lines, OT = Optically thick lines ($\tau$ $\rm \geq$ 1)}
\end{table*}

\begin{table*}[]
	\small
	\centering
	\caption{Summary of derived values of $N_{\textrm{s}}$ and $T_{\textrm{ex}}$.}
	\label{tab:line_models}
	\begin{tabular}{lcccccc}
		\toprule
		Species & $T_{\textrm{ex}}$ & \multicolumn{3}{c}{$N_{\textrm{s}}$} & X/CH$_3$OH & X/H$_{2}$ \\
		& [K] & \multicolumn{3}{c}{[cm$^{-1}$]} & [\%] & $\times$10$^{-8}$ \\
		\midrule
			$^{13}$CH$_3$OH & 120 $\pm$ 15\phantom{1} & \multicolumn{3}{c}{(9.2 $\pm$ 0.6)$\times$10$^{16}$} & 1.7 $\pm$ 0.2 & 23.0 $\pm$ 1.5 \\		
			CH$_3$C$_2$H & 320 $\pm$ 90\phantom{1} &  \multicolumn{3}{c}{(3.8 $\pm$ 0.7)$\times$10$^{16}$}  &  0.7 $\pm$ 0.2 & \phantom{0}9.5 $\pm$ 1.8 \\
			CH$_3$CN & 270 $\pm$ 40\phantom{1} &  \multicolumn{3}{c}{(3.4 $\pm$ 0.3)$\times$10$^{16}$} &  0.62 $\pm$ 0.07 & \phantom{0}8.5 $\pm$ 0.8 \\		
			CH$_3$CN, v$_8$=1 & 220 $\pm$ 40\phantom{1}  &  \multicolumn{3}{c}{(4.3 $\pm$ 0.6)$\times$10$^{16}$} & 0.8 $\pm$ 0.2 & 10.8 $\pm$ 1.5 \\		
			CH$_3$CHO & 160 $\pm$ 125 &  \multicolumn{3}{c}{(1.5 $\pm$ 0.8)$\times$10$^{16}$} & 0.3 $\pm$ 0.2 & \phantom{0}3.8 $\pm$ 2.0 \\	
			NH$_2$CHO & 190 $\pm$ 35\phantom{1} &  \multicolumn{3}{c}{\phantom{0}(1.0 $\pm$ 0.09)$\times$10$^{16}$} & 0.18 $\pm$ 0.02 & \phantom{0}2.5 $\pm$ 0.3 \\
			H$_2$CS & 160 $\pm$ 10\phantom{1} &  \multicolumn{3}{c}{(4.3 $\pm$ 0.2)$\times$10$^{16}$} & 0.78 $\pm$ 0.07 & 10.8 $\pm$ 0.5 \\				
			CH$_3$OCH$_3$ & 160 $\pm$ 15\phantom{1} &  \multicolumn{3}{c}{(1.3 $\pm$ 0.1)$\times$10$^{17}$} & 2.4 $\pm$ 0.3 & 32.5 $\pm$ 2.5 \\			
			C$_2$H$_5$OH & 120 $\pm$ 45\phantom{1} &  \multicolumn{3}{c}{(7.6 $\pm$ 1.8)$\times$10$^{16}$} & 1.4 $\pm$ 0.4 & 19.0 $\pm$ 4.5 \\
			C$_2$H$_3$CN & 240 $\pm$ 105 &  \multicolumn{3}{c}{(6.2 $\pm$ 1.1)$\times$10$^{15}$} & 0.11 $\pm$ 0.02 & \phantom{0}1.6 $\pm$ 0.3 \\
			CH$_3$OCHO & 310 $\pm$ 75\phantom{1} &  \multicolumn{3}{c}{(1.7 $\pm$ 0.3)$\times$10$^{17}$} & 3.1 $\pm$ 0.6 & 42.5 $\pm$ 7.5 \\		
			\textit{aGg'}(CH$_2$OH)$_2$ & 160 $\pm$ 130 &  \multicolumn{3}{c}{(3.0 $\pm$ 0.5)$\times$10$^{16}$} & 0.6 $\pm$ 0.1 & \phantom{0}7.5 $\pm$ 1.3 \\
			\textit{gGg'}(CH$_2$OH)$_2$ & 140 $\pm$ 120 &  \multicolumn{3}{c}{(2.6 $\pm$ 0.6)$\times$10$^{16}$} & 0.5 $\pm$ 0.2 & \phantom{0}6.5 $\pm$ 1.5 \\		
			SO$_2$ & 150 $\pm$ 30\phantom{1} &  \multicolumn{3}{c}{(8.1 $\pm$ 1.9)$\times$10$^{17}$} & 14.7 $\pm$ 3.6\phantom{0} & \phantom{.0}203 $\pm$ 47.5 \\
			\midrule
			CH$_3$OH & [120] & \multicolumn{3}{c}{[\phantom{0}(5.5 $\pm$ 0.4)$\times$10$^{18}$]\tablefootmark{a}} & $\equiv$100 & \phantom{0}1375 $\pm$ 100\phantom{0} \\
			$^{33}$SO$_2$ & [150] & \multicolumn{3}{c}{\phantom{0}(1.3 $\pm$ 0.8)$\times$10$^{16}$\tablefootmark{b}}  & 0.3 $\pm$ 0.2 & \phantom{0}3.3 $\pm$ 2.0 \\	
			$^{34}$SO$_2$ & [150] & \multicolumn{3}{c}{(8.1 $\pm$ 1.8)$\times$10$^{16}$} & 1.5 $\pm$ 0.4 & 20.3 $\pm$ 4.5 \\			
			SO$^{18}$O & [150] & \multicolumn{3}{c}{(7.6 $\pm$ 1.0)$\times$10$^{15}$} & 0.14 $\pm$ 0.02 & \phantom{0}1.9 $\pm$ 0.3 \\[0.2cm]
			& & \multicolumn{3}{c}{$T_{\textrm{ex}}$ [K]} & $T_{\textrm{ex}}$ [K] \\
			\cmidrule(l{2pt}r{2pt}){3-5}
			\cmidrule(l{2pt}r{2pt}){6-6}
			& & [130] & [200] & [270] & [200] \\
			\cmidrule(l{2pt}r{2pt}){3-5}
			\cmidrule(l{2pt}r{2pt}){6-6}
			H$_2$CCO & -- & 6.4$\times$10$^{15}$ & (9.2 $\pm$ 1.5)$\times$10$^{15}$ &  1.3$\times$10$^{16}$ & 0.17 $\pm$ 0.03 & \phantom{0}2.3 $\pm$ 0.4 \\	
			HNCO & -- & 5.5$\times$10$^{16}$  & (6.2 $\pm$ 1.3)$\times$10$^{16}$ & 7.8$\times$10$^{16}$ & 1.2 $\pm$ 0.3 & 15.5 $\pm$ 3.3 \\	
			NS & -- & 9.2$\times$10$^{15}$ & (1.2 $\pm$ 0.1)$\times$10$^{16}$ & 1.7$\times$10$^{16}$ & 0.22 $\pm$ 0.02 & \phantom{0}3.0 $\pm$ 0.3 \\		
			C$^{34}$S & -- & 7.2$\times$10$^{15}$ & (8.1 $\pm$ 1.6)$\times$10$^{15}$ & 1.0$\times$10$^{16}$ & 0.15 $\pm$ 0.03 & \phantom{0}2.1 $\pm$ 0.4 \\		
			t-HCOOH & -- & 2.6$\times$10$^{16}$ & (4.3 $\pm$ 0.6)$\times$10$^{16}$ & 5.5$\times$10$^{16}$ & 0.8 $\pm$ 0.2 & 10.8 $\pm$ 1.5 \\
			SO & -- & 6.2$\times$10$^{17}$ & (7.0 $\pm$ 1.5)$\times$10$^{17}$ & 7.8$\times$10$^{17}$ & 12.7 $\pm$ 2.9\phantom{0} & \phantom{0.}175 $\pm$ 37.5 \\			
			$^{34}$SO & -- & 1.4$\times$10$^{16}$ & (1.8 $\pm$ 0.5)$\times$10$^{16}$ & 2.1$\times$10$^{16}$ & 0.4 $\pm$ 0.1& \phantom{0}4.5 $\pm$ 1.3 \\	
			HC$_3$N, v=0 & --  & 1.0$\times$10$^{16}$ & (6.2 $\pm$ 2.2)$\times$10$^{15}$ & 5.5$\times$10$^{15}$ & 0.11 $\pm$ 0.04 & \phantom{0}1.6 $\pm$ 0.6 \\
			HC$_3$N, v$_7$=2 & -- & \phantom{0}1.4$\times$10$^{17}$\tablefootmark{c}  & \phantom{0}(2.5 $\pm$ 0.6)$\times$10$^{16}$\tablefootmark{c} & \phantom{0}1.2$\times$10$^{16}$\tablefootmark{c} & 0.5 $\pm$ 0.2 & \phantom{0}6.3 $\pm$ 1.5 \\
			HCC$^{13}$CN & -- & 4.3$\times$10$^{14}$ &  (3.8 $\pm$ 0.8)$\times$10$^{14}$ & 4.3$\times$10$^{14}$ & 0.007 $\pm$ 0.002 & \phantom{0}0.10 $\pm$ 0.02 \\
			HCCC$^{15}$N & -- & 2.1$\times$10$^{14}$ & (1.8 $\pm$ 0.4)$\times$10$^{14}$ & 1.8$\times$10$^{14}$ & 0.003 $\pm$ 0.001 & \phantom{0}0.05 $\pm$ 0.01 \\
			HCC$^{13}$CN, v$_7$=1 & -- & \phantom{0}1.9$\times$10$^{15}$\tablefootmark{d} & \phantom{0}(8.1 $\pm$ 1.2)$\times$10$^{14}$\tablefootmark{d} & \phantom{0}5.5$\times$10$^{14}$\tablefootmark{d} & 0.010 $\pm$ 0.002 & \phantom{0}0.21 $\pm$ 0.03 \\
			C$_2$H$_5$CN & -- & \phantom{0}5.6$\times$10$^{15}$\tablefootmark{b} & \phantom{0}(6.4 $\pm$ 0.8)$\times$10$^{15}$\tablefootmark{b} & \phantom{0}8.1$\times$10$^{15}$\tablefootmark{b} & 0.12 $\pm$ 0.02 & \phantom{0}1.6 $\pm$ 0.2 \\
			CH$_3$COCH$_3$ & -- & \phantom{0}2.3$\times$10$^{16}$\tablefootmark{b} & \phantom{0}(5.3 $\pm$ 0.5)$\times$10$^{16}$\tablefootmark{b} & \phantom{0}1.4$\times$10$^{17}$\tablefootmark{b} & 1.0 $\pm$ 0.2 & 13.3 $\pm$ 1.3 \\	
			CH$_2$(OH)CHO & -- & $\leq$8.4$\times$10$^{15}$\tablefootmark{e}\phantom{0}  & $\leq$1.1$\times$10$^{16}$\tablefootmark{e} & $\leq$1.5$\times$10$^{16}$\tablefootmark{e} \phantom{0} & $\leq$0.2 & $\leq$2.8 \\
			O$^{13}$CS & -- & 4.8$\times$10$^{15}$ & (5.5 $\pm$ 0.7)$\times$10$^{15}$ & 6.2$\times$10$^{15}$ & 0.10 $\pm$ 0.01 & \phantom{0}1.4 $\pm$ 0.2 \\				
			\bottomrule
	\end{tabular}
	\tablefoot{Values in square brackets are fixed. $N_{\textrm{s}}$ is the synthesised beam-averaged column density. \tablefoottext{a}{Based on $^{13}$CH$_3$OH, assuming a $^{12}$C/$^{13}$C ratio of 60.} \tablefoottext{b}{Column density derived assuming fixed $T_{\textrm{ex}}$ due to insufficient range of $E_{\textrm{up}}$ of unblended lines.} \tablefoottext{c}{Includes the vibrational correction factor of 1.17.} \tablefoottext{d}{Includes the vibrational correction factor of 1.48.}  \tablefoottext{e}{Upper limit due to low signal-to-noise.}}
\end{table*}

\begin{figure*}[]
	\centering
	\begin{subfigure}{1\textwidth}
		\centering
		\includegraphics[width=0.87\textwidth, trim={0 0 0 0}, clip]{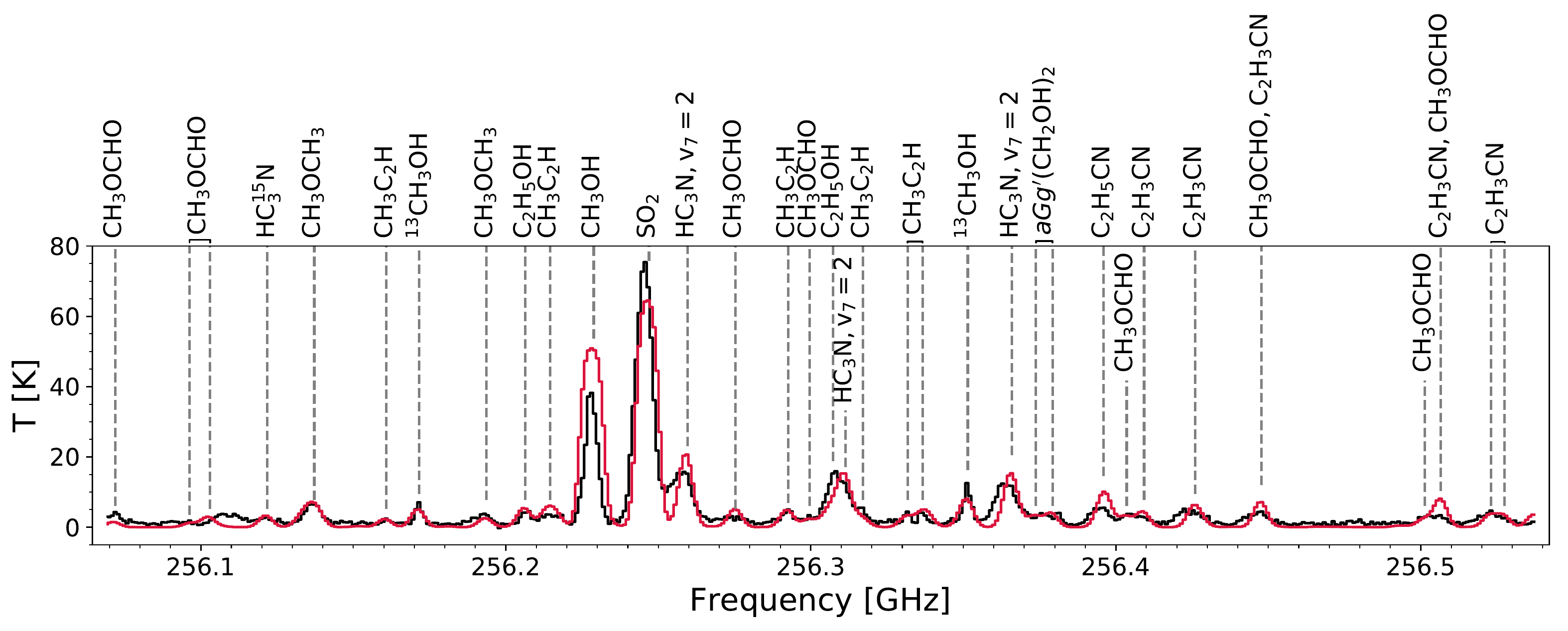}  
		\caption*{}
		\label{fig:spw0_sub}
	\end{subfigure}
	\begin{subfigure}{1\textwidth}
		\vspace{-0.7cm}
		\centering
		\includegraphics[width=0.87\textwidth, trim={0 0 0 0}, clip]{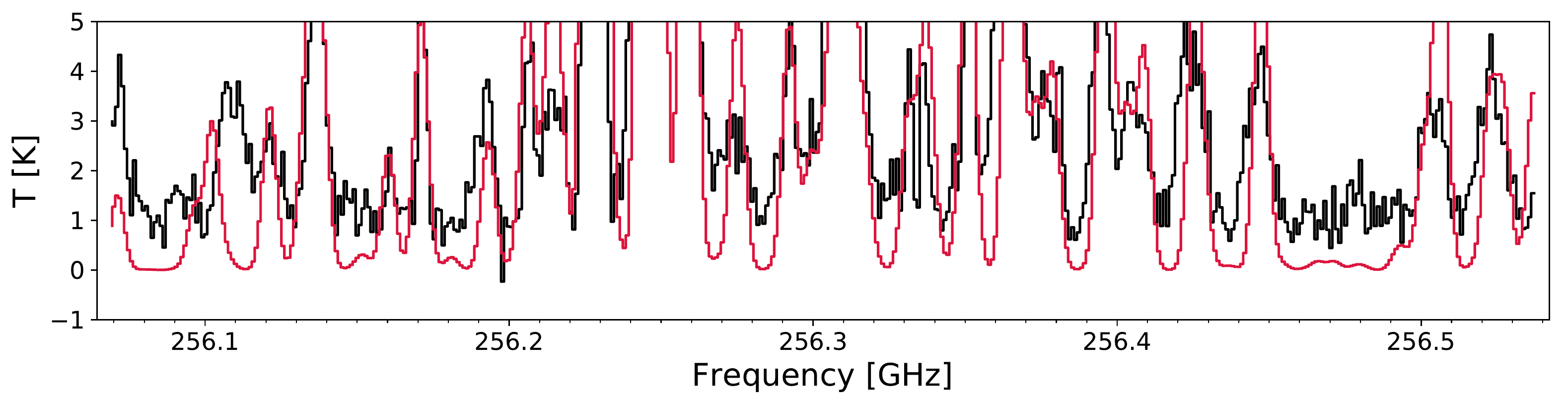}
		\caption*{}
		\label{fig:spw0_zoomed}
	\end{subfigure}
	\caption{\textit{Top panel:} Full model (red), i.e., the sum of synthetic spectra, for all species detected towards AFGL~4176 in the spectral window centred at 256.3 GHz. Frequencies are shifted to the systemic velocity of the region. The data are shown in black. \textit{Bottom panel:} Zoom-in of top panel to highlight weak lines.}
	\label{fig:spw0}
\end{figure*}

The derived excitation temperatures range between 120 and 320 K with an average of 200 K, consistent with the range of temperatures derived for CH$_3$CN by \citet{Johnston2015}. The highest column density is 5.5$\times$10$^{18}$ cm$^{-2}$, derived for CH$_3$OH (based on $^{13}$CH$_3$OH and corrected with the $^{12}$C/$^{13}$C ratio), while the lowest values is 1.8$\times$10$^{14}$ cm$^{-2}$, derived for HC$_3^{15}$N. While these column densities span more than four orders of magnitude, the majority of species have column densities between 10$^{16}$ and 10$^{17}$ cm$^{-2}$. The species with the highest number of detected lines, 41 in total, is CH$_3$OH. These lines also span the largest range of upper state energies ranging from 20 to 950 K. Similarly, the range of upper state energies of the $^{13}$C-isotopologue of methanol span 60 to 640 K, though only twelve lines associated with this isotopologue are detected. For the remaining species, the number of detected lines range from a single line up to 30 lines in the case of \textit{aGg'}(CH$_2$OH)$_2$, though it should be noted that more than half of these are blended with emission from other species. On average the upper state energies of species span a range of 300 K.  

The column density calculation above assumes that the dust is optically thin. We can test if this is indeed the case for AFGL~4176, by converting the continuum intensity into a brightness temperature $T_{\rm B}$ = 5.7 K. If we compare this to the excitation temperature of the gas ($\sim$200 K), we find that $T_{\rm B}$ is much less than the physical temperature of the material. Therefore, the optical depth is likely low. We therefore conclude that, averaged over the beam, continuum opacity is negligible. Only if the emitting material is distributed over a region with a size six times smaller than the beam, would dust opacity become a factor. The effect would be that part of the gas is ‘hidden’ by the dust, and real column densities are higher; however, since all lines would be equally affected, the ratios of species are unaffected.

By far the highest column density ratios with respect to methanol are derived for SO$_2$ and SO with values of 14.7$\%$ and 12.7$\%$, respectively. The remaining species have column density ratios with respect to methanol ranging between 0.003$\%$ and 3$\%$, with most species showing ratios of the order of 0.1$\%$. On average, the column density ratios derived for O-bearing species are a factor of three higher than those derived for N-bearing species. This trend will be further discussed in sect. \ref{sec:discussion}.

Vibrationally excited transitions are detected for two species, CH$_3$CN and HC$_3$N. In both cases, the column density ratio derived from the vibrationally excited transitions are higher than those derived for the ground-state vibrational transitions; 0.8\% vs 0.6\% for CH$_3$CN/CH$_3$OH and 0.5\% vs 0.1\% for HC$_3$N/CH$_3$OH. Note that for the latter species only one and three lines are detected for the ground and vibrationally excited states respectively. That the column densities derived based on the vibrationally excited states are higher than those derived from the ground-state vibrational transitions is most likely not representative of the actual distribution of molecules but rather due to the fact that the vibrationally excited transitions are excited via shocks or infrared pumping and therefore not in LTE. The column densities derived from these transitions can therefore not be trusted.   

\subsection{Upper limit on the column density of glycolaldehyde}
Glycolaldehyde (CH$_2$(OH)CHO) is of prebiotic interest because of its structural similarities with sugars and the fact that it therefore could be at the basis of the formation of more complex sugar compounds, such as ribose. It was first detected in the interstellar medium towards Sgr~B2(N) \citep{Hollis2000} and subsequently towards various high- and low-mass hot cores \citep[e.g.][]{Beltran2009,Jorgensen2012,Coutens2015,Taquet2015a,Jorgensen2016}.

A number of transitions of glycolaldehyde are covered by the data although only three of these lines, one of which is blended, are considered detected. These lines have signal-to-noise ratios of approximately four. The remaining lines identified as likely to be due to glycolaldehyde, five in total, are detected with signal-to-noise ratios of between two and three and are therefore not included in the line list in Table \ref{tab:line_summary}. Due to the generally low signal-to-noise of the glycolaldehyde lines, we report an upper limit column density for this species. The upper limit is derived based on the two unblended lines and assumes an excitation temperature of 200 K. The column density upper limit is $\leq$1.1$\times$10$^{16}$ cm$^{-2}$, equivalent to $\leq$0.2\% with respect to methanol. 

The formation of glycolaldehyde has been investigated both in the laboratory and with chemical models \citep{Bennett2007,Woods2012,Woods2013}. Recently, a laboratory study by \cite{Chuang2017} found that the relative abundance of this and other complex species, can be used as a diagnostic tool to derive the processing history of the ice in which the species formed. In particular, the ratio of glycolaldehyde to ethylene glycol provides a good handle on distinguishing ices processed purely by atom-addition (hydrogenation), ices processed purely by UV irradiation or ice processed by both. In the case of AFGL~4176, the ratio of glycolaldehyde to ethylene glycol is $\leq$0.4. This ratio is consistent with the relative abundance of the species formed in experiments where ice analogues are exposed to both UV irradiation and hydrogenation.

At the same time, observational studies have shown a trend in the glycolaldehyde to ethylene glycol ratio based on source luminosities. \citet{Rivilla2017} found that this ratio decreases with source luminosity, with ratios $\lessapprox$0.1 for high-mass sources with a luminosity similar to AFGL~4176. Of course, the ratio found in AFGL~4176 is an upper limit and the actual ratio can thus either follow this trend or deviate from it. Apart from the parameters listed above, also density can affect the glycolaldehyde to ethylene glycol ratio \citep{Coutens2018}.

\subsection{Isotopologues with only blended lines}
Three isotopologues of HC$_3$N and one of CH$_3$CN are detected towards AFGL~4176, although only through blended lines. For completeness, these isotopologues are included in the full model. Since no column density and excitation temperature can be derived from the blended lines these are adopted from the isotopologues for which unblended lines are detected. That is, for HC$^{13}$CCN the column density derived for HCC$^{13}$CN is adopted, while for vibrationally excited H$^{13}$CCCN and HC$^{13}$CCN the column density derived for vibrationally excited HCC$^{13}$CN is adopted. In the case of CH$_3^{13}$CN, the column density derived for the main isotopologue has been corrected by the $^{12}$C/$^{13}$C ratio of 60. The isotopologues and the adopted column densities and excitation temperatures are listed in Table \ref{tab:only_blended}. It should be noted that no lines of either H$^{13}$CCCN, $^{13}$CH$_3$CN or CH$_3$C$^{15}$N, were covered by the data. A couple of transitions of vibrationally excited CH$_3^{13}$CN are within the data range but since these are all weak and highly blended they could not be modelled. 

\subsection{Isotope ratios} \label{subsec:isotope_ratios}
From the three detected isotopologues of HC$_3$N, it is possible to derive $^{12}$C/$^{13}$C and $^{14}$N/$^{15}$N isotope ratios. These are found to be $\sim$16 and $\sim$34, respectively. At the galactocentric distance of AFGL~4176 ($d_{\textrm{GC}}$ = 6.64 kpc), $^{12}$C/$^{13}$C = 60  and $^{14}$N/$^{15}$N = 399 according to \citet{Milam2005} and \citet{Colzi2018a}, respectively. Therefore both isotope ratios found in AFGL~4176 are significantly lower. However, it should be noted that all three HC$_{3}$N isotopologues are tentative detections and more importantly that the detected line belonging to the main HC$_{3}$N isotopologue is optically thick; both issues could cause a severe isotope ratio deviation. Furthermore, isotope fractionation could be the result of specific reactions HC$_{3}$N is involved in. Clearly, isotope ratios need to be determined from addition molecules (or other lines of HC$_{3}$N and its isotopologues) in order to verify or disprove the deviation from the galactocentric trends found in this work.

\begin{table*}[]
\begin{small}	
	\centering
	\caption{Summary of isotopologues with only blended lines.}
	\label{tab:only_blended}
	\begin{tabular}{lcccccccc}
		\toprule
		Species & \multicolumn{3}{c}{N$_{\textrm{lines}}$} & $E_{\textrm{up}}$ & $A_{\textrm{ij}}$ & Catalogue & $N_{\textrm{s}}$ & $T_{\textrm{ex}}$ \\
		\cline{2-4}
		& Unblended & Blended & Total & [K] & $\times$10$^{-5}$ [s$^{-1}$] & & [cm$^{-2}$] & [K] \\
		\midrule
		CH$_3^{13}$CN & 0 & 6 & 6 & \phantom{0}80 -- 259 & 100 -- 118 & JPL & [5.7$\times$10$^{14}$]\tablefootmark{a} & [270] \\
		HC$^{13}$CCN & 0 & 1 & 1 & 177 & 130 & CDMS & [3.8$\times$10$^{14}$]\phantom{a} & [200] \\
		H$^{13}$CCCN, v$_7$=1 & 0 & 2 & 2 & 479 -- 503 & 108 -- 133 & CDMS & [8.1$\times$10$^{14}$]\phantom{a} & [200] \\	
		HC$^{13}$CCN, v$_7$=1 & 0 & 2 & 2 & 493 & 130 -- 131 & CDMS & [8.1$\times$10$^{14}$]\phantom{a}& [200] \\								
		\bottomrule
	\end{tabular}
	\tablefoot{Values in square brackets are fixed. \tablefoottext{a}{Based on CH$_3$CN, assuming a $^{12}$C/$^{13}$C ratio of 60.}}
\end{small}
\end{table*}

\subsection{Spatial distribution of selected species} 
Two line maps are produced for each of the species for which five or more unblended lines are detected. The imaged lines are chosen so that both high and low upper state energy transitions are represented, in order to investigate whether these occupy different spatial regions. Also, only lines which are relatively isolated, that is to say whose peak frequency is separated by at least 3 km~s$^{-1}$ from neighbouring peaks, are imaged. This is done in order to minimise line confusion. After suitable lines have been identified for each species, the zero- and first-moment maps, that is the velocity integrated intensity map and intensity-weighted velocity map, respectively, are produced. The spatial extent of each species is determined by fitting a 2D Gaussian to the zero-moment maps (fit parameters are listed in Table \ref{tab:spatial_extent}). As a representative sample of these maps, the lines of CH$_3$OH, NH$_2$CHO and CH$_3$C$_2$H are shown in Fig. \ref{fig:moment_maps}; maps for the remaining species are presented in Appendix \ref{app:line_maps}. 

There are no large differences between the spatial distribution of O- and N-bearing species. Except for CH$_3$C$_2$H, all species have emission peaks near the position of the continuum peak. The N-bearing species (CH$_3$CN, C$_2$H$_3$CN and C$_2$H$_5$CN) peak very close to the continuum peak, while some O-bearing species (e.g. CH$_3$OH and CH$_3$OCH$_3$) peak up to 0$\overset{\second}{.}$2 away from the continuum peak. Although this scale is of the same order as the size of the synthesised beam, the signal-to-noise of these maps (30 -- 90) is large enough to make these spatial differences significant. 

\begin{figure}[]
	\centering
	\includegraphics[width=0.5\textwidth, trim={0 0 0 0}, clip]{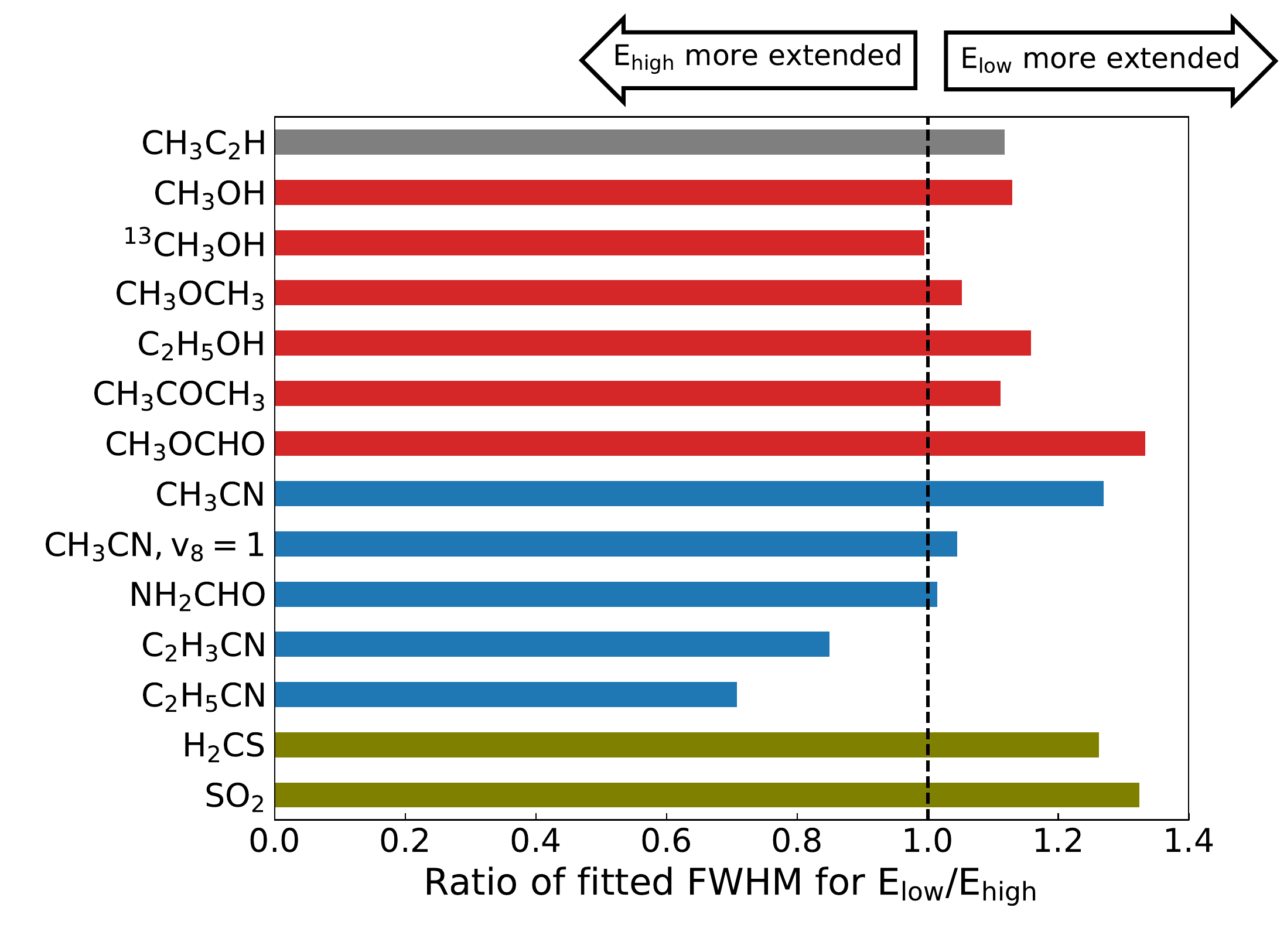}  
	\caption{Ratio between the fitted FWHM of the low and high upper state energy transitions listed in Table \ref{tab:spatial_extent}. A ratio lager than 1 indicates that the spatial extent of the low upper state energy transition is larger than that of the high upper state energy transition.}
	\label{fig:e_histogram}
\end{figure}

Noticeable differences in spatial distributions exist between transitions of the same species with low and high upper state energies. These differences are illustrated in Fig. \ref{fig:e_histogram} where the ratio between the spatial extent (as measured by the fitted FWHM) of the low and high upper state energy transitions are plotted. In this figure, a ratio above 1 indicates that the spatial extent of the low $E_{\textrm{up}}$ transition is larger than that of the high $E_{\textrm{up}}$ transition, while a ratio below 1 indicates that the spatial extent of the high $E_{\textrm{up}}$ transition is larger than that of the low $E_{\textrm{up}}$ transition. For the majority of the imaged species, the spatial extent of the low upper state energy transition is larger than that of the high upper state transition. This trends is especially clear in the case of the S-bearing species H$_2$CS and SO$_2$, the O-bearing species CH$_3$OCHO and the N-bearing species CH$_3$CN. For these species, the spatial extent of the low upper state energy transition is $\sim$30\% larger than that of the high $E_{\textrm{up}}$ transition. The larger spatial extent of the low $E_{\textrm{up}}$ transitions indicates that these species are present in colder gas. This is consistent with the relatively low excitation temperatures derived for SO$_2$ and H$_2$CS of 150 K and 160 K, respectively, but contradicts the high excitation temperatures derived for CH$_3$CN and CH$_3$OCHO of 270 K and 310 K, respectively. For C$_2$H$_3$CN and C$_2$H$_5$CN the trend is reversed, with the spatial extent of low $E_{\textrm{up}}$ transitions smaller than that of high $E_{\textrm{up}}$ transitions by up to 40\% (as measured by the fitted FWHM). The large differences between the spatial extent of low and high $E_{\textrm{up}}$ transitions seen in the case of H$_2$CS, SO$_2$, CH$_3$OCHO and CH$_3$CN indicate that these species trace both the warm central region and the cooler outer region of the hot core, while the majority of the remaining complex organic molecules (e.g. CH$_3$OH, CH$_3$OCH$_3$, C$_2$H$_5$OH, CH$_3$COCH$_3$ and NH$_2$CHO) are likely only excited in the central parts of the core since these species show only limited differences between the low and high $E_{\textrm{up}}$ transitions. Finally, as noted above, CH$_3$C$_2$H is the only species whose emission is not concentrated at the location of the continuum peak emission. Instead this species shows two peaks, of approximately similar intensity, with one roughly coinciding with that of the continuum and a second at the location of the secondary continuum peak, mm2. The spatially more diffuse emission of this species is consistent with trends observed towards other hot cores \citep[see e.g.][]{Fayolle2015}. This indicates that a cold gas-phase formation mechanism likely dominates the formation of CH$_3$C$_2$H.

For a number of species a velocity gradient is detected across the source. The gradient is most pronounced in the case of NH$_2$CHO but also visible in the high upper state energy lines of CH$_3$CN, CH$_3$OCHO, C$_2$H$_3$CN and C$_2$H$_5$CN. The presence of a velocity gradient across the sources is consistent with the result of \citet{Johnston2015} who model the emission of CH$_3$CN and find this to be consistent with a Keplerian-like disk. A few species, namely CH$_{3}$OH, C$_{2}$H$_{5}$OH, CH$_{3}$OCH$_{3}$ and H$_{2}$CS, seem to have a velocity gradient that differs from CH$_{3}$CN.  

\begin{figure*}[]
	\centering
	\begin{subfigure}{.8\textwidth}
		\centering
		\includegraphics[width=1\textwidth, trim={0 0 0 0}, clip]{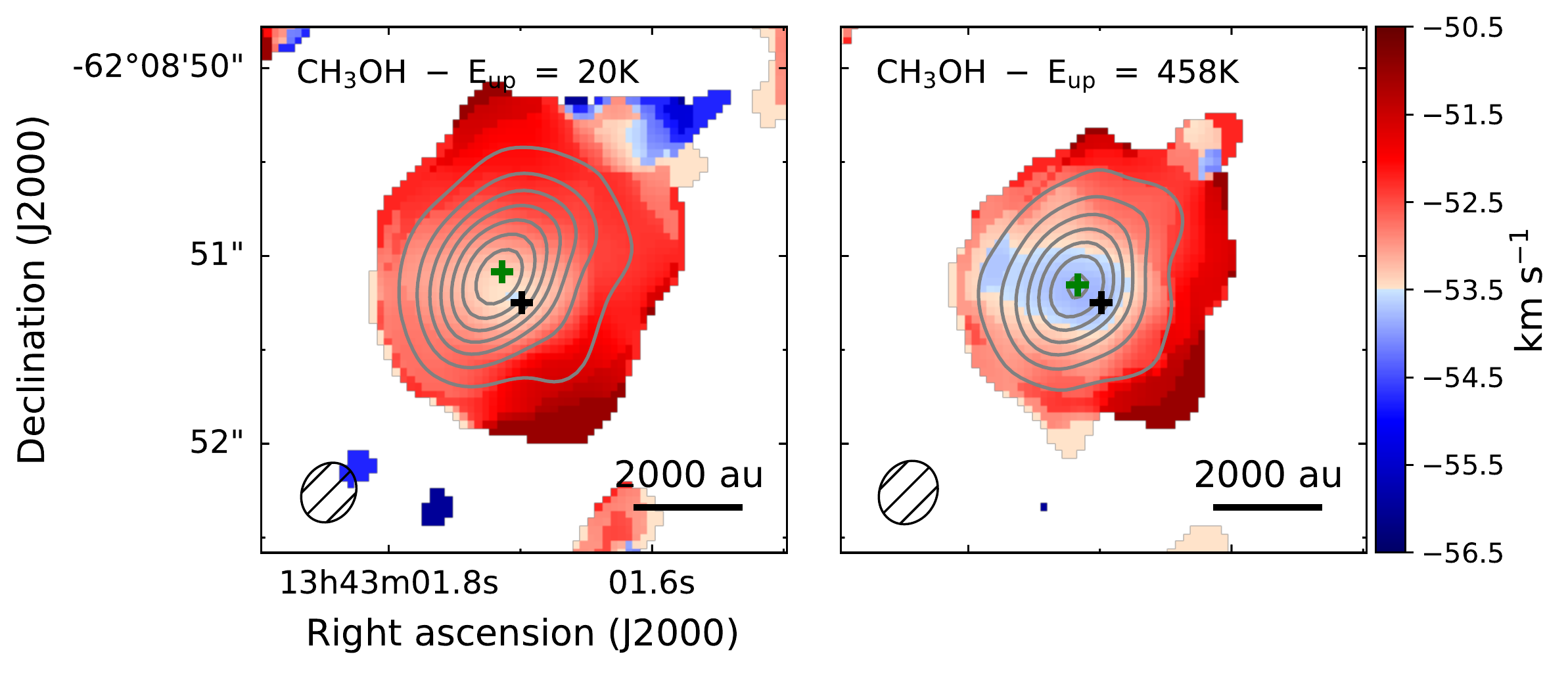}
		\caption*{}
	\end{subfigure}
	\begin{subfigure}{.8\textwidth}
		\centering
		\includegraphics[width=1\textwidth, trim={0 0 0 0}, clip]{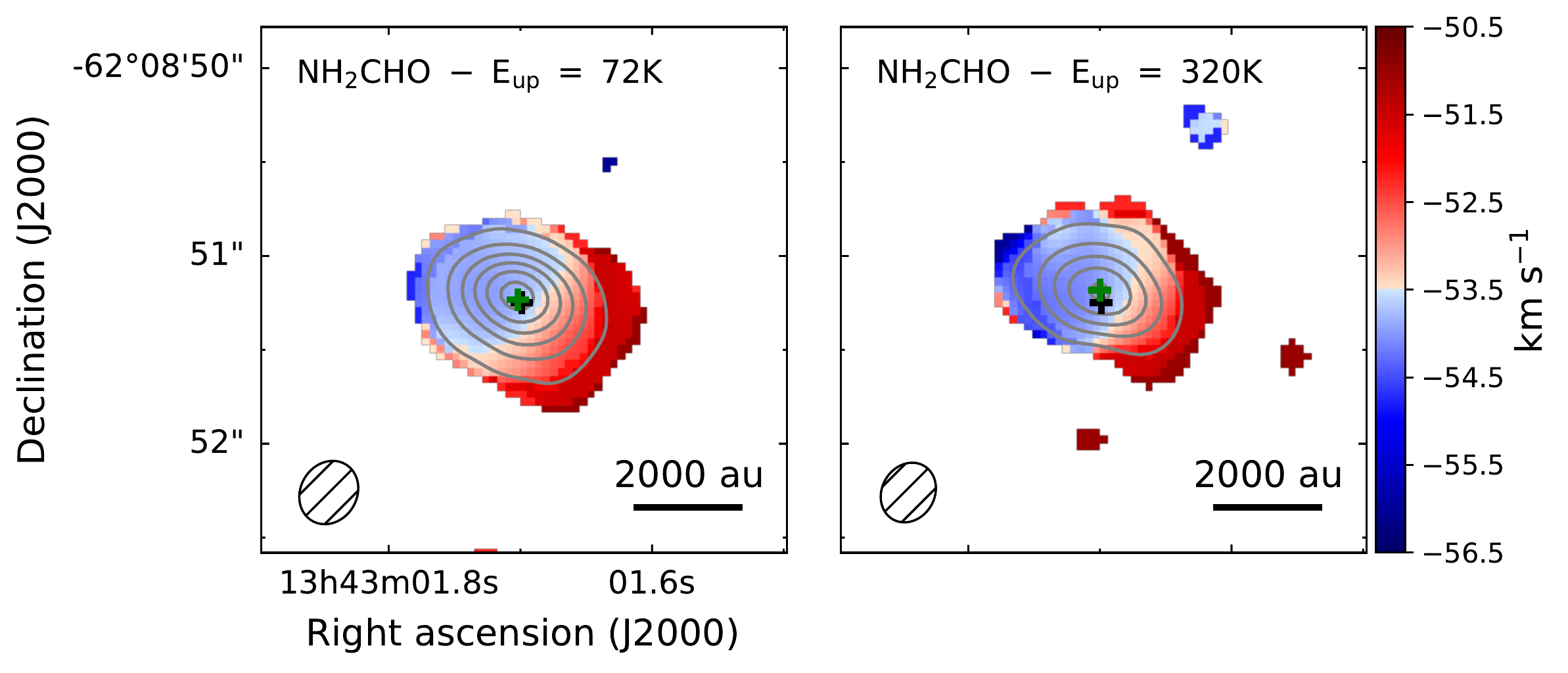}
		\caption*{}
	\end{subfigure}
	\begin{subfigure}{.8\textwidth}
		\centering
		\includegraphics[width=1\textwidth, trim={0 0 0 0}, clip]{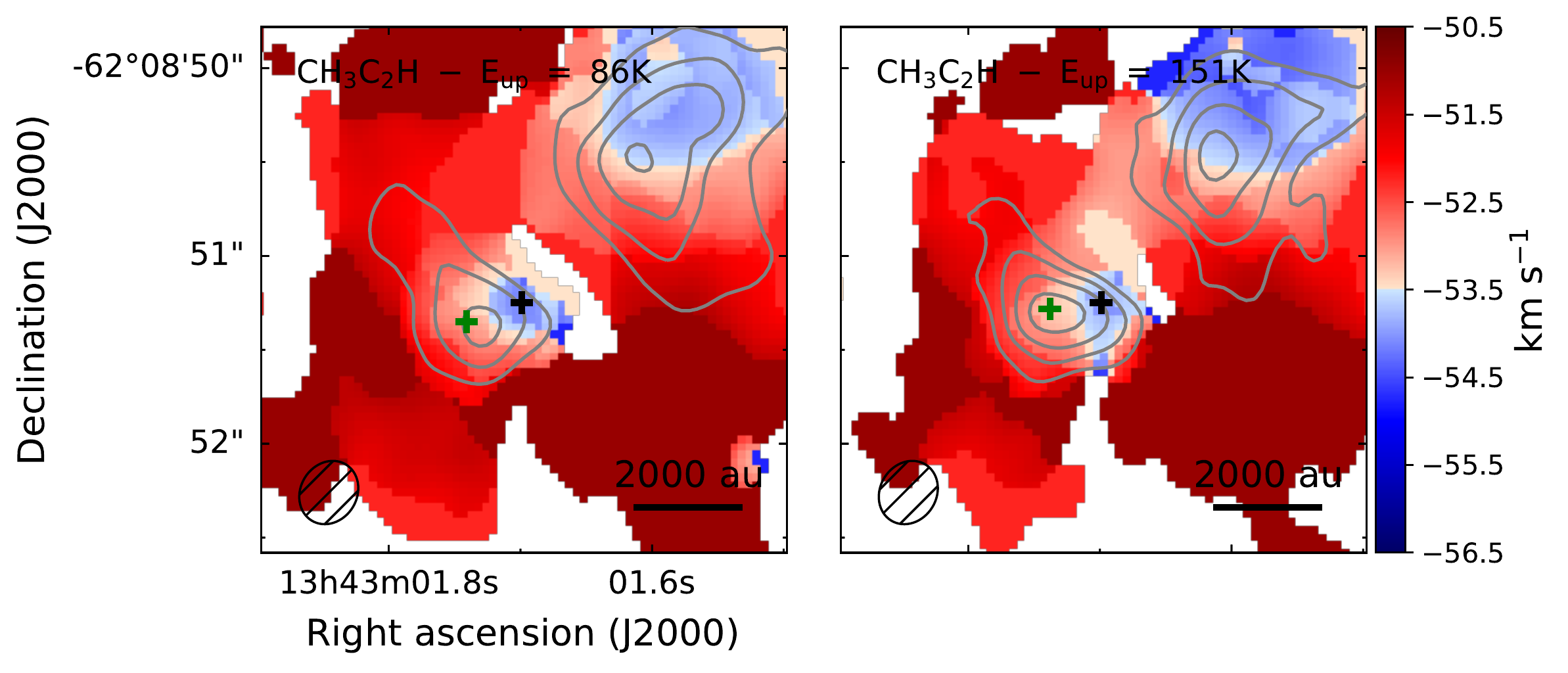}
		\caption*{}
	\end{subfigure}
	\caption{\textit{Top panels:} First-moment (intensity-weighted velocity) map of CH$_3$OH lines at 254.0153~GHz (left) and 241.2679~GHz (right). Pixels with signal-to-noise ratios of less than three are masked out. The zero-moment (integrated intensity) map for each line is overlaid in grey contours. Contours start at 9$\sigma$ and are in steps of 12$\sigma$, with $\sigma$ = 1.34$\times$10$^{-2}$ and 7.16$\times$10$^{-3}$ Jy~beam$^{-1}$~km~s$^{-1}$, for the left and right panel, respectively. The black and green crosses mark the locations of the peak continuum emission and peak integrated line intensity, respectively. \textit{Middle panels:} Same as top panels but for NH$_2$CHO lines at 239.952~GHz (left) and 254.727~GHz (right). Contours start at 9$\sigma$ and are in steps of 12$\sigma$, with $\sigma$ = 6.17$\times$10$^{-3}$ and 6.25$\times$10$^{-3}$ Jy~beam$^{-1}$~km~s$^{-1}$, for the left and right panel, respectively. \textit{Bottom panels:} Same as top panels but for CH$_3$C$_2$H lines at 239.2523~GHz (left) and 239.2112~GHz (right). Contours start at 6$\sigma$ and are in steps of 3$\sigma$, with $\sigma$ = 7.29$\times$10$^{-3}$ and 6.35$\times$10$^{-3}$ Jy~beam$^{-1}$~km~s$^{-1}$, for the left and right panel, respectively.}
	\label{fig:moment_maps}
\end{figure*}

\section{Discussion} \label{sec:discussion}
Table \ref{tab:O_and_N-bearing} presents an overview of all detected O-, N- and S-bearing species (isotopologues not included) towards AFGL~4176. These detected species are common in regions of star-formation \citep[see review by][and references therein]{Herbst2009}.

On average, the column density ratios with respect to methanol derived for the O-bearing species are a factor of three higher than the ratios derived for the N-bearing species. Also, the excitation temperatures derived for the O-bearing species are generally low, 120~--~160 K, while the excitation temperatures derived for the N-bearing species are generally high, 190~--~240 K. This differentiation of species with excitation temperature is similar to trends observed in the Orion molecular cloud \citep{Blake1987, Crockett2015} and in the high-mass star-forming complex G19.62-0.23 \citep{Qin2010}. Similar trends are also reported by \citet{Suzuki2018} who carry out a survey of N- and O-bearing species towards eight high-mass star forming regions, including Orion KL and G19.62-0.23, using the 45m radio telescope at the Nobeyama Radio Observatory. When comparing these observations with chemical models, \citet{Suzuki2018} conclude that the correlations between fractional abundances of different groups of species can be explained by a combination of different temperature structures inside the cores and different evolutionary phases of the studied regions. As examples of a younger, less evolved source and an older, more evolved source, \citet{Suzuki2018} discuss NGC 6334F (also known as NGC 6334I) and G10.47+0.03, respectively. While G10.47+0.03 displays a relatively high fractional abundance of N-bearing species, $\sim$2--10$\%$ with respect to methanol, the fractional abundances of the same species detected towards NGC~6334F are only $\sim$0.1$\%$. By assuming a more dominant high-temperature region ($\sim$200 K) and later evolutionary stage for G10.47+0.03 with respect to NGC 6334F, \citet{Suzuki2018} reproduce the observed trends. The trend of younger regions being characterised by lower abundances of N-bearing species with respect to O-bearing species may be a consequence of gas-phase nitrogen chemistry taking longer to initiate compared with the chemistry of O-bearing species \citep{Charnley1992}. 

\begin{table}[]
	\centering
	\caption{Overview of oxygen-, nitrogen and sulphur-bearing species detected towards AFGL~4176.}
	\label{tab:O_and_N-bearing}
	\begin{tabular}{lll}
		\toprule
		\textbf{O-bearing} & \textbf{N-bearing} & \textbf{S-bearing} \\
		\midrule
		CH$_3$OH & CH$_3$CN &   H$_2$CS\\
		\cmidrule{3-3}
		CH$_3$CHO & C$_2$H$_3$CN & \textbf{ O, N-bearing}  \\
		\cmidrule{3-3}
		C$_2$H$_5$OH & C$_2$H$_5$CN & NH$_2$CHO  \\	
		CH$_3$OCH$_3$ & \textcolor{red}{HC$_3$N} & \textcolor{red}{HNCO} \\
		\cmidrule{2-3}
		CH$_3$COCH$_3$ & \textbf{Hydrocarbons}  & \textbf{O, S-bearing }  \\
		\cmidrule{2-3}
		CH$_3$OCHO & CH$_3$C$_2$H & SO$_2$\\
		(CH$_2$OH)$_2$ &  & \textcolor{red}{SO} \\
		\cmidrule{3-3}
		\textcolor{red}{H$_2$CCO}  &  &  \textbf{N, S-bearing} \\
			\cmidrule{3-3}
		\textcolor{red}{t-HCOOH} &  & NS  \\
		\bottomrule
	\end{tabular}
	\tablefoot{Not including isotopologues. Species in red have fewer than five detected transitions and are considered tentative.} 
\end{table}

The studies discussed above are primarily based on single-dish observations and therefore, mostly, spatially unresolved. Whereas single-dish telescopes often cover both the inner hot envelope around the centrally forming star and the emission of its cooler outer envelope, interferometric observations are able to filter out the extended emission and focus solely on the hot core. Also, since single-dish telescopes are generally less sensitive when compared with interferometric observations and especially ALMA, it may not be possible to identify enough optically thin lines from which excitation conditions can be derived. The generally larger beam sizes of single-dish telescopes may also result in underestimated column densities of molecular species if effects of beam dilution are not accounted for correctly. 

\begin{figure*}[]
	\centering
	\begin{subfigure}{0.8\textwidth}
		\centering
		\includegraphics[width=1\textwidth, trim={0 0 0 0}, clip]{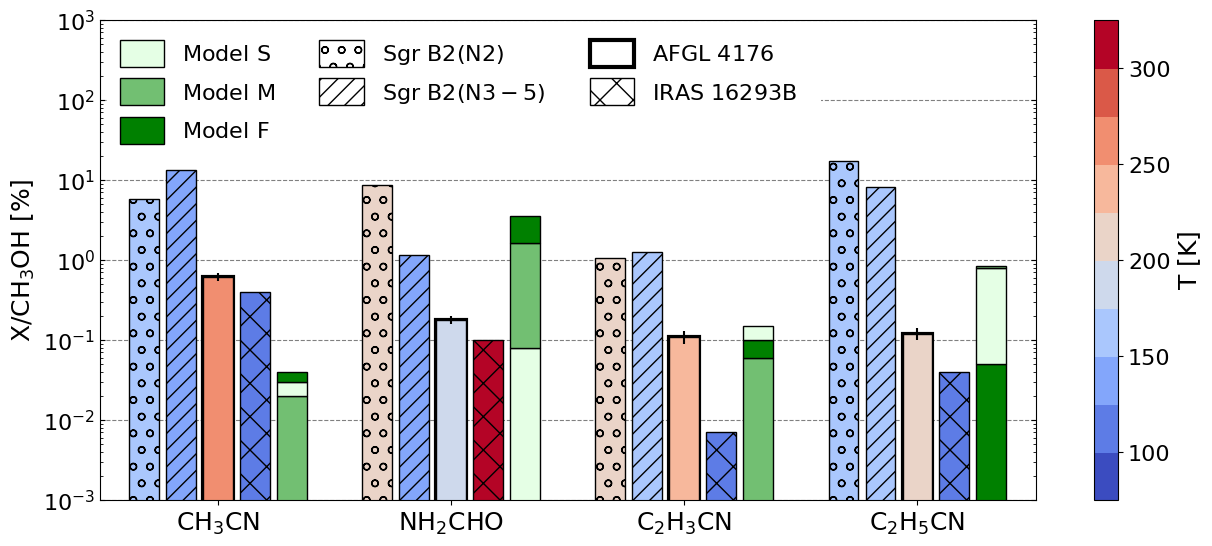}
		\caption*{}
	\end{subfigure}
	\begin{subfigure}{0.8\textwidth}
		\centering
		\includegraphics[width=1\textwidth, trim={0 0 0 0}, clip]{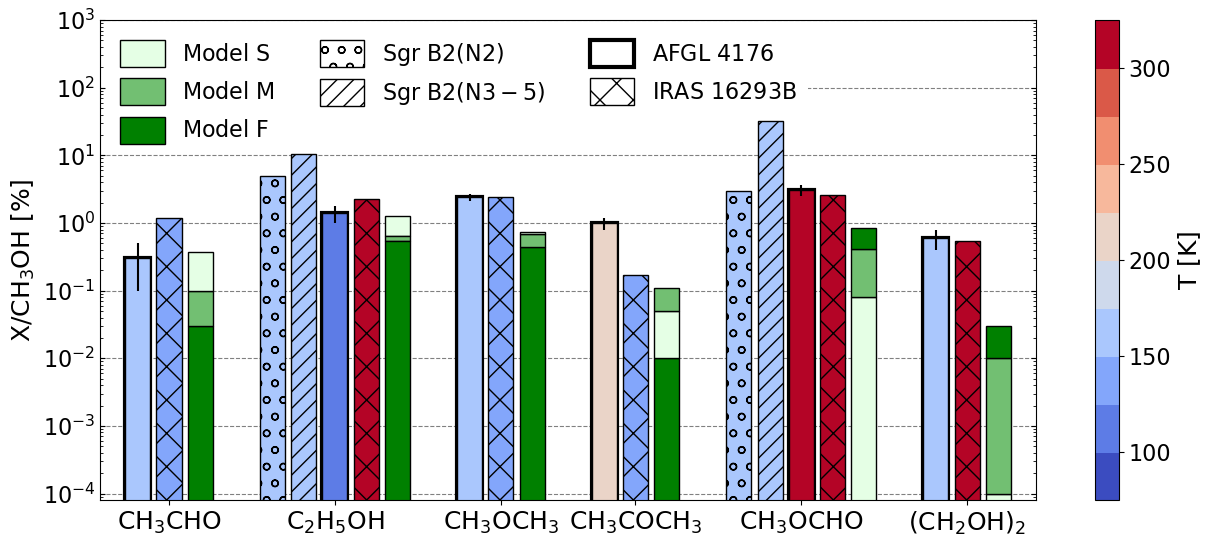}
		\caption*{}
	\end{subfigure}
	\caption{\textit{Top panel:} Relative abundances of N-bearing species predicted by models and detected towards AFGL~4176, Sgr B2(N) and IRAS~16293B. For AFGL~4176, Sgr B2(N) and IRAS~16293B the colours of bars indicate the excitation temperature derived for each species. \textit{Bottom panel:} Same as top panel but for O-bearing species.}
	\label{fig:histrogram}
\end{figure*}

In the following, we compare our ALMA results for AFGL~4176 with a selection of ALMA studies, that suffer less from sensitivity and resolution limitations, but do cover sources located at a range of different distances, therefore sampling different spatial scales, ranging from $\sim$70 to $\sim$13300 au. These studies focus on Sgr B2(N), Orion KL and the low-mass protostellar binary IRAS~16293. We conclude the section with a comparison to chemical models. Table \ref{tab:summary_of_column_density_ratios} and Fig. \ref{fig:histrogram} summarises the comparisons.

\subsection{Comparison with the high-mass star-forming regions in Sgr B2(N) and Orion KL}
\textbf{Sgr B2(N):} Located in the Galactic central region, the Sgr B2 molecular cloud hosts some of the most active sites of high-mass star-formation in the galaxy. One of these sites, Sgr B2~(N), is the subject of the ALMA line survey EMoCA \citep[Exploring Molecular Complexity with ALMA,][]{Belloche2016} aimed at characterising the molecular content of the region. Due to the high spatial resolution of the observations, $\sim$1.6$^{\second}$, probing scales down to 0.06 pc ($\sim$13300 au assuming a distance of 8.34 kpc), \citet{Bonfand2017} were able to identify three new hot cores towards Sgr B2, labelled N3, N4 and N5, in addition to the previously identified cores N1 and N2. \citet{Bonfand2017} find that the chemical composition of these new cores are very similar to each other and very different from that of the N2 core. Derived C$_{2}$H$_{3}$CN / C$_{2}$H$_{5}$CN and CH$_{3}$CN / C$_{2}$H$_{5}$CN ratios suggest that the N2 core is chemically less evolved than the three new cores.

For the hot cores in Sgr B2, the column density ratio with respect to methanol of N-bearing species (CH$_3$CN, NH$_2$CHO, C$_2$H$_3$CN and C$_2$H$_5$CN) are higher than those derived towards AFGL~4176 by up to two orders of magnitude. For the O-bearing species (C$_2$H$_5$OH and CH$_3$OCHO), the variations are smaller, though still up to an order of magnitude higher in Sgr B2 compared with AFGL~4176
 
In addition to the main isotopologues, a number of $^{13}$C- and $^{15}$N-isotopologues, as well as deuterated and vibrationally excited species, have been detected towards Sgr B2(N2) \citep{Belloche2016}. No deuterated species are detected towards AFGL~4176, despite a number of strong transitions belonging to deuterated molecules, primarily DC$_{3}$N and deuterated NH$_{2}$CHO, being covered by the spectra. A number of lines of $^{13}$C-cyanoacetylene are detected in addition to $^{13}$CH$_3$OH and some blended lines of CH$_3^{13}$CN. $^{13}$C doubly substituted cyanoacetylene isotopologues are also detected towards Sgr B2(N2). Towards AFGL~4176, two transitions of $^{13}$C doubly substituted cyanoacetylene are covered but both are weak and highly blended. The tentative detection of HC$_3^{15}$N towards Sgr B2(N2) results in a ratio with respect to methanol identical to the value derived for AFGL 4176. It should be noted however, that only a very limited number of lines of these isotopologues are detected and therefore their detection is considered tentative and their ratios should be seen as indicative of trends rather than definite values.  
 
\textbf{Orion KL:} Due to its proximity, $\sim$414 pc from the Sun \citep{Menten2007}, the high-mass star-forming regions associated with the Orion molecular clouds are some of the most studied. Using ALMA observations, the morphology and molecular composition of the region was recently studied by \citet{Pagani2017} who, thanks to the high angular resolution of their data of 1.7$\arcsec$, probing scales of $\sim$700~au, were able to separate the region into a number of components, including the hot core, plateau and extended ridge, but also a variety of molecular clumps, and report a complex velocity structure. However, due to the lack of zero-spacing data to recover the extended emission of many species, \citet{Pagani2017} do not derive column densities or excitation temperatures for the detected species and limit their analysis to line identification and determination of line velocity and line widths. Orion KL may therefore be qualitatively compared with AFGL~4176 though no quantitative comparison is possible. 

With the exception of CH$_3$C$_2$H and NS, all species detected towards AFGL~4176 are also detected towards Orion KL by \citet{Pagani2017}, including vibrationally excited HC$_3$N \citep[also investigated by][]{Peng2017} and its $^{13}$C singly substituted isotopologues. The highest energy vibrationally excited state is the $v_6$ = $v_7$ = 1 state, detected towards the Orion KL hot core region. In contrast, only the first two vibration states (HC$_3$N, $v_7$=2 and H$^{13}$CCCN, $v_7$=1, HC$^{13}$CCN, $v_7$=1 and HCC$^{13}$CN, $v_7$=1 ) were tentatively detected towards AFGL~4176.

\citet{Tercero2018} also investigate the inventory of complex molecules towards Orion KL, this timed focussing on O-bearing species. With a resolution of $\sim$1.5$\arcsec$, they prope spatial scales of $\sim$620 au. \citet{Tercero2018} derive molecular abundances at three locations. These are selected based on where each of the species methylformate, ethylene glycol and ethanol peak. When comparing the relative abundances with respect to CH$_{3}$OH for C$_{2}$H$_{5}$OH, CH$_{3}$OCH$_{3}$, CH$_{3}$COCH$_{3}$, CH$_{3}$OCHO, and (CH$_{2}$OH)$_{2}$,  we find that the relative abundances of CH$_{3}$OCH$_{3}$ and CH$_{3}$OCHO are consistently lower in AFGL~4176 (1.4 -- 3.9 and 1.5 -- 7.2 times lower, respectively) than in Orion KL. However, the relative abundances of CH$_{3}$COCH$_{3}$ and (CH$_{2}$OH)$_{2}$ are consistently higher in AFGL~4176 compared with Orion KL (6.7 -- 20 and 1.2 -- 6 times higher, respectively), while C$_{2}$H$_{5}$OH has a roughly equal relative abundance in the two sources. What causes the enhancement of certain species remains unclear.
 
As for AFGL~4176, both conformers of ethylene glycol are detected towards Orion KL \citep{Favre2017}. This is interesting since previously only the more stable of the two, \textit{aGg'}(CH$_2$OH)$_2$, was detected towards high-mass star-forming regions \citep[see e.g.][and references therein]{Lykke2015, Brouillet2015,Rivilla2017}, while the \textit{gGg'}(CH$_2$OH)$_2$ conformer was only detected towards the low-mass system IRAS~16293 \citep{Jorgensen2016}. The ratio between the \textit{aGg'} and \textit{gGg'} ethylene glycol conformers is 1.2 in AFGL~4176, within the errors of the value of 1.1 derived for IRAS 16293, and half the values of 2.3 and 2.5 derived for the 5 and 8 km s$^{-1}$ components of Orion~KL respectively \citep{Favre2017}. 

\subsection{Comparison with the low-mass protobinary IRAS~16293--2422}
The low-mass protobinary system IRAS~16293, located at a distance of 141 pc \citep{Dzib2018}, was observed in the ALMA Protostellar Interferometric Line Survey \citep[PILS, see][for overview and first results]{Jorgensen2016}. The survey covers a total of 33.7 GHz between 329 and 363 GHz with spectral and angular resolutions of 0.2 km s$^{-1}$ and 0$\overset{\second}{.}$5 ($\sim$ 70 au) respectively. The IRAS~16293 system is composed of two main components, IRAS~16293A and IRAS~16293B, with the narrow lines associated with the B source, $\sim$1 km s$^{-1}$, making it ideal for line identification.

\begin{figure}[]
	\centering
	\includegraphics[width=0.5\textwidth, trim={0 0 0 0}, clip]{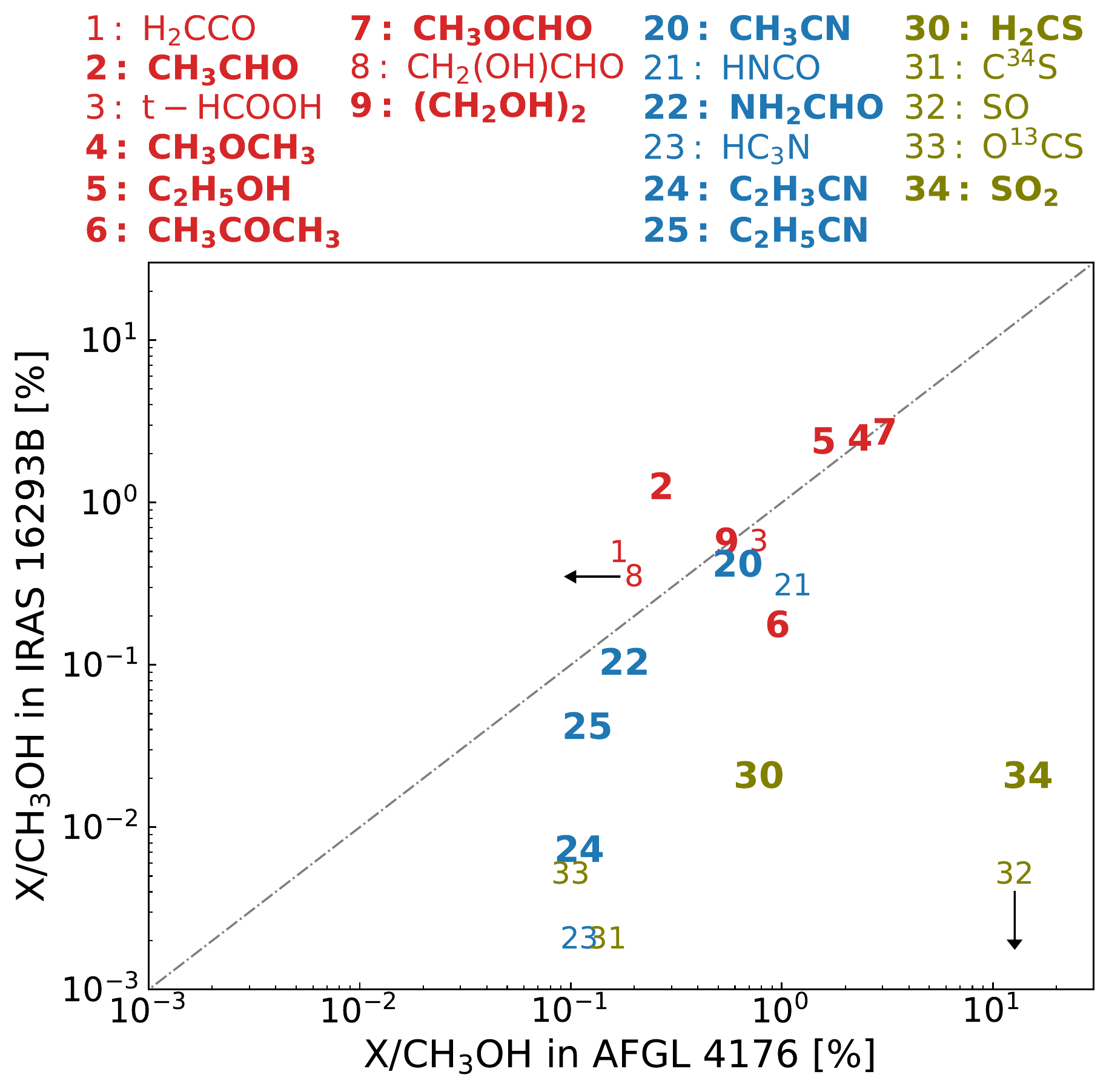}  
	\caption{Relative abundances of species detected towards AFGL~4176 and IRAS~16293. Labels in bold refer to species with five or more detected transitions towards AFGL~4176. Oxygen-bearing species are labelled by numbers 1 through 9 in red, N-bearing species are labelled by numbers 20 through 25 in blue and S-bearing species are labelled by numbers 30 through 34 in yellow. Black arrows indicate upper limits on the relative abundance of CH$_2$(OH)CHO in AFGL~4176 and of SO in IRAS~16293B. The dashed grey line indicates the 1:1 ratio of relative abundances in AFGL~4176 and IRAS~16293B.}
	\label{fig:afgl_vs_iras}
\end{figure}

Of the species for which five or more lines are detected towards AFGL~4176, all have also been identified towards IRAS~16293, although CH$_3$C$_2$H and NS are not reported in the PILS survey \citep{vanDishoeck1995, Caux2011, Coutens2016, Jorgensen2016, Lykke2017, Calcutt2018, Drozdovskaya2018, Jorgensen2018}. In addition, eight species with fewer than five transitions detected towards AFGL~4176 (CS and OCS only via their isotopologues) are also common between the sources. Figure \ref{fig:afgl_vs_iras} presents an overview of the relative abundances of all species detected towards AFGL~4176 and IRAS~16293B. 

Overall, the composition of AFGL 4176 is more similar to that of IRAS~16293B than to the high-mass star-forming regions in the Galactic centre. Specifically, the relative column densities derived for the O-bearing species C$_2$H$_5$OH, CH$_3$OCH$_3$, CH$_3$OCHO and (CH$_2$OH)$_2$ towards IRAS~16293B are within a factor of two of the values derived for AFGL~4176. The remaining species show slightly larger variations with the ratio of CH$_3$CHO to CH$_3$OH being a factor of four higher in IRAS~16293B compared with AFGL~4176, and the ratio of CH$_3$COCH$_3$ to CH$_3$OH a factor of six lower. For N-bearing species, similar abundances are derived for CH$_3$CN and NH$_2$CHO, with variations within a factor of two between the sources. In contrast, lower ratios of both C$_2$H$_3$CN and C$_2$H$_5$CN, are reported towards IRAS~16293B compared to AFGL~4176, by factors of 16 and three, respectively. By far the largest variations between the sources are seen in the ratios of the S-bearing species with SO$_2$ being close to three orders of magnitude higher in AFGL~4176 compared to IRAS~16293B, and H$_2$CS higher by a factor 39. However, \cite{Drozdovskaya2018} note that the SO$_2$ emission detected towards IRAS~16293 is likely not homogeneously distributed within the 0$\overset{\second}{.}$5 PILS beam, which also misses a large extended component, and therefore the large difference between the sources could in part be explained by local variations in the distribution of the species towards IRAS~16293. 

\subsection{Comparison with chemical models}
The chemistry of hot cores is commonly divided into three main phases: 1) a cold collapse phase dominated by reactions on grain surfaces involving the diffusion of light species (i.e. H); 2) a warm-up phase where relatively complex species can be formed in the ice and subsequently released into the gas phase; 3) a hot core phase dominated by gas-phase reactions triggered by the evaporation of the icy content. The models presented by \cite{Garrod2013} couple all of these phases to trace the gas-phase, grain-surface and bulk ice chemistry throughout the evolution of the core. The physical model adopted by \citet{Garrod2013} consists of a collapse phase followed by a gradual warm-up of the gas and dust. Three warm-up timescales are adopted: a 'fast' scale with warm-up to 200~K in 5$\times$10$^4$~yr, a 'medium' scale reaching 200~K in 2$\times$10$^5$~yr and a 'slow' scale taking 1$\times$10$^6$~yr to reach 200~K. Listed in Table \ref{tab:summary_of_column_density_ratios} are the predicted peak gas-phase abundance ratios for each of these models.

In the models presented by \citet{Garrod2013}, the formation of the complex O-bearing species detected in this study (i.e. CH$_3$CHO, C$_2$H$_5$OH, CH$_3$OCH$_3$, CH$_3$COCH$_3$, CH$_3$OCHO and (CH$_2$OH)$_2$) is closely related to CH$_3$OH. Within the framework of the models, O-bearing species are formed primarily in warm (i.e. 30~K~<~T~<~80~K) ices from the recombination of radicals formed through the UV-photodissociation of CH$_3$OH along with other simple solid species such as H$_2$CO and H$_2$O \citep[see also][]{Garrod2006,Garrod2008}. Here, the UV field is internally generated by cosmic ray collisions with H$_{2}$. While the models match the abundances of CH$_3$CHO, C$_2$H$_5$OH and CH$_3$OCH$_3$ in AFGL~4176 within a factor of three, the abundances of CH$_3$COCH$_3$ and (CH$_2$OH)$_2$ are underpredicted by at least an order of magnitude. For CH$_3$OCHO, the most abundant O-bearing species after CH$_3$OH, with abundances of $\sim$2--3$\%$ in both AFGL~4176 and IRAS~16293B, the models predict values which are lower than the observed ones by a factor of four in the case of the fast model and a factor of 39 in the case of the slow model. A possible explanation for the mismatch between models and derived abundances of this species is poorly understood ice chemistry as well as missing gas-phase routes or included gas-phase routes which are more efficient than presumed. Indeed, a number of studies suggest that gas-phase chemistry could provide an important contribution to the formation of the molecule, either through ion-neutral or neutral-neutral reactions, following the sublimation of precursors from the ice \citep[see e.g.][]{Neill2011, Cole2012, Vasyunin2013,Balucani2015,Taquet2016}. A significant contribution from gas-phase reactions to the formation of CH$_3$COCH$_3$, as suggested by \citet{Charnley2001}, is also possible. 

The abundances of complex N-bearing species detected in this study (i.e. CH$_3$CN, NH$_2$CHO, C$_2$H$_3$CN and C$_2$H$_5$CN) are likely the result of a combination of grain-surface and gas-phase reactions. For instance, CH$_3$CN can be formed on ices from the radical recombination of CH$_3$ and CN but also in the gas-phase via ion reactions between HCN and CH$_3^+$, followed by recombination with an electron or a proton transfer reaction. In the models by \citet{Garrod2013}, CH$_3$CN is formed primarily in ices via CH$_3$~+~CN~$\rightarrow$~CH$_3$CN, though a formation channel via hydrogenation of C$_2$N on grains at early times is also included. C$_2$H$_3$CN and C$_2$H$_5$CN are formed mainly during the warm-up stage as a result of a series of both gas-phase and grain-surface reactions including the formation, recombination and hydrogenation of the CN radical as well as C$_2$H$_2$, C$_2$H$_4$ and HC$_3$N \citep{Garrod2017}. The formation of NH$_2$CHO is likely ice-dominated and proceeds via radical recombination reactions between NH$_2$ and HCO \citep[e.g.][]{Jones2011, Fedoseev2016}. Alternatively, hydrogenation of HNCO may also lead to NH$_2$CHO. This idea is based on the strong correlation between the abundances of HNCO and NH$_2$CHO reported by \cite{Bisschop2007} and \citet{Lopez-Sepulcre2015} for samples of pre-stellar and protostellar objects, indicating a connection between these species. However, such a connection contradicts the results of experimental works which show that solid-state HNCO hydrogenation does not produce NH$_2$CHO \citep{Noble2015}. A similar conclusion is reached by \cite{Ligterink2018} who find that HNCO is likely not at the basis of the formation of NH$_2$CHO but rather formed simultaneously. Finally, a gas-phase formation route for NH$_2$CHO via reactions between NH$_2$ and H$_2$CO, has also been proposed \citep{Barone2015, Codella2017}. Chemical models by \citet{Quenard2018} show that both solid-state and gas-phase reactions are important for the formation of NH$_{2}$CHO and they claim its correlation with HNCO is the result of chemical reactions leading to both species responding in very similar ways to physical parameters, such as temperature.

For the detected N-bearing species, only the abundance ratio of C$_2$H$_3$CN is well-matched by the models. In contrast, the abundance ratios predicted by the models for CH$_3$CN, NH$_2$CHO and C$_2$H$_5$CN are all about an order of magnitude off with respect to the values derived for AFGL~4176. The abundances in AFGL~4176 do not favour either of the fast, medium or slow models over the other.   

In summary, the presence of most of the detected complex molecules could be explained by warm ice chemistry triggered by cosmic-ray UV photons. However, for some species, O- as well as N-bearing, formation routes via gas-phase reactions cannot be neglected. The chemical similarity between IRAS~16293B and AFGL~4176 hints that the chemical composition of complex species may already be set in the cold cloud stage. In this context, the lack of chemical differences between the two sources, despite the large difference in luminosity, implies that AFGL~4176 is a very young source where little processing of the chemical inventory by the protostar has occurred.

\begin{table*}[]
	\centering
	\caption{Summary of column density ratios with respect to methanol predicted by models and derived towards AFGL~4176, Sgr~B2(N), Orion KL, and IRAS~16293B.}
	\label{tab:summary_of_column_density_ratios}
	\begin{tabular}{llcccccccc}
		\toprule
		& & \multicolumn{8}{c}{X/CH$_3$OH [\%]} \\
		\cmidrule{2-10}
		& & \multicolumn{2}{c}{Sgr B2} & Orion KL & AFGL~4176 & IRAS 16293B\tablefootmark{a} & \multicolumn{3}{c}{Model} \\
		\cmidrule{3-4}
		\cmidrule{8-10}
		& & (N2) & (N3-5) & & & & F & M & S \\
		\midrule
		Hydrocarbons & CH$_3$C$_2$H & -- & -- & -- & 0.7 $\pm$ 0.2 & -- 
		& -- & -- & -- \\ 
		
		O-bearing %
		& CH$_3$CHO & -- & -- & -- & 0.3 $\pm$ 0.2 & 1.20 & 0.03 & 0.10 & 0.37 \\
		& C$_2$H$_5$OH & 5.0 & 7.6 -- 10.4 & 0.37 -- 1.98 & 1.4 $\pm$ 0.4 & 2.3 & 0.54 & 0.64 & 1.29 \\
		& CH$_3$OCH$_3$ & -- & -- & 3.40 -- 9.26 & 2.4 $\pm$ 0.3 & 2.4 & 0.44 & 0.69 & 0.74 \\
		& CH$_3$COCH$_3$ & -- & -- & 0.05 -- 0.15 & 1.0 $\pm$ 0.2 & 0.17 & 0.01 & 0.11 & 0.05 \\
		& CH$_3$OCHO & 3.0 & 18.9 -- 32.0 & 4.76 -- 22.2 & 3.1 $\pm$ 0.6 & 2.6 & 0.84 & 0.41 & 0.08 \\
		& (CH$_2$OH)$_2$ & -- & -- & 0.10 -- 0.42 & 0.5 $\pm$ 0.2 -- 0.6 $\pm$ 0.1 & 0.5 -- 0.55\tablefootmark{b} & 0.03 & 0.01 & 10$^{-4}$\\
		
		N-bearing %
		& CH$_3$CN & 5.78 & 10.6 -- 13.4 & -- & 0.62 $\pm$ 0.07 & 0.4 & 0.04 & 0.02 & 0.03 \\
		& NH$_2$CHO & 8.75 & 0.83 -- 1.16\tablefootmark{c} & -- & 0.18 $\pm$ 0.02 & 0.1 & 3.55 & 1.65 & 0.08 \\
		& C$_2$H$_3$CN & 1.05 & 0.6 -- 1.25 & -- & 0.11 $\pm$ 0.02 & 0.007 & 0.1 & 0.06 & 0.15 \\
		& C$_2$H$_5$CN & 17.25 & 4.8 -- 8.11 & -- & 0.12 $\pm$ 0.02 & 0.04 & 0.05 & 0.85 & 0.79 \\		
		
		S-bearing %
		& NS & -- & -- & -- & 0.22 $\pm$ 0.02 & -- & -- & -- & -- \\
		& H$_2$CS & -- & -- & -- & 0.78 $\pm$ 0.07 & 0.02 & -- & -- & -- \\
		& SO$_2$ & -- & -- & -- & 14.7 $\pm$ 3.6 & 0.02 & 0.67 & 0.84 & 2.05 \\		
		\midrule
		Reference & & \multicolumn{2}{c}{1, 2} & 3 & this work & 4, 5, 6, 7, 8, 9 & \multicolumn{3}{c}{10} \\
		\bottomrule
	\end{tabular}
	\tablefoot{F = Fast model, M = Medium model, S = Slow model. \tablefoottext{a}{Listed values are derived at 0.5\arcsec (one beam) offset from IRAS~16293B.} 
		\tablefoottext{b}{Derived at 0.25\arcsec (half beam) offset from IRAS~16293B.}
		\tablefoottext{c}{Excluding the upper limit of $\leq$0.56\% derived for N4.}}
	\tablebib{(1)~\cite{Belloche2016}; (2) \cite{Bonfand2017}; (3) \cite{Tercero2018}; (4) \cite{Coutens2016}; (5) \cite{Jorgensen2016}; (6) \cite{Lykke2017}; (7) \cite{Calcutt2018}; (8) \cite{Drozdovskaya2018}; (9) \cite{Jorgensen2018}; (10) \cite{Garrod2013}}
\end{table*}
 
\section{Summary} \label{sec:summary}
This paper presents a comprehensive study of the chemical inventory of the high-mass protostar AFGL~4176. Due to the exceptional resolving power offered by ALMA, the source is analysed on disk scales, making it possible to probe only the emission coming from the inner hot core region around the forming star while avoiding that of its large-scale cooler envelope. The source displays a rich chemistry consisting of 23 different molecular species, of which more than half, fourteen in total, are defined as complex, that is, consisting of six or more atoms. Of the detected species, the majority are oxygen-bearing while fewer contain nitrogen, sulphur or a combination thereof. 

Assuming LTE, the column density is derived for all species detected towards AFGL~4176. With respect to methanol, the O-bearing species are approximately a factor of three more abundant than N-bearing species. This may indicate that AFGL~4176 is a relatively young source since nitrogen chemistry generally takes longer to evolve in the gas-phase compared with the chemistry of O-bearing species. Alternatively, if the composition of complex species is already set in the cold cloud stage, this gives an indication that formation of O-bearing species is favoured over that of N-bearing molecules. This may be due to inefficient reactions pathways (i.e. high diffusion energies and reactions barriers) for formation of N-bearing species, or due to a dominant nitrogen loss channel, for example into N$_{2}$. 

With the exception of CH$_3$C$_2$H, which shows two emission peaks, the spatial distribution of N- and O-bearing species is roughly similar. Generally, all species show emission peaks near the position of the continuum peak though some O-bearing species peak up to 0$\overset{\second}{.}$2 away. Differences in the spatial extent of transitions of the same species with low and high $E_{\textrm{up}}$ are seen for the majority of species. For a number of species a velocity gradient is detected across the source consistent with the Keplerian-like disk reported by \citet{Johnston2015}.

Overall, the chemical composition of AFGL~4176 is more similar to that of the low-mass protostar IRAS~16293B than to that of the high-mass star-forming region Sgr~B2(N). Taking methanol as a reference, the abundances of C$_2$H$_5$OH, CH$_3$OCH$_3$, CH$_3$OCHO and (CH$_2$OH)$_2$ in IRAS~16293B are within a factor of two of the abundances in AFGL~4176. In contrast, the C$_2$H$_5$OH/CH$_3$OH and CH$_3$OCHO/CH$_3$OH values derived for Sgr~B2(N) are higher than those in AFGL~4176 by up to an order of magnitude. For the N-bearing species, the abundances of CH$_3$CN, C$_2$H$_5$CN and NH$_2$CHO are similar between AFGL4176 and IRAS~16293B while C$_2$H$_3$CN/CH$_3$OH is lower by a factor of 16 in IRAS~16293B. For Sgr~B2(N) the abundances of all N-bearing species are significantly higher than in AFGL~4176. The similarity between abundances in AFGL~4176 and IRAS~16293B indicates that the production of complex species does not depend strongly on the luminosity of sources, but may be universal despite differences in physical conditions or that the composition of species is set already in the ice during the cold cloud stage.

\begin{acknowledgements} 
We thank the anonymous referee for their thorough review and for contributing many clarifications which helped us improve our manuscript. This paper makes use of the following ALMA data: ADS/JAO.ALMA\#2012.1.00469.S. ALMA is a partnership of ESO (representing its member states), NSF (USA) and NINS (Japan), together with NRC (Canada) and NSC and ASIAA (Taiwan) and KASI (Republic of Korea), in cooperation with the Republic of Chile. The Joint ALMA Observatory is operated by ESO, AUI/NRAO and NAOJ.
\end{acknowledgements}

\bibliographystyle{aa}
\bibliography{AFGL4176} 

\newpage
\appendix

\section{Model grids}
\begin{table*}[]
	\centering
	\caption{Overview of model grids.}
	\label{tab:model_grid}
	\begin{tabular}{lcccccc}
		\toprule
		$N_{\textrm{s}}$ range [cm$^{-2}$]: & 10$^{15}$ -- 10$^{16}$ & 5$\times$10$^{15}$ -- 5$\times$10$^{16}$ & 10$^{16}$ -- 10$^{17}$ & 5$\times$10$^{16}$ -- 5$\times$10$^{17}$ & 10$^{16}$ -- 5$\times$10$^{17}$ & 5$\times$10$^{17}$ -- 5$\times$10$^{18}$ \\
		\midrule
		\multirow{6}{*}{Species} 
		& C$_2$H$_3$CN & CH$_3$CHO & CH$_3$C$_2$H & $^{13}$CH$_3$OH & C$_2$H$_5$OH & SO$_2$ \\
		& & NH$_2$CHO & CH$_3$CN & CH$_3$OCH$_3$ &  & \\
		& & & CH$_3$CN, v$_8$=1 & CH$_3$OCHO & & \\
		& & & H$_2$CS & & & \\
		& & & (CH$_2$OH)$_2$ & & & \\			
		\bottomrule
	\end{tabular}
	\tablefoot{All grids have $T_{\textrm{ex}}$ spanning 50 -- 500 K in steps of 10 K and $N_{\textrm{s}}$ sampled by 20 logarithmically spaced steps, apart from the grid for C$_2$H$_5$OH which has 30 logarithmically spaced steps.}
\end{table*}

\section{Unidentified lines} \label{app:unidentified_lines}
\begin{table}[]
	\centering
	\caption{List of unidentified lines.}
	\label{tab:unidentified_lines}
	\begin{tabular}{cc}
		\toprule
		Rest frequency\tablefootmark{a} & Peak intensity \\
		$[$GHz$]$ & [K] \\
		\midrule
		238.806 & 1.9 \\
		239.078 & 4.2 \\
		239.082 & 3.6 \\
		\midrule
		239.651 & 1.9 \\
		239.655 & 2.7 \\
		239.926 & 1.9 \\
		240.129 & 2.0 \\
		240.132 & 2.0 \\
		240.155 & 2.1 \\
		240.517 & 2.5 \\
		240.562 & 2.0 \\
		240.565 & 2.2 \\
		240.574 & 1.9 \\
		240.579 & 2.2 \\
		240.702 & 2.1 \\
		240.718 & 2.0 \\
		240.753 & 2.0 \\
		241.102 & 1.4 \\
		241.127 & 1.5 \\
		\midrule
		254.192 & 2.9 \\
		254.340 & 2.0 \\
		254.346 & 2.6 \\
		254.581 & 3.9 \\
		254.669 & 3.0 \\
		254.686 & 7.0 \\
		254.713 & 5.7 \\
		254.716 & 5.2 \\
		254.858 & 4.6 \\
		\midrule
		256.107 & 3.8 \\
		256.111 & 3.8 \\
		\bottomrule
	\end{tabular}
	\tablefoot{\tablefoottext{a}{Assuming a source velocity of -53.5 km s$^{-1}$.}}
\end{table}

\newpage
\section{Full model}\label{app:full_models}
\begin{figure*}[]
	\centering
	\begin{subfigure}{1.\textwidth}
		\centering
		\includegraphics[width=0.87\textwidth, trim={0 0 0 0}, clip]{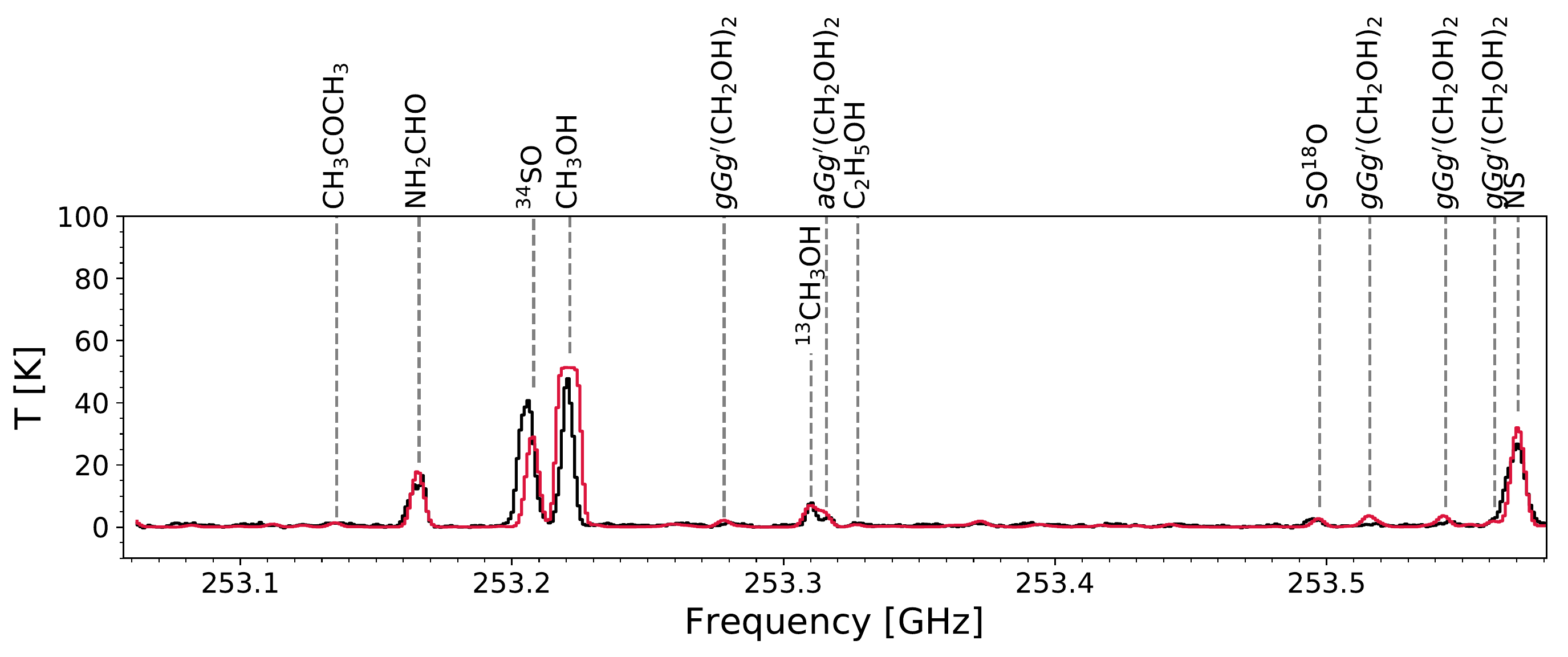}
		\caption*{}
		\label{fig:spw1a}
	\end{subfigure}
	\begin{subfigure}{1.\textwidth}
		\vspace{-0.7cm}
		\centering
		\includegraphics[width=0.87\textwidth, trim={0 0 0 0}, clip]{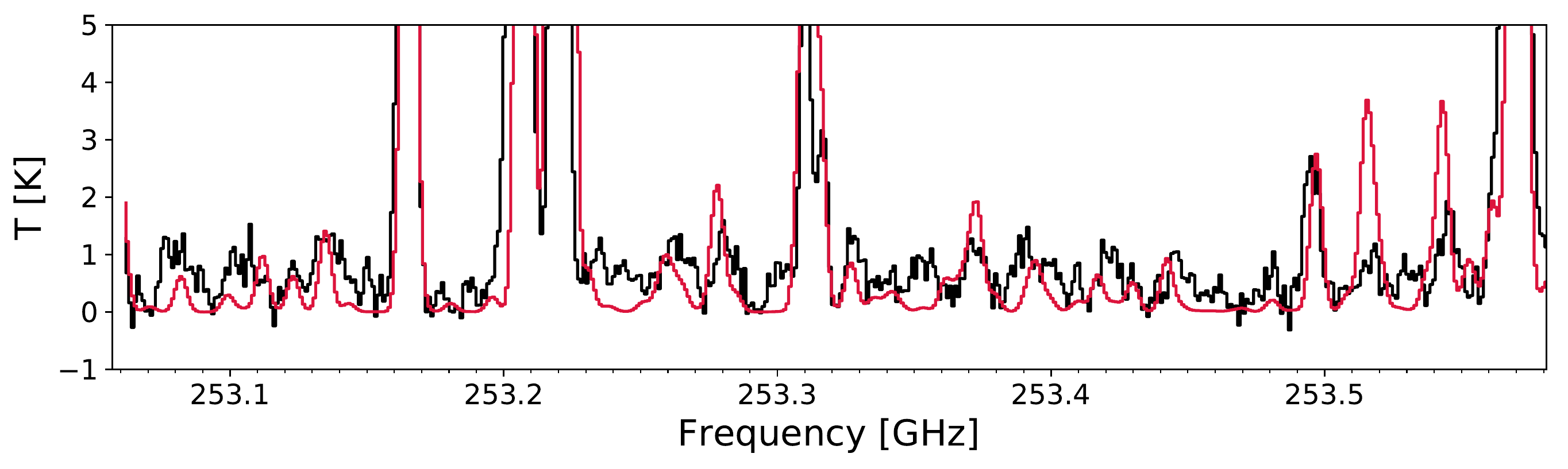}
		\caption*{}
		\label{fig:spw1a_zoomed}
	\end{subfigure}
	\begin{subfigure}{1.\textwidth}
		\centering
		\includegraphics[width=0.87\textwidth, trim={0 0 0 0}, clip]{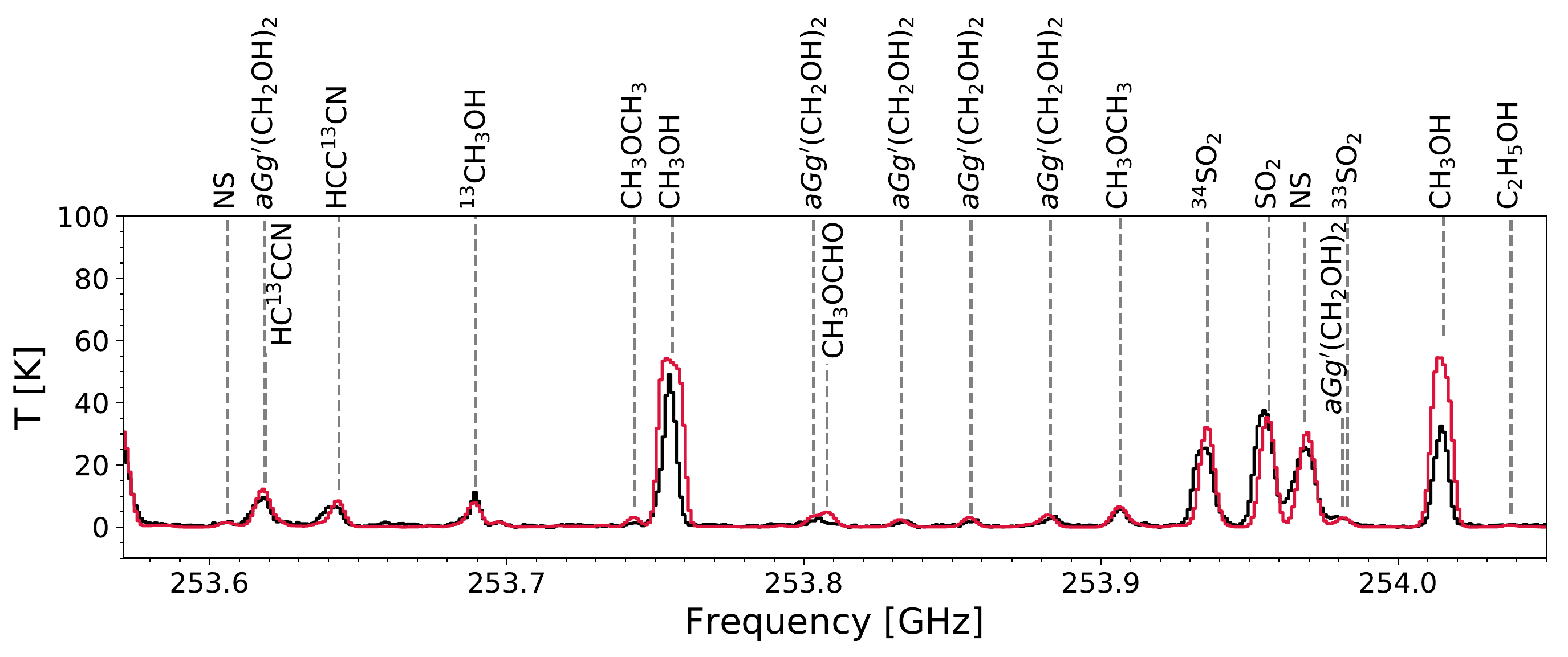}
		\caption*{}
		\label{fig:spw1b}
	\end{subfigure}
	\begin{subfigure}{1.\textwidth}
		\vspace{-0.7cm}
		\centering
		\includegraphics[width=0.87\textwidth, trim={0 0 0 0}, clip]{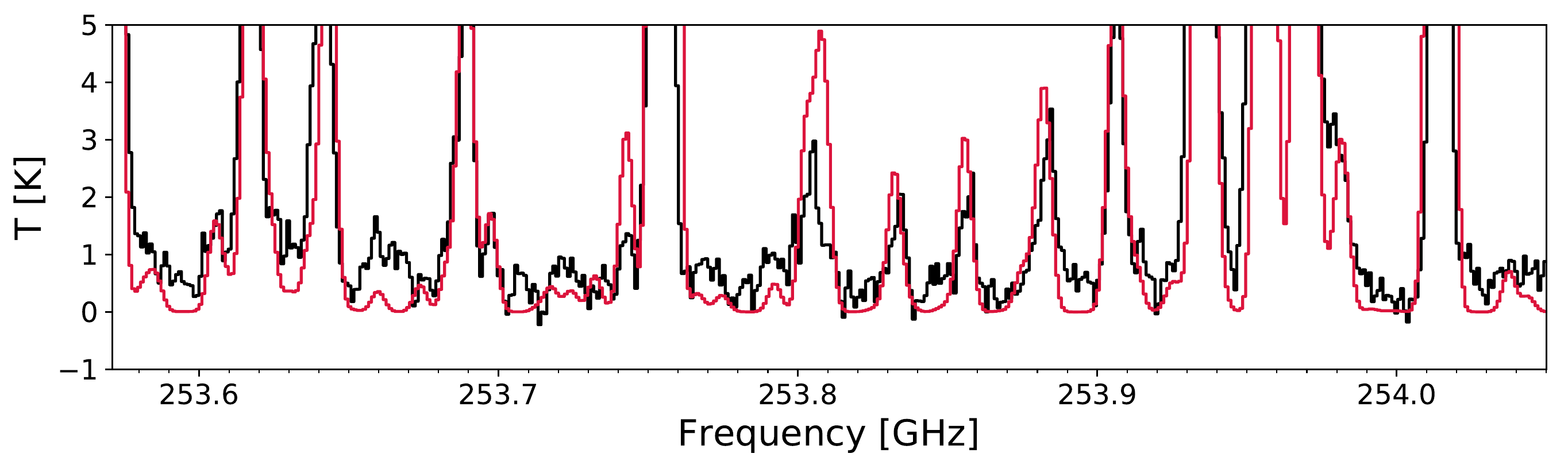}
		\caption*{}
		\label{fig:spw1b_zoomed}
	\end{subfigure}
\end{figure*}
\begin{figure*}
	\begin{subfigure}{1.\textwidth}
		\ContinuedFloat
		\centering
		\includegraphics[width=0.87\textwidth, trim={0 0 0 0}, clip]{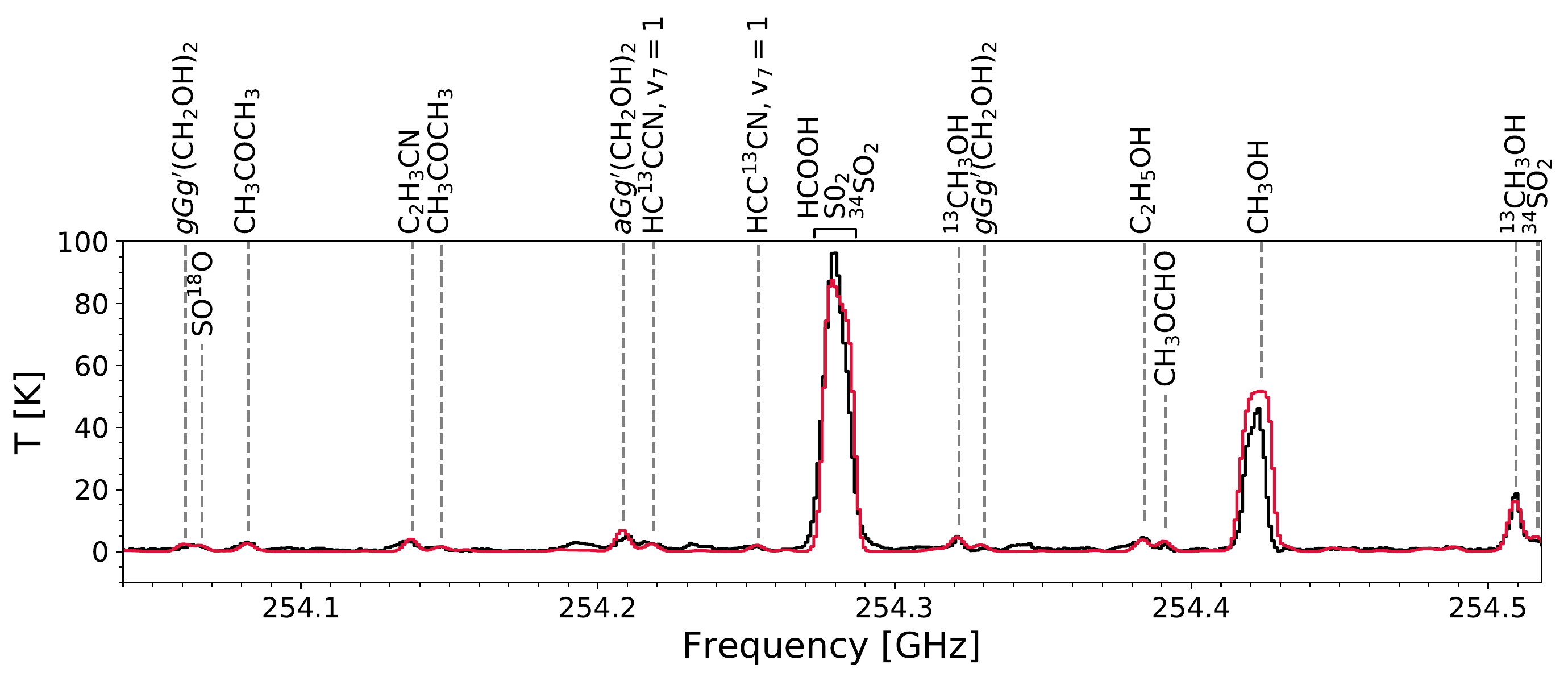}
		\caption*{}
		\label{fig:spw1c}
	\end{subfigure}
	\begin{subfigure}{1.\textwidth}
		\vspace{-0.7cm}
		\centering
		\includegraphics[width=0.87\textwidth, trim={0 0 0 0}, clip]{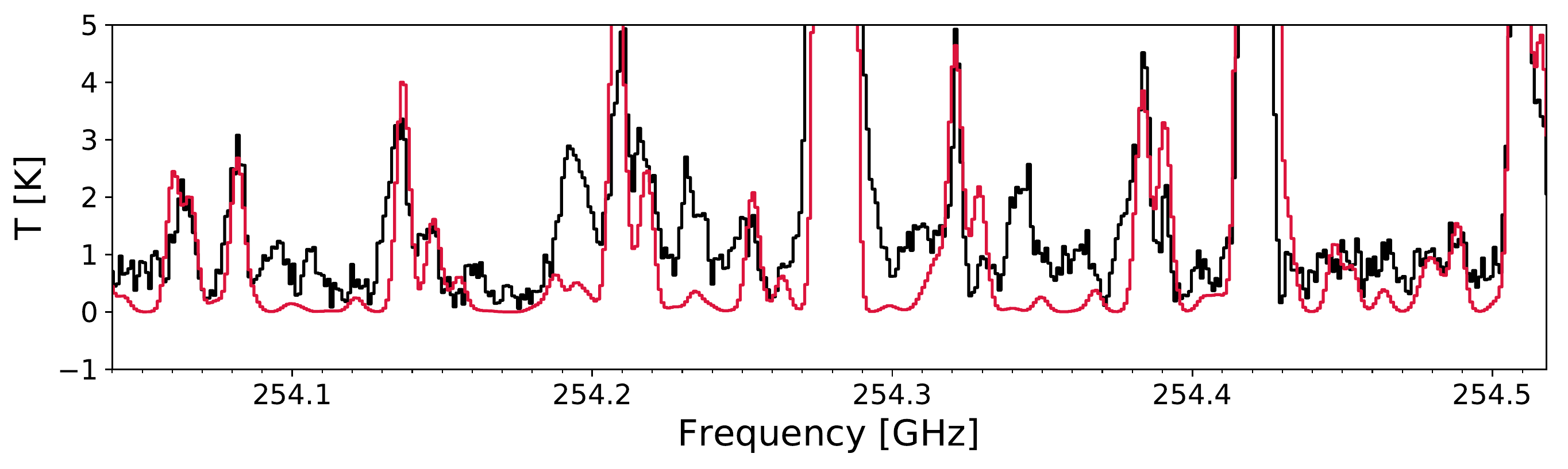}
		\caption*{}
		\label{fig:spw1c_zoomed}
	\end{subfigure}
	\begin{subfigure}{1.\textwidth}
		\ContinuedFloat
		\centering
		\includegraphics[width=0.87\textwidth, trim={0 0 0 0}, clip]{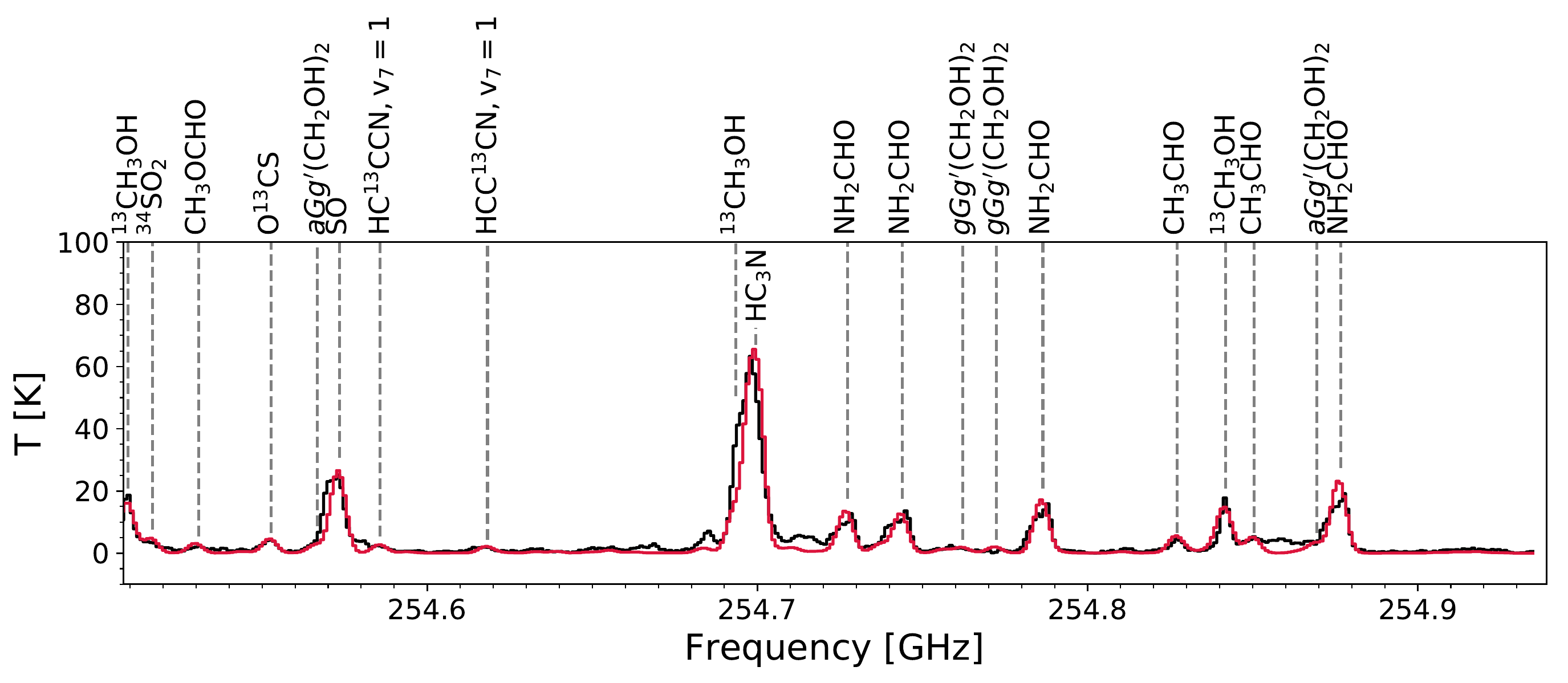}
		\caption*{}
		\label{fig:spw1d}
	\end{subfigure}
	\begin{subfigure}{1.\textwidth}
		\vspace{-0.7cm}
		\centering
		\includegraphics[width=0.87\textwidth, trim={0 0 0 0}, clip]{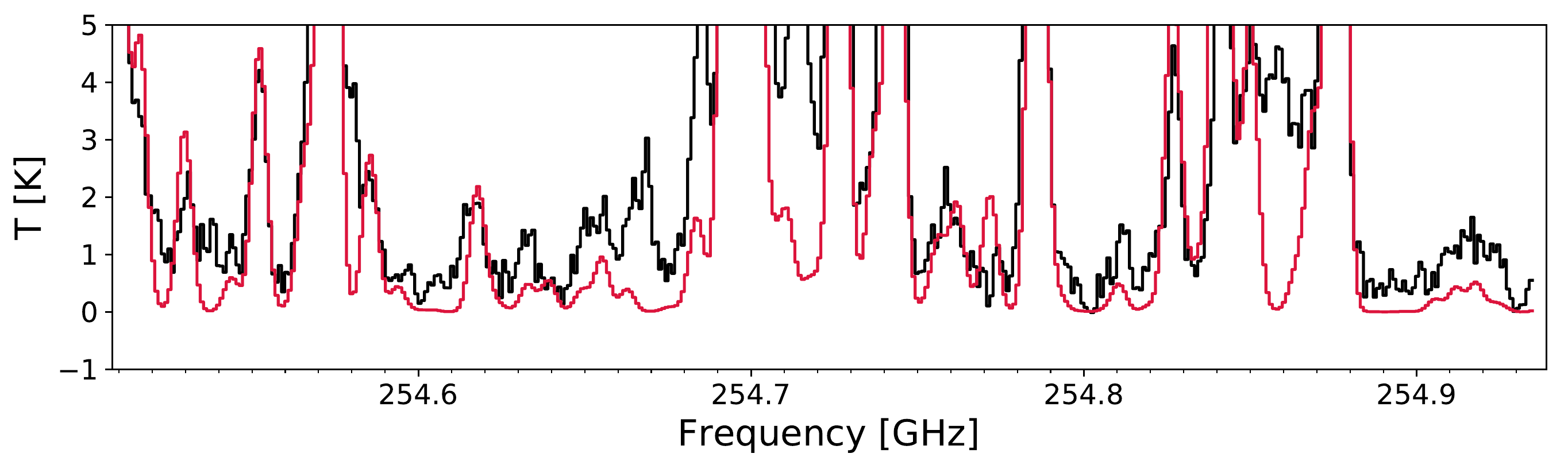}
		\caption*{}
		\label{fig:spw1d_zoomed}
	\end{subfigure}
	\caption{Same as Fig. \ref{fig:spw0} for species detected towards AFGL~4176 in the spectral window centred at 254.0 GHz.}
	\label{fig:spw1}
\end{figure*}

\begin{figure*}[]
	\centering
	\begin{subfigure}{1\textwidth}
		\centering
		\includegraphics[width=0.87\textwidth, trim={0 0 0 0}, clip]{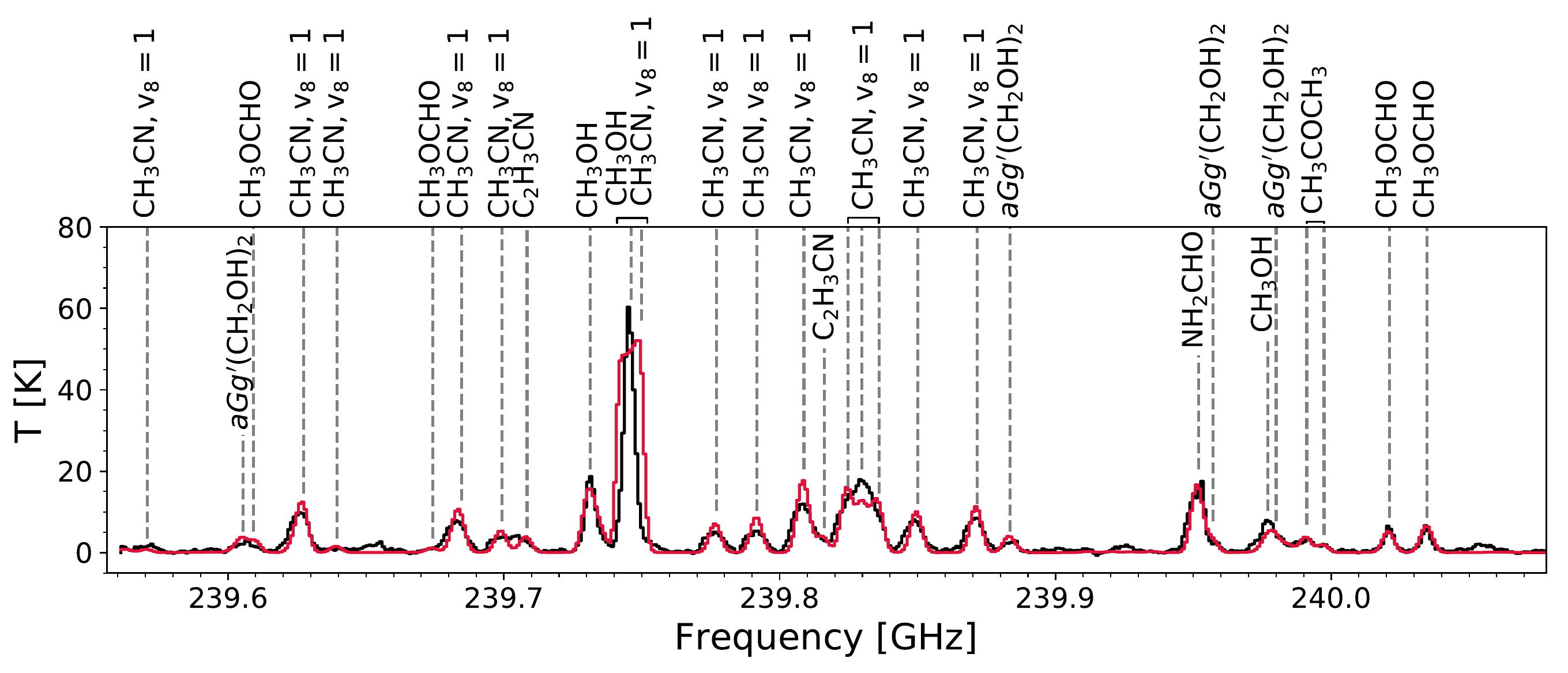}
		\caption*{}
		\label{fig:spw2a}
	\end{subfigure}
	\begin{subfigure}{1\textwidth}
		\vspace{-0.7cm}
		\centering
		\includegraphics[width=0.87\textwidth, trim={0 0 0 0}, clip]{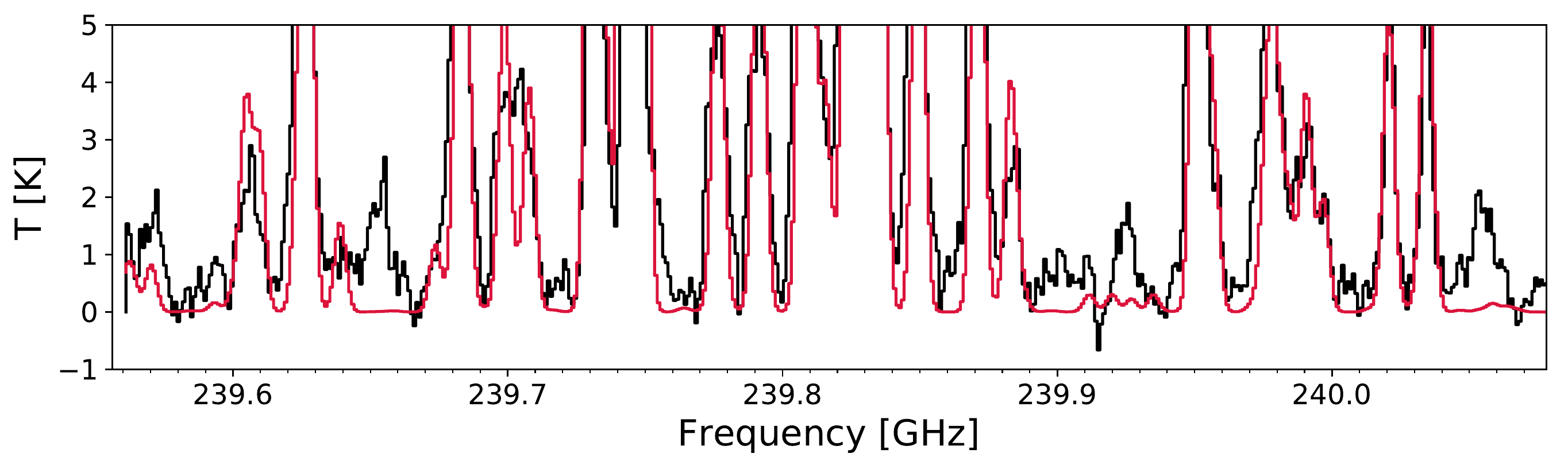}
		\caption*{}
		\label{fig:spw2a_zoomed}
	\end{subfigure}
	\begin{subfigure}{1\textwidth}
		\centering
		\includegraphics[width=0.87\textwidth, trim={0 0 0 0}, clip]{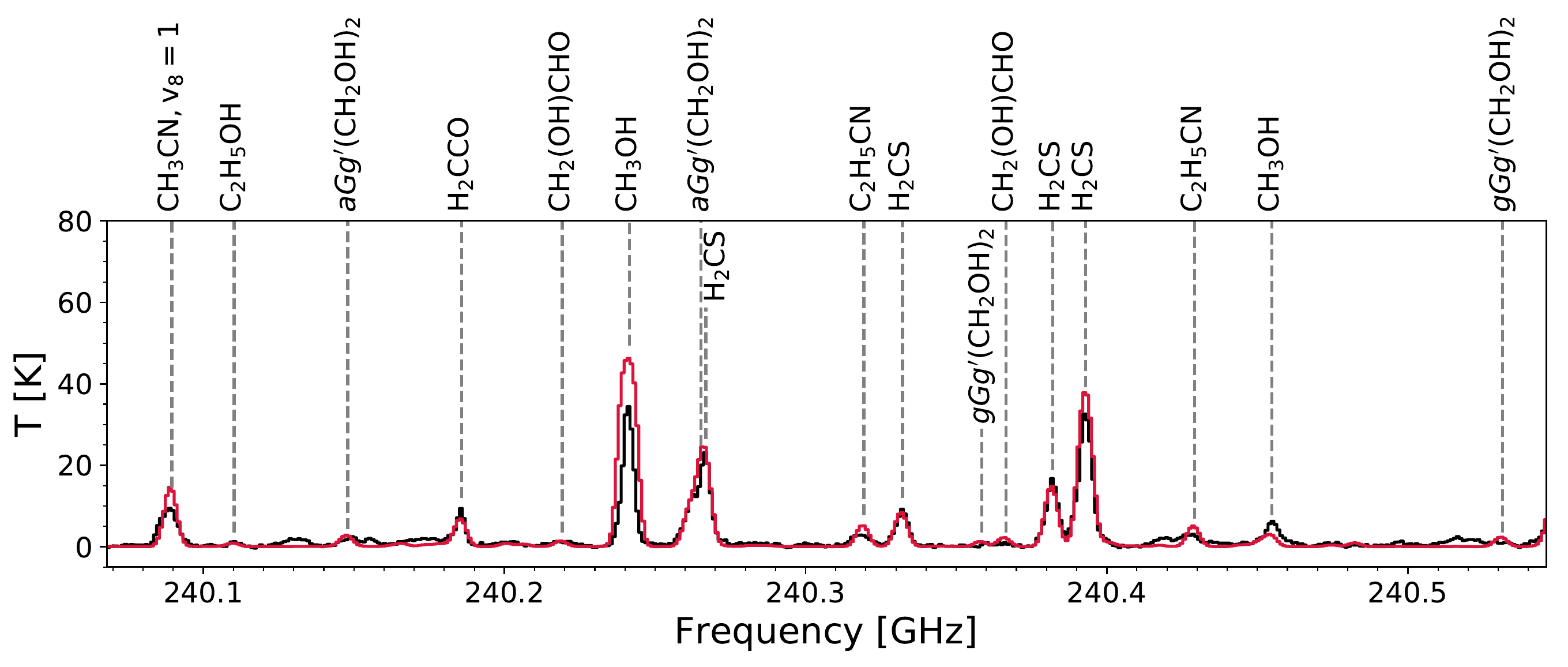}
		\caption*{}
		\label{fig:spw2b}
	\end{subfigure}
	\begin{subfigure}{1\textwidth}
		\vspace{-0.7cm}
		\centering
		\includegraphics[width=0.87\textwidth, trim={0 0 0 0}, clip]{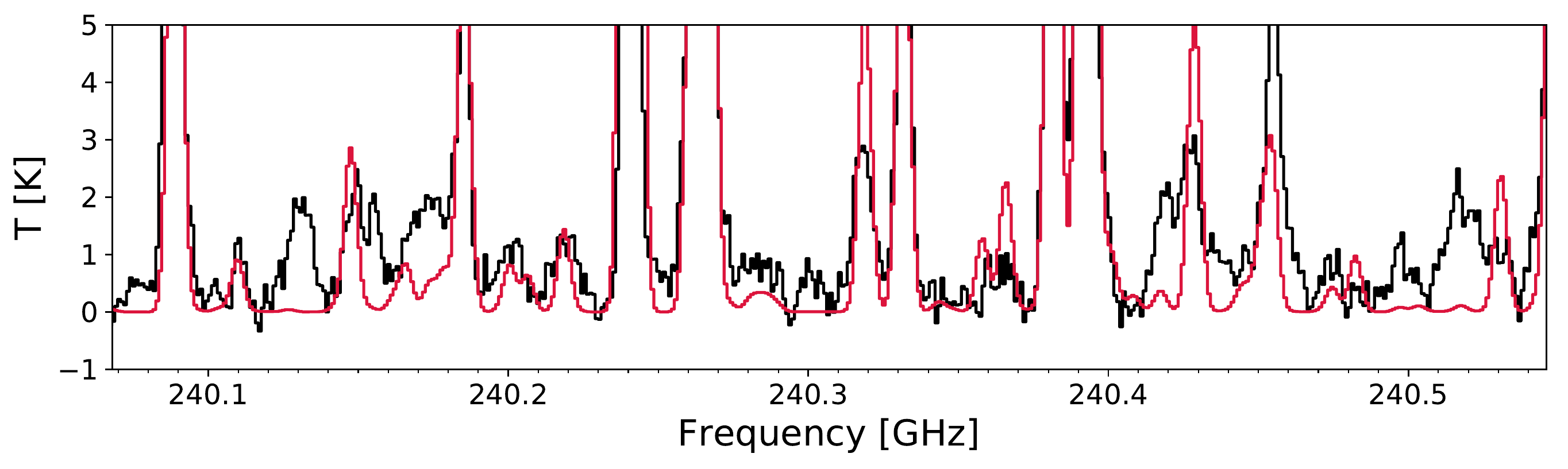}
		\caption*{}
		\label{fig:spw2b_zoomed}
	\end{subfigure}
\end{figure*}
\begin{figure*}
	\begin{subfigure}{1\textwidth}
		\ContinuedFloat
		\centering
		\includegraphics[width=0.87\textwidth, trim={0 0 0 0}, clip]{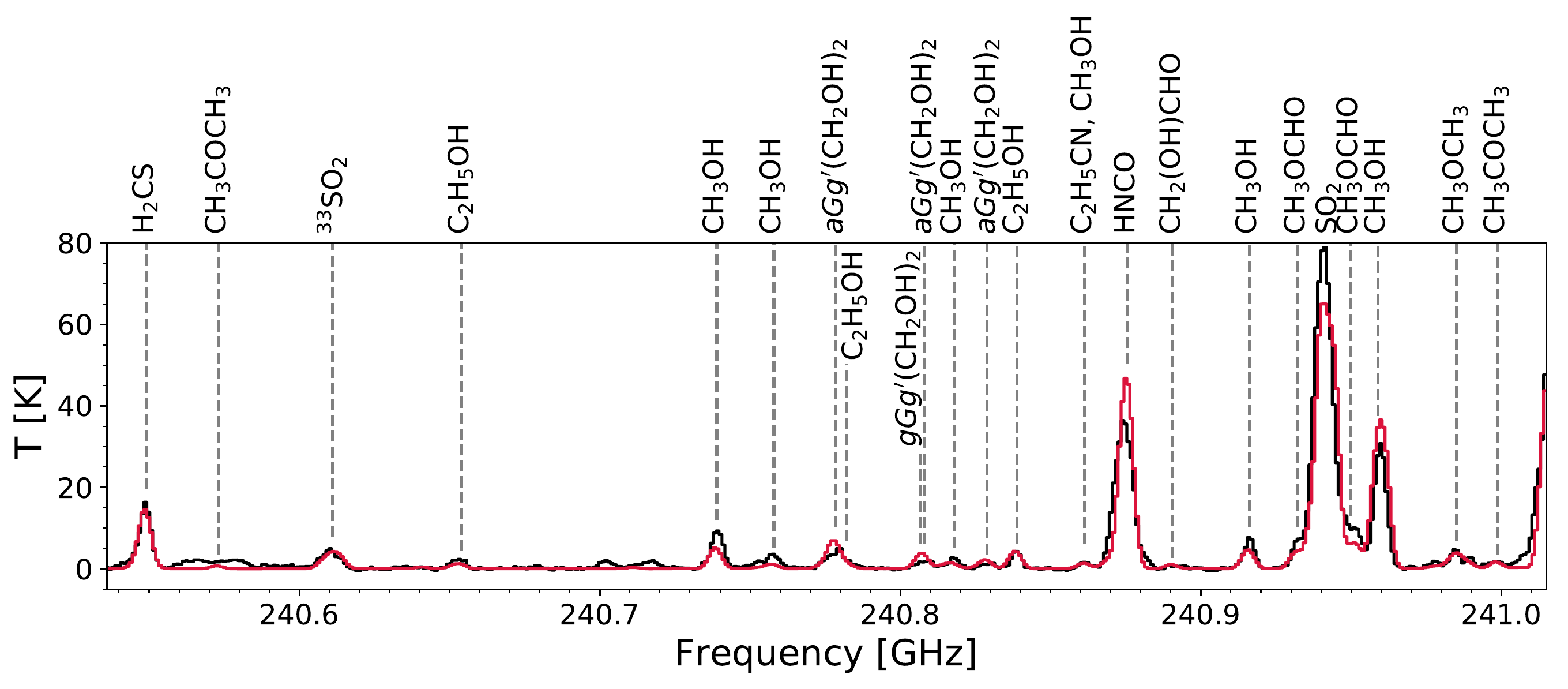}
		\caption*{}
		\label{fig:spw2c}
	\end{subfigure}
	\begin{subfigure}{1\textwidth}
		\vspace{-0.7cm}
		\centering
		\includegraphics[width=0.87\textwidth, trim={0 0 0 0}, clip]{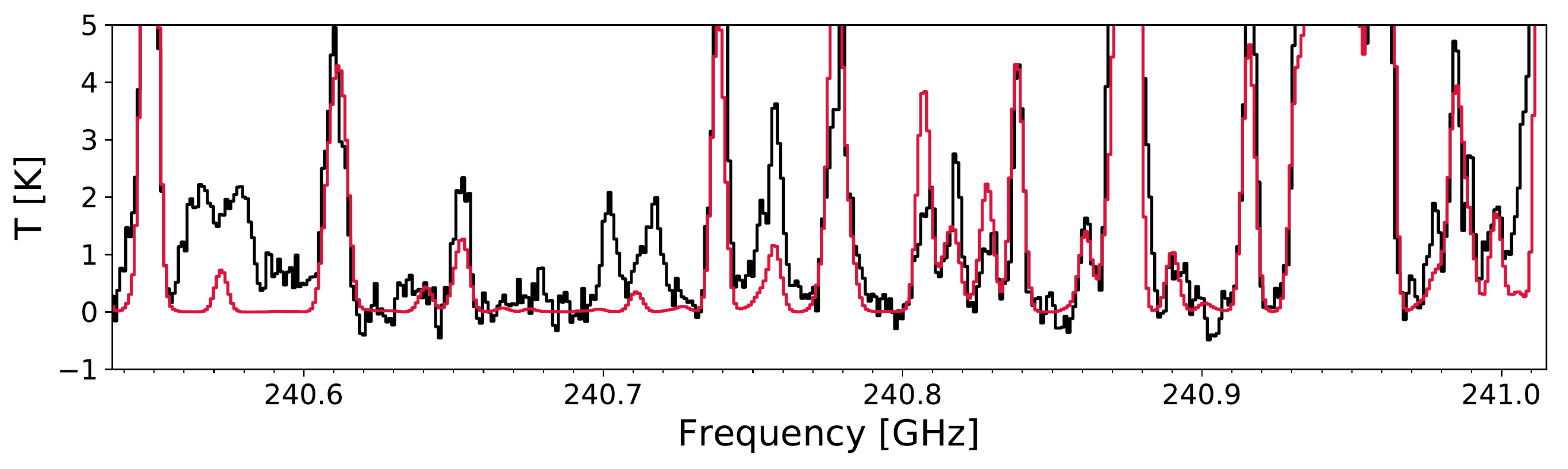}
		\caption*{}
		\label{fig:spw2c_zoomed}
	\end{subfigure}
	\begin{subfigure}{1\textwidth}
		\centering
		\includegraphics[width=0.87\textwidth, trim={0 0 0 0}, clip]{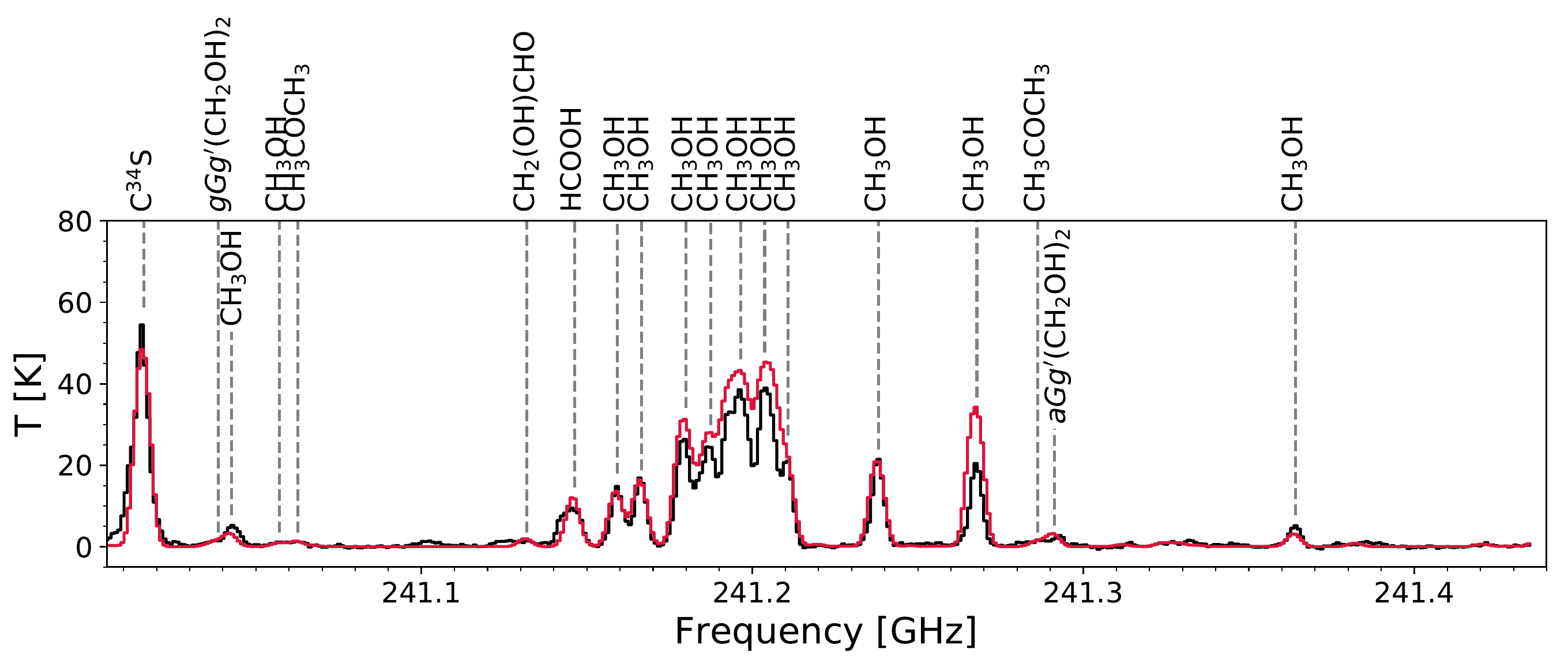}
		\caption*{}
		\label{fig:spw2d}
	\end{subfigure}
	\begin{subfigure}{1\textwidth}
		\vspace{-0.7cm}
		\centering
		\includegraphics[width=0.87\textwidth, trim={0 0 0 0}, clip]{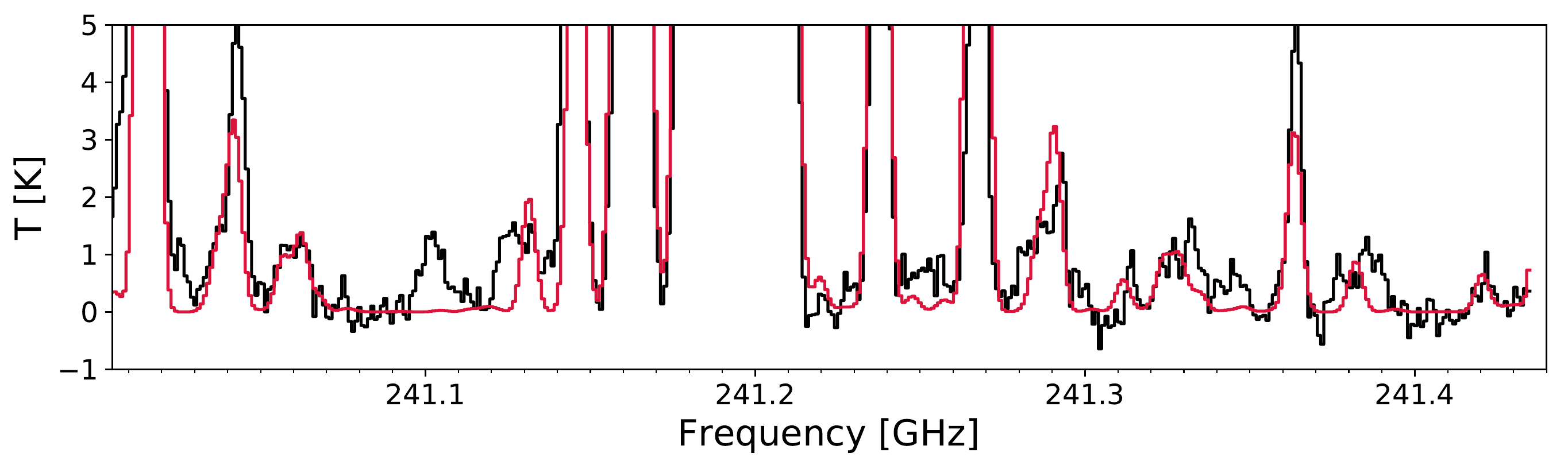}
		\caption*{}
		\label{fig:spw2d_zoomed}
	\end{subfigure}
	\caption{Same as Fig. \ref{fig:spw0} for species detected towards AFGL~4176 in the spectral window centred at 240.5 GHz.}
	\label{fig:spw2}
\end{figure*}

\begin{figure*}[]
	\centering
	\begin{subfigure}{1\textwidth}
		\centering
		\includegraphics[width=0.87\textwidth, trim={0 0 0 0}, clip]{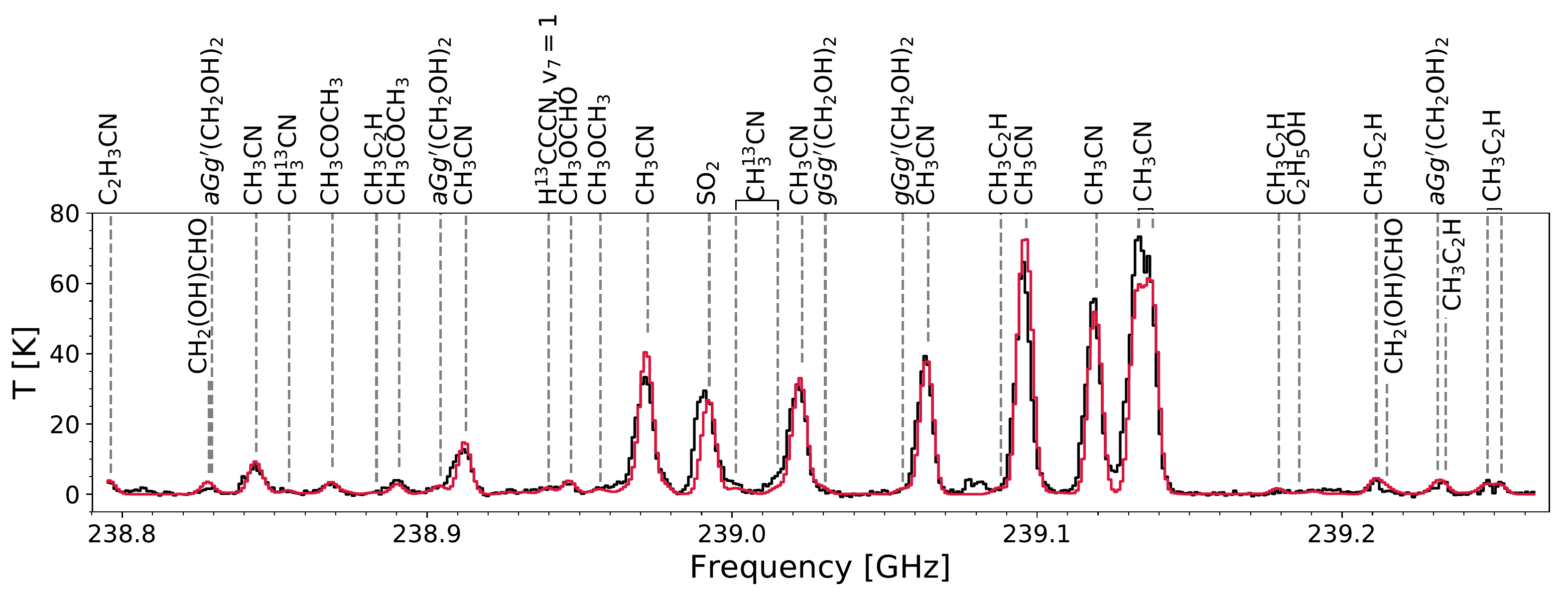}
		\caption*{}
		\label{fig:spw3_sub}
	\end{subfigure}
	\begin{subfigure}{1\textwidth}
		\vspace{-0.7cm}
		\centering
		\includegraphics[width=0.87\textwidth, trim={0 0 0 0}, clip]{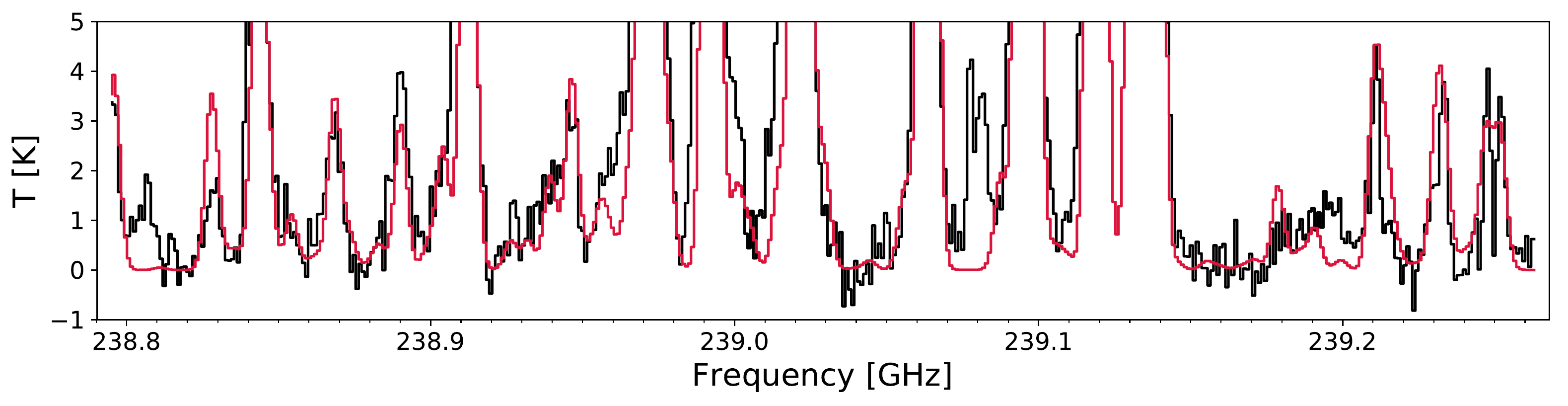}
		\caption*{}
		\label{fig:spw3_zoomed}
	\end{subfigure}
	\caption{Same as Fig. \ref{fig:spw0} for species detected towards AFGL~4176 in the spectral window centred at 239.0 GHz.}
	\label{fig:spw3}
\end{figure*}

\section{Line maps} \label{app:line_maps}
\begin{figure*} 
	\centering
	\includegraphics[width=.8\textwidth]{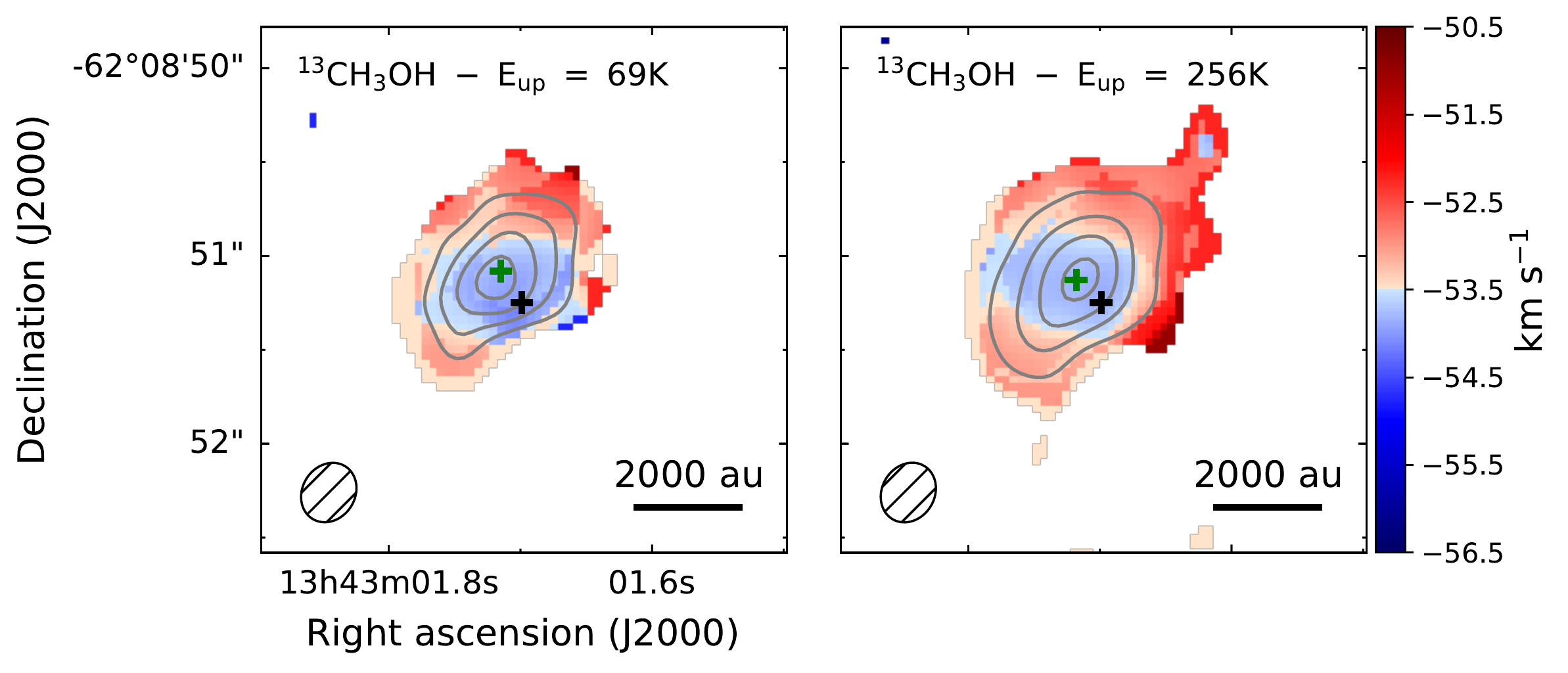}
	\caption[]{Same as Fig. \ref{fig:moment_maps} for $^{13}$CH$_3$OH lines at 256.1716 GHz (left) and 253.6895 GHz (right). Contours start at 9$\sigma$ and are in steps of 6$\sigma$ and 12$\sigma$, with $\sigma$ = 6.27$\times$10$^{-3}$ and 6.87$\times$10$^{-3}$ Jy beam$^{-1}$ km s$^{-1}$, for the left and right panel, respectively.}
	\label{fig:mom_13CH3OH}
\end{figure*}

\begin{figure*} 
	\centering
	\includegraphics[width=.8\textwidth]{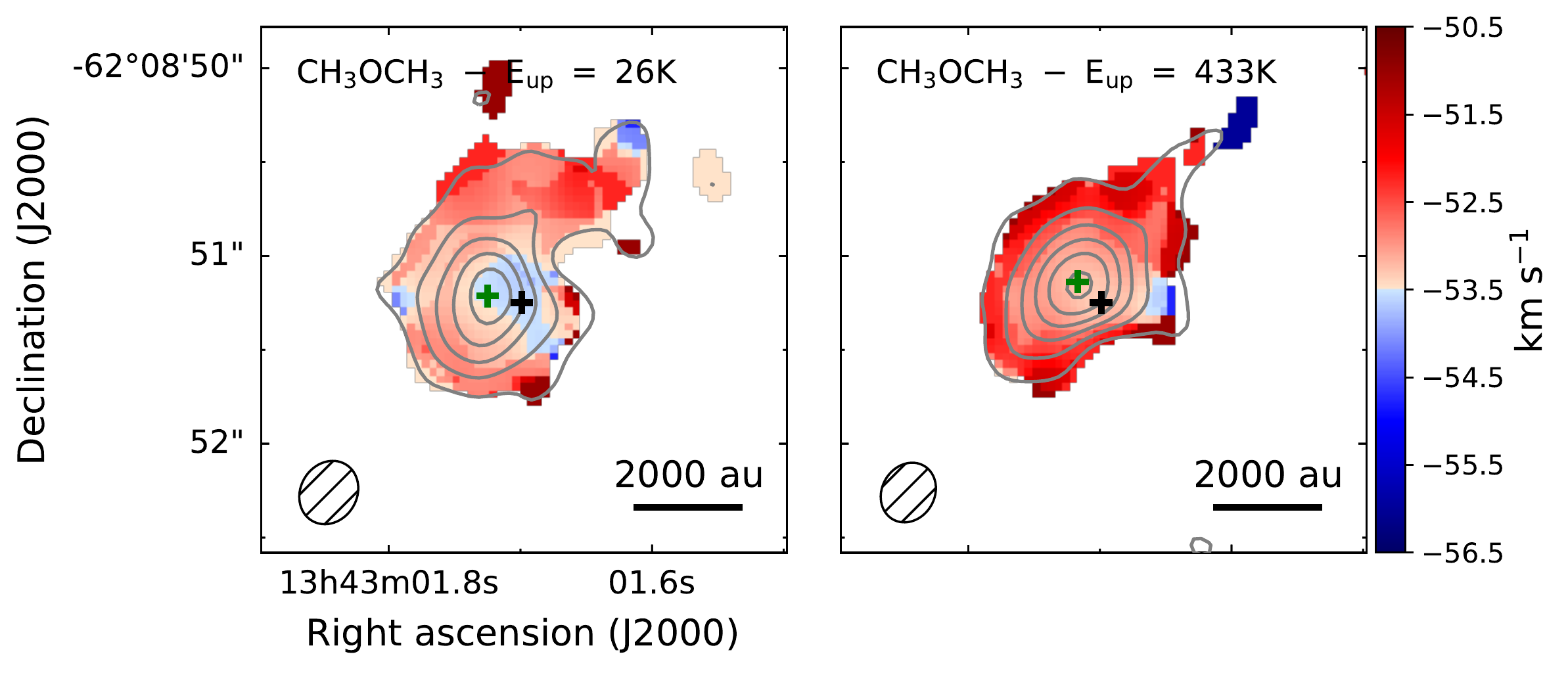}
	\caption[]{Same as Fig. \ref{fig:moment_maps} for CH$_3$OCH$_3$ lines at 240.9851 GHz (left) and 253.9075 GHz (right). Contours start at 3$\sigma$ and are in steps of 6$\sigma$, with $\sigma$ = 4.95$\times$10$^{-3}$ and 6.08$\times$10$^{-3}$ Jy beam$^{-1}$ km s$^{-1}$, for the left and right panel, respectively.}
	\label{fig:mom_CH3OCH3}
\end{figure*}

\begin{figure*} 
	\centering
	\includegraphics[width=.8\textwidth]{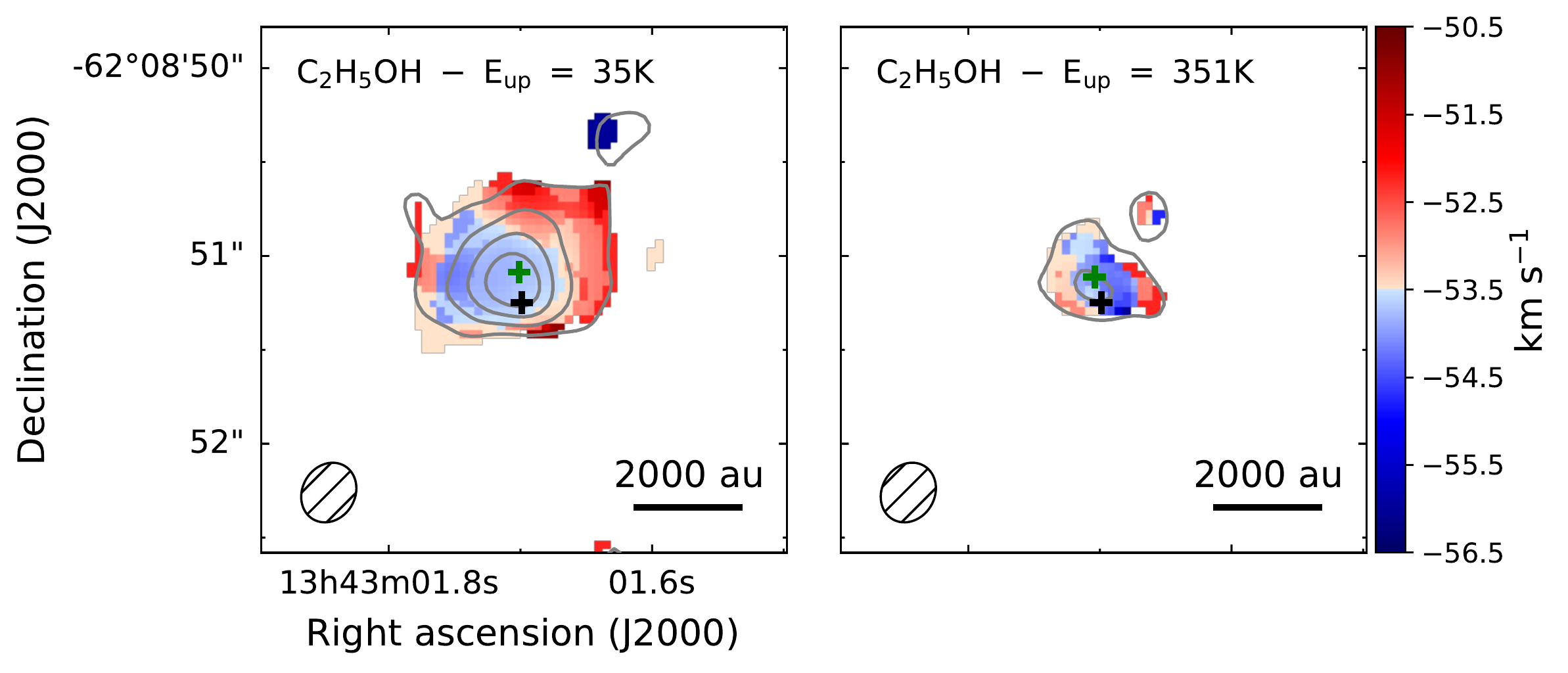}
	\caption[]{Same as Fig. \ref{fig:moment_maps} for C$_{2}$H$_5$OH lines at 254.3841 GHz (left) and 253.3274 GHz (right). Contours start at 3$\sigma$ and are in steps of 6$\sigma$, with $\sigma$ = 5.58$\times$10$^{-3}$ and 5.39$\times$10$^{-3}$ Jy beam$^{-1}$ km s$^{-1}$, for the left and right panel, respectively.}
	\label{fig:mom_C2H5OH}
\end{figure*}

\begin{figure*} 
	\centering
	\includegraphics[width=.8\textwidth]{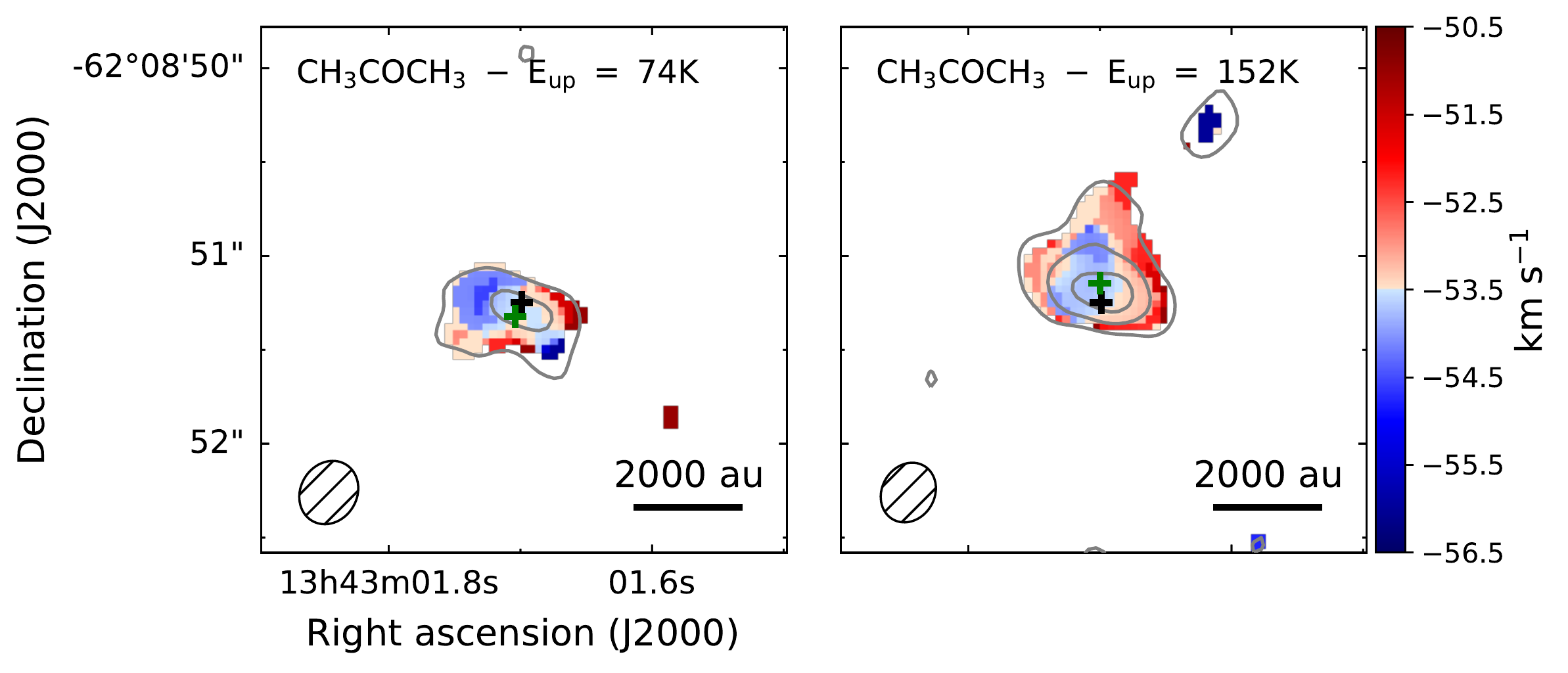}
	\caption[]{Same as Fig. \ref{fig:moment_maps} for CH$_3$COCH$_3$ lines at 240.9988 GHz (left) and 254.0822 GHz (right). Contours start at 3$\sigma$ and are in steps of 6$\sigma$, with $\sigma$ = 4.96$\times$10$^{-3}$ and 5.32$\times$10$^{-3}$ Jy beam$^{-1}$ km s$^{-1}$, for the left and right panel, respectively.}
	\label{fig:mom_CH3COCH3}
\end{figure*}

\clearpage
\begin{figure*} 
	\centering
	\includegraphics[width=.8\textwidth]{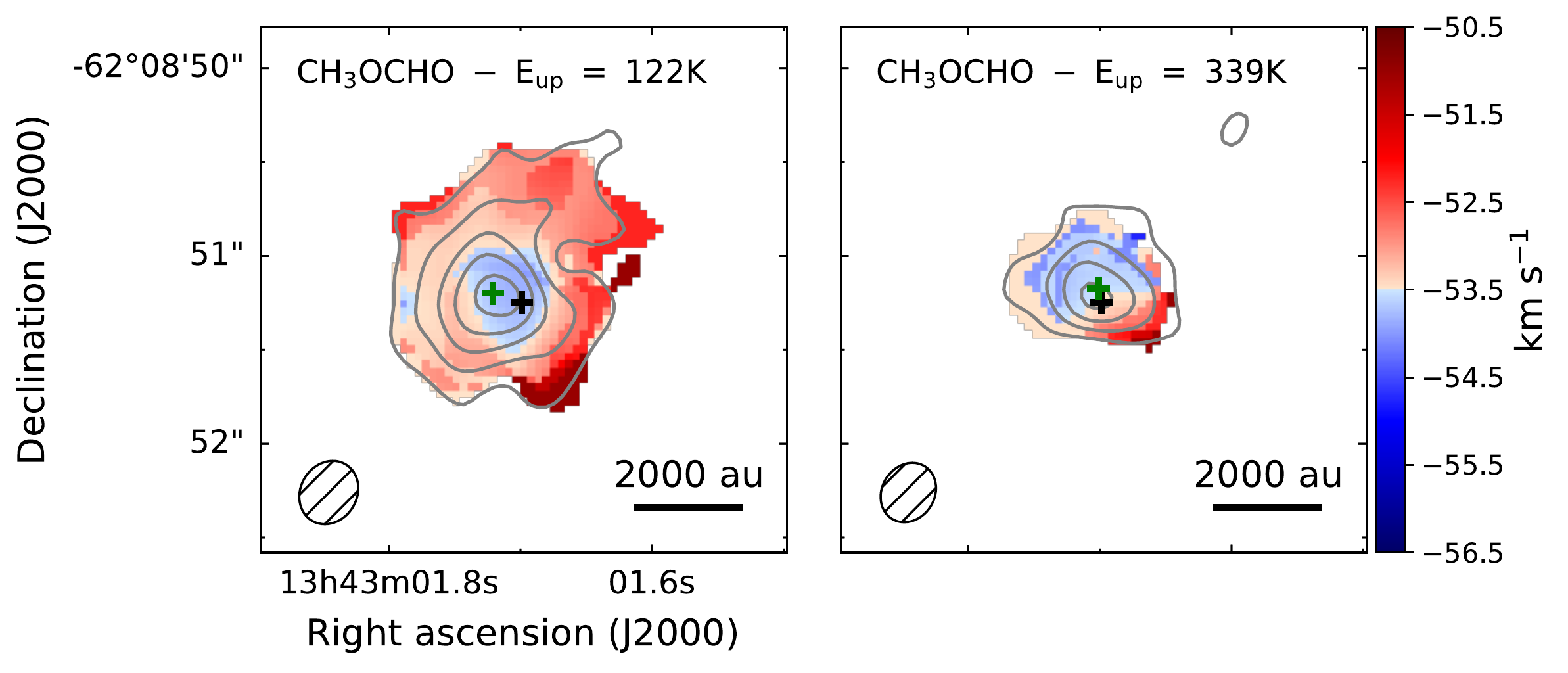}
	\caption[]{Same as Fig. \ref{fig:moment_maps} for CH$_3$OCHO lines at 240.0211 GHz (left) and 256.4478 GHz (right). Contours start at 3$\sigma$ and are in steps of 6$\sigma$, with $\sigma$ = 5.36$\times$10$^{-3}$ and 5.96$\times$10$^{-3}$ Jy beam$^{-1}$ km s$^{-1}$, for the left and right panel, respectively.}
	\label{fig:mom_CH3OCHO}
\end{figure*}

\begin{figure*} 
	\centering
	\includegraphics[width=.8\textwidth]{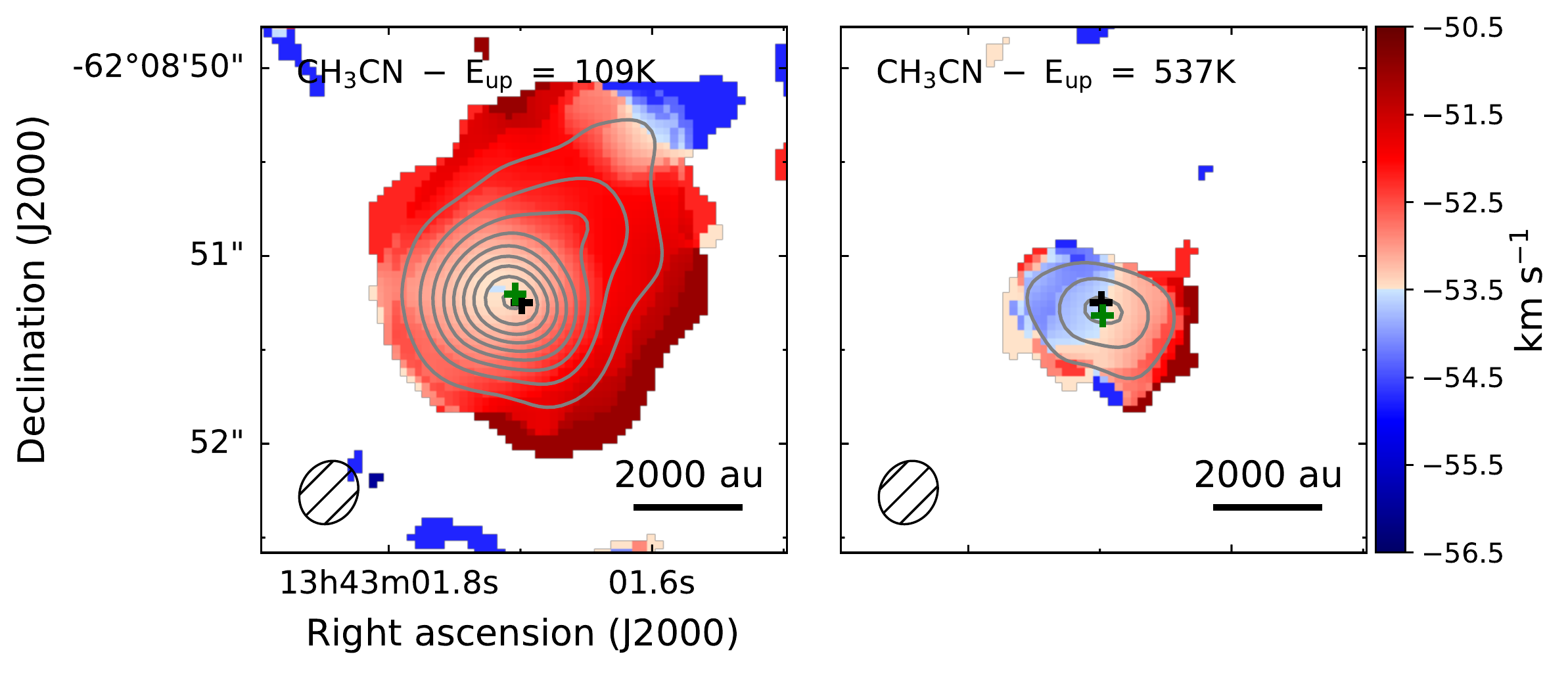}
	\caption[]{Same as Fig. \ref{fig:moment_maps} for CH$_3$CN lines at 239.0965 GHz (left) and 238.8439 GHz (right). Contours start at 9$\sigma$ and are in steps of 12$\sigma$, with $\sigma$ = 1.26$\times$10$^{-2}$ and 5.78$\times$10$^{-3}$ Jy beam$^{-1}$ km s$^{-1}$, for the left and right panel, respectively.}
	\label{fig:mom_CH3OH}
\end{figure*}

\begin{figure*} 
	\centering
	\includegraphics[width=.8\textwidth]{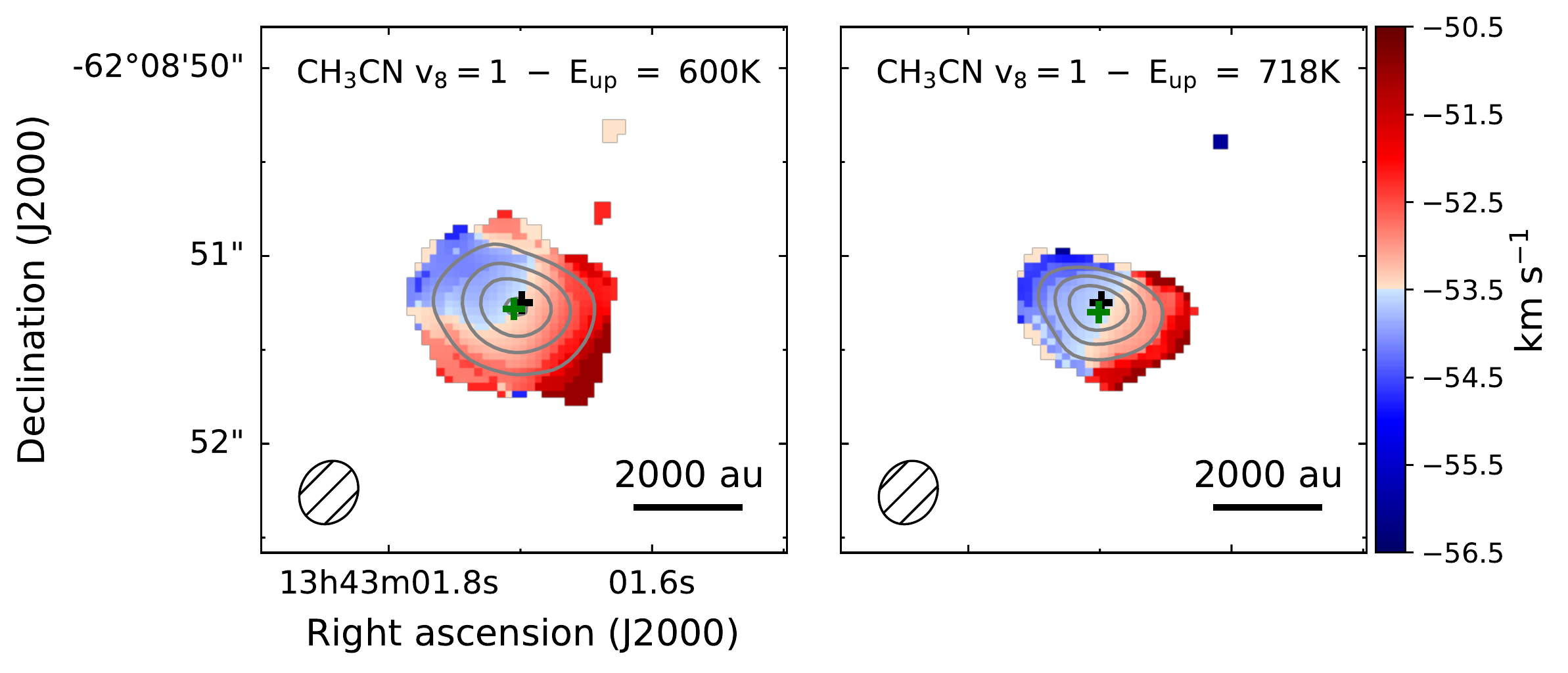}
	\caption[]{Same as Fig. \ref{fig:moment_maps} for vibrationally excited CH$_3$CN lines at 240.0895 GHz (left) and 239.7917 GHz (right). Contours start at 9$\sigma$ and are in steps of 12$\sigma$ and 6$\sigma$, with $\sigma$ = 5.41$\times$10$^{-3}$, for the left and right panel, respectively.}
	\label{fig:mom_CH3CN_v8}
\end{figure*}

\begin{figure*} 
	\centering
	\includegraphics[width=.8\textwidth]{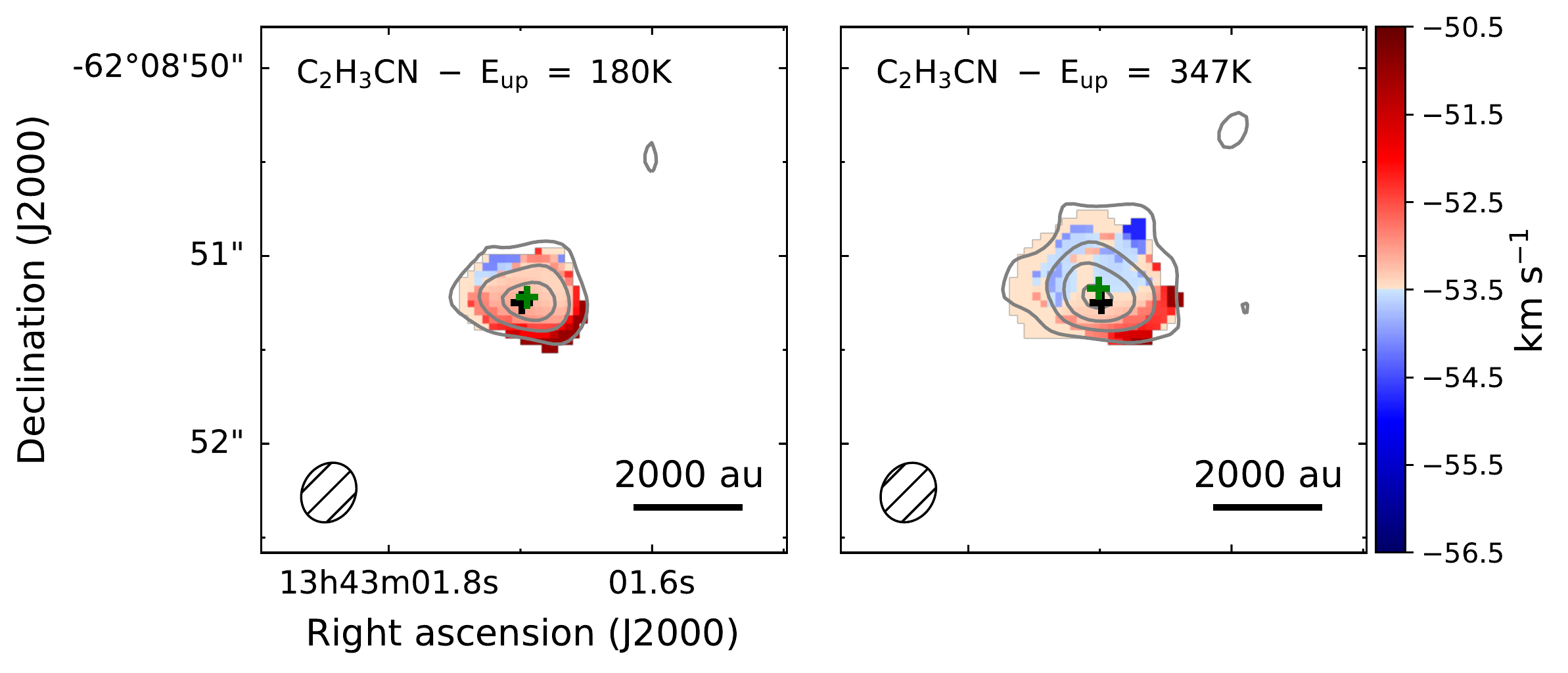}
	\caption[]{Same as Fig. \ref{fig:moment_maps} for C$_{2}$H$_3$CN lines at 254.1375 GHz (left) and 256.4480 GHz (right). Contours start at 3$\sigma$ and are in steps of 6$\sigma$, with $\sigma$ = 5.29$\times$10$^{-3}$ and 6.0$\times$10$^{-3}$ Jy beam$^{-1}$ km s$^{-1}$, for the left and right panel, respectively.}
	\label{fig:mom_C2H3CN}
\end{figure*}

\begin{figure*} 
	\centering
	\includegraphics[width=.8\textwidth]{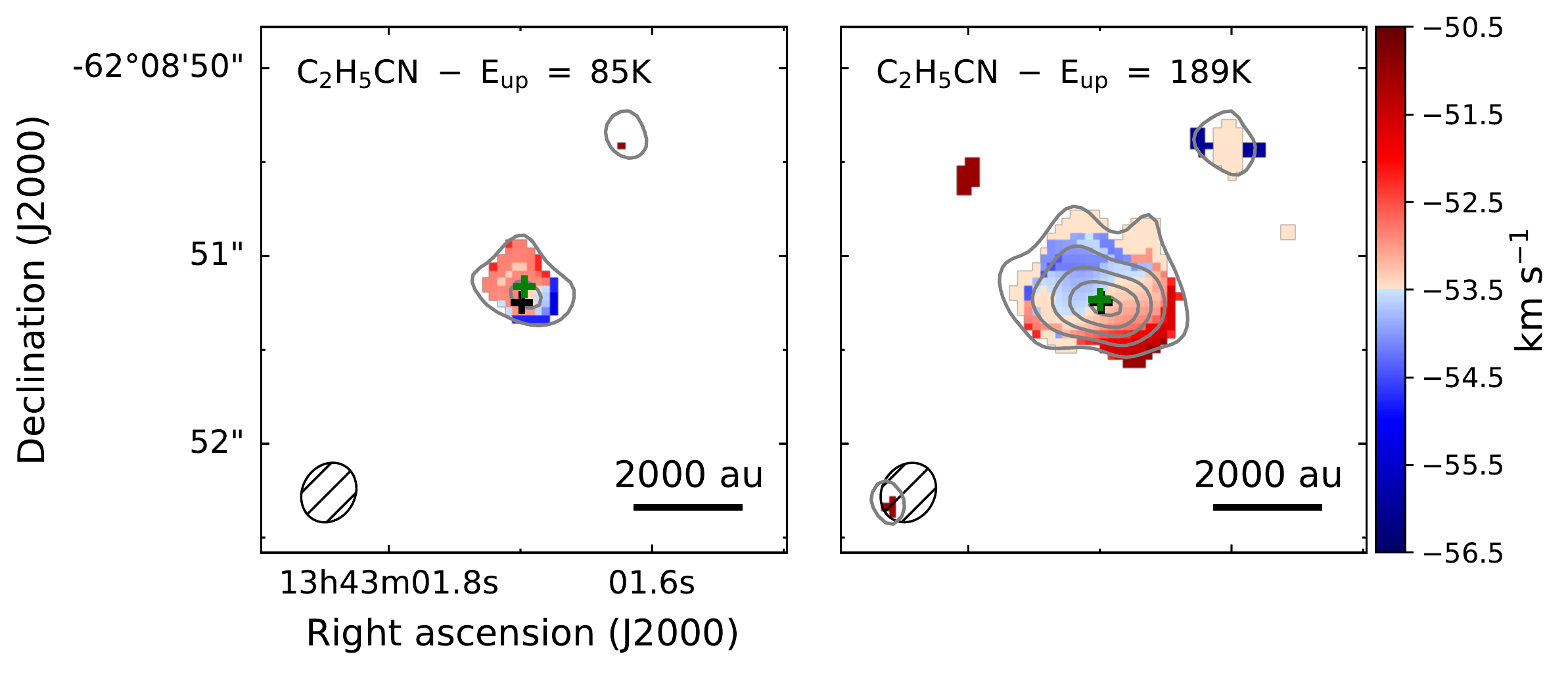}
	\caption[]{Same as Fig. \ref{fig:moment_maps} for C$_2$H$_5$CN lines at 254.6336 GHz (left) and 256.3959 GHz (right). Contours start at 3$\sigma$ and are in steps of 6$\sigma$, with $\sigma$ = 5.18$\times$10$^{-3}$ and 5.7$\times$10$^{-3}$ Jy beam$^{-1}$ km s$^{-1}$, for the left and right panel, respectively.}
	\label{fig:mom_C2H5CN}
\end{figure*}

\begin{figure*} 
	\centering
	\includegraphics[width=.8\textwidth]{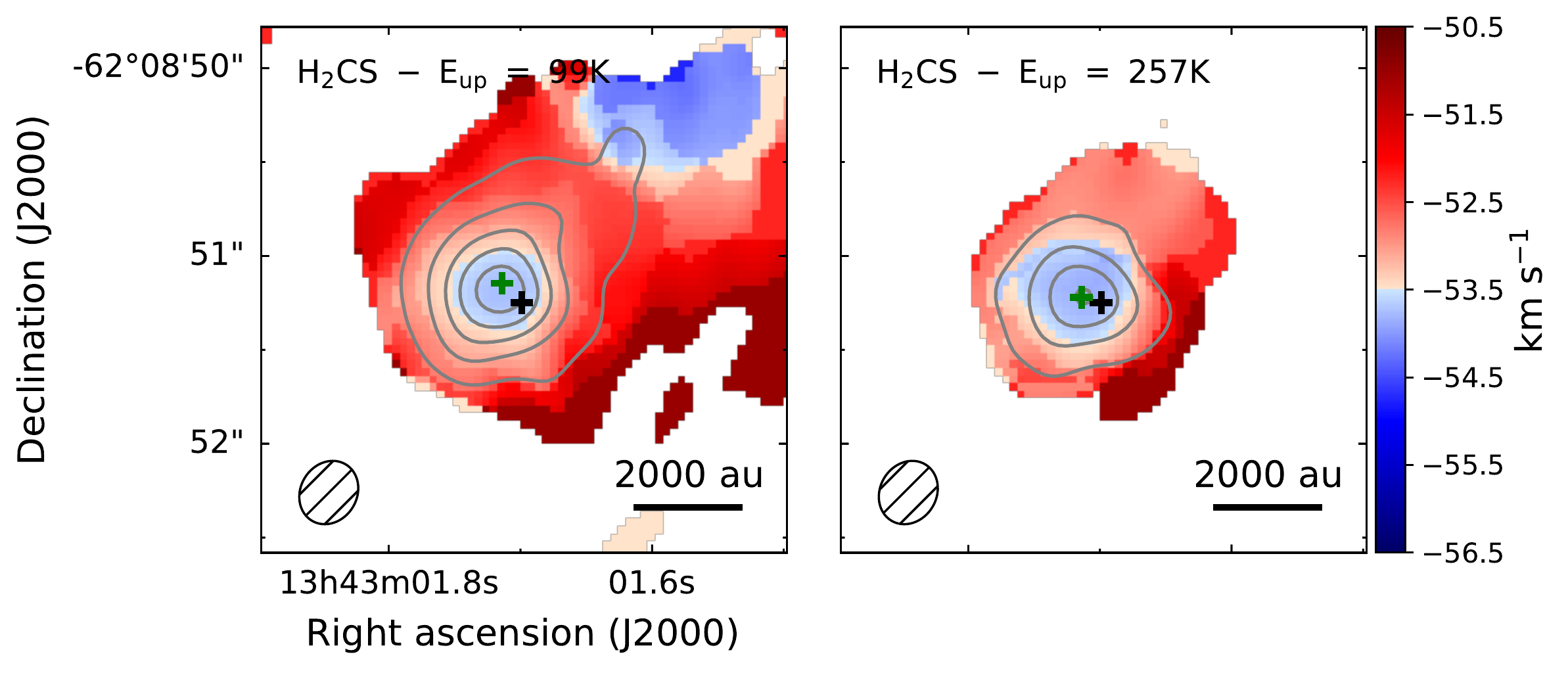}
	\caption[]{Same as Fig. \ref{fig:moment_maps} for H$_2$CS lines at 240.5491 GHz (left) and 240.3322 GHz (right). Contours start at 9$\sigma$ and are in steps of 12$\sigma$, with $\sigma$ = 6.81$\times$10$^{-3}$ and 5.27$\times$10$^{-3}$ Jy beam$^{-1}$ km s$^{-1}$, for the left and right panel, respectively.}
	\label{fig:mom_H2CS}
\end{figure*}

\begin{figure*} 
	\centering
	\includegraphics[width=.8\textwidth]{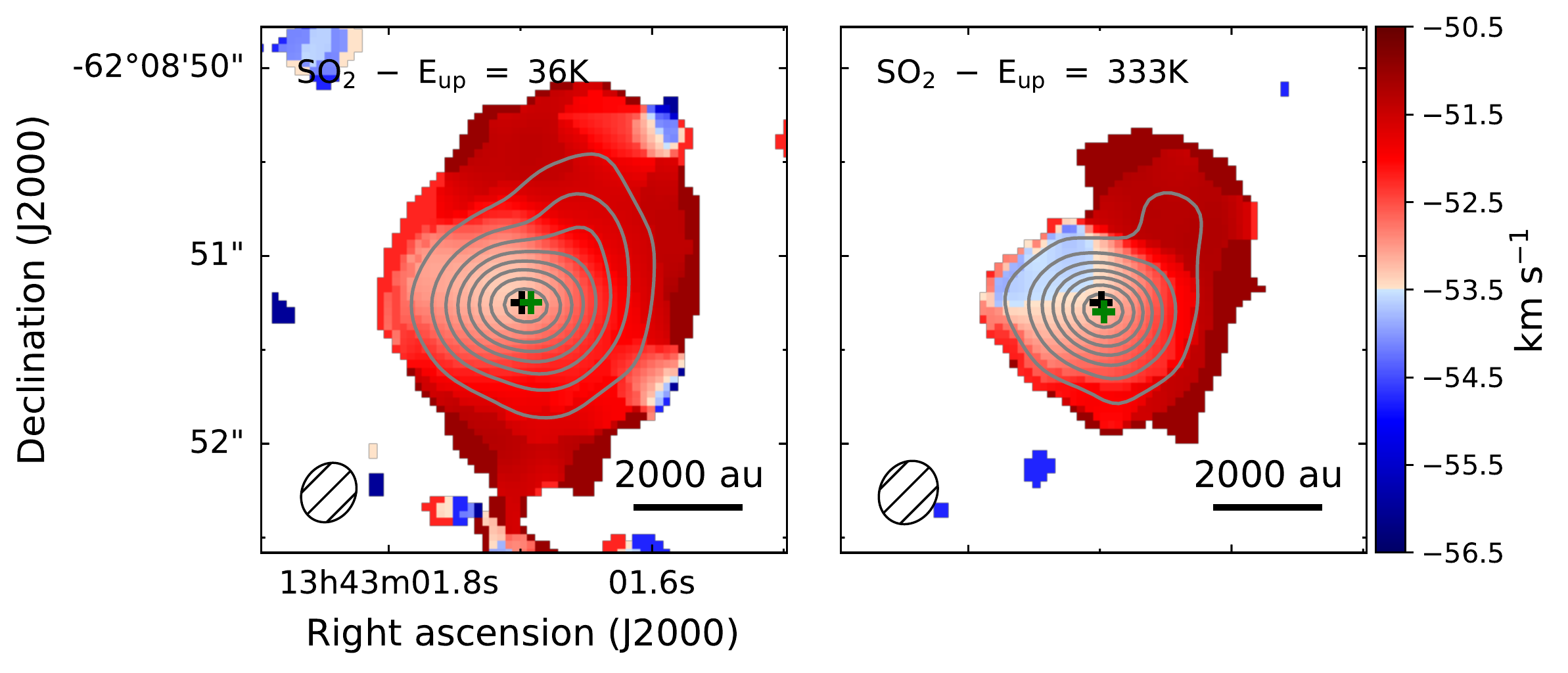}
	\caption[]{Same as Fig. \ref{fig:moment_maps} for SO$_2$ lines at 256.2469 GHz (left) and 238.9925 GHz (right). Contours start at 9$\sigma$ and are in steps of 12$\sigma$, with $\sigma$ = 1.96$\times$10$^{-2}$ and 8.63$\times$10$^{-3}$ Jy beam$^{-1}$ km s$^{-1}$, for the left and right panel, respectively.}
	\label{fig:mom_SO2}
\end{figure*}

\begin{table*}[]
	\small
	\centering
	\caption{Spatial extent of molecules mapped towards AFGL~4176}
	\label{tab:spatial_extent}
	\begin{tabular}{lccccc}
		\toprule
		Species & $E_{\textrm{up}}$ & RA & Dec & FWHM$_{\textrm{X}}$ & FWHM$_{\textrm{Y}}$ \\
				& & & & \multicolumn{2}{c}{[mas]} \\
		\midrule 
			CH$_{3}$C$_{2}$H & low & 13:43:01.748 $\pm$ 0.008 & -62.08.51.30 $\pm$ 0.07 & 1055 $\pm$ 172 & 659 $\pm$ 95 \\
			& high & 13:43:01.739 $\pm$ 0.007 & -62.08.51.29 $\pm$ 0.05 & \phantom{0}944 $\pm$ 127 & 560 $\pm$ 63 \\
		
			CH$_{3}$OH & low & 13:43:01.7138 $\pm$ 0.0009 & -62.08.51.086 $\pm$ 0.007 & \phantom{0}784 $\pm$ 18\phantom{0} & 564 $\pm$ 12 \\
			& high & 13:43:01.7172 $\pm$ 0.0007 & -62.08.51.154 $\pm$ 0.006 & \phantom{0}694 $\pm$ 16\phantom{0} & 535 $\pm$ 11 \\  
			$^{13}$CH$_3$OH & low & 13:43:01.715 $\pm$ 0.002 & -62.08.51.09 $\pm$ 0.02
			& \phantom{0}711 $\pm$ 37\phantom{0} & 489 $\pm$ 22 \\
			& high & 13:43:01.718 $\pm$ 0.001 & -62.08.51.129 $\pm$ 0.008 & \phantom{0}715 $\pm$ 22\phantom{0} & 479 $\pm$ 12 \\
			CH$_{3}$OCH$_{3}$ & low & 13:43:01.726 $\pm$ 0.002 & -62.08.51.22 $\pm$ 0.02 & \phantom{0}687 $\pm$ 41\phantom{0} & 535 $\pm$ 29 \\
			& high & 13:43:01.718 $\pm$ 0.002 & -62.08.51.14 $\pm$ 0.02 & \phantom{0}653 $\pm$ 35\phantom{0} & 460 $\pm$ 21 \\
			C$_{2}$H$_{5}$OH & low & 13:43:01.706 $\pm$ 0.003 & -62.08.51.09 $\pm$ 0.02 & \phantom{0}580 $\pm$ 47\phantom{0} & 464 $\pm$ 33 \\
			& high & 13:43:01.704 $\pm$ 0.008 & -62.08.51.12 $\pm$ 0.03 & \phantom{0}501 $\pm$ 123 & 321 $\pm$ 57 \\
			CH$_{3}$COCH$_{3}$ & low & 13:43:01.704 $\pm$ 0.007 & -62.08.51.33 $\pm$ 0.02 & \phantom{0}590 $\pm$ 117 & 305 $\pm$ 37 \\
			& high & 13:43:01.701 $\pm$ 0.005 & -62.08.51.15 $\pm$ 0.02 & \phantom{0}531 $\pm$ 73\phantom{0} & 345 $\pm$ 35 \\
			CH$_{3}$OCHO & low & 13:43:01.722 $\pm$ 0.002 & -62.08.51.20 $\pm$ 0.02 & \phantom{0}701 $\pm$ 38\phantom{0} & 616 $\pm$ 32 \\
			& high & 13:43:01.701 $\pm$ 0.003 & -62.08.51.18 $\pm$ 0.02 & \phantom{0}526 $\pm$ 3\phantom{10} & 373 $\pm$ 28 \\

			CH$_{3}$CN & low & 13:43:01.704 $\pm$ 0.003 & -62.08.51.21 $\pm$ 0.02 & \phantom{0}726 $\pm$ 37\phantom{0} & 654 $\pm$ 32 \\
			& high & 13:43:01.698 $\pm$ 0.002 & -62.08.51.318 $\pm$ 0.007 & \phantom{0}572 $\pm$ 26\phantom{0} & 391 $\pm$ 14 \\
			CH$_{3}$CN, v8 = 1 & low & 13:43:01.705 $\pm$ 0.001 & -62.08.51.283 $\pm$ 0.005 & \phantom{0}559 $\pm$ 16\phantom{0} & 430 $\pm$ 11 \\
			& high & 13:43:01.701 $\pm$ 0.003 & -62.08.51.31 $\pm$ 0.01 & \phantom{0}535 $\pm$ 40\phantom{0} & 348 $\pm$ 19 \\			 
			NH$_{2}$CHO & low & 13:43:01.7020 $\pm$ 0.0007 & -62.08.51.234 $\pm$ 0.004 & \phantom{0}567 $\pm$ 12\phantom{0} & 434 $\pm$ 8\phantom{0} \\
			& high & 13:43:01.6997 $\pm$ 0.0009 & -62.08.51.183 $\pm$ 0.005 & \phantom{0}559 $\pm$ 16\phantom{0} & 392 $\pm$ 9\phantom{0} \\
			C$_{2}$H$_{3}$CN & low & 13:43:01.696 $\pm$ 0.004 & -62.08.51.23 $\pm$ 0.02 & \phantom{0}441 $\pm$ 56\phantom{0} & 295 $\pm$ 26 \\
			& high & 13:43:01.701 $\pm$ 0.003 & -62.08.51.18 $\pm$ 0.02 & \phantom{0}519 $\pm$ 51\phantom{0} & 387 $\pm$ 32 \\
			C$_{2}$H$_{5}$CN & low & 13:43:01.697 $\pm$ 0.009 & -62.08.51.17 $\pm$ 0.04 & \phantom{0}400 $\pm$ 141 & 256 $\pm$ 59 \\
			& high & 13:43:01.700 $\pm$ 0.003 & -62.08.51.24 $\pm$ 0.01 & \phantom{0}565 $\pm$ 41\phantom{0} & 362 $\pm$ 20 \\
			
			H$_{2}$CS & low & 13:43:01.714 $\pm$ 0.003 & -62.08.51.15 $\pm$ 0.02 & \phantom{0}756 $\pm$ 40\phantom{0} & 625 $\pm$ 31 \\
			& high & 13:43:01.715 $\pm$ 0.002 & -62.08.51.23 $\pm$ 0.02 & \phantom{0}599 $\pm$ 26\phantom{0} & 562 $\pm$ 24 \\				
			SO$_{2}$ & low & 13:43:01.692 $\pm$ 0.002 & -62.08.51.25 $\pm$ 0.02 & \phantom{0}728 $\pm$ 29\phantom{0} & 624 $\pm$ 24 \\
			 & high & 13:43:01.6972 $\pm$ 0.0009 & -62.08.51.299 $\pm$ 0.006 & \phantom{0}550 $\pm$ 15\phantom{0} & 470 $\pm$ 12 \\
			\bottomrule
	\end{tabular}
	\tablefoot{Fit parameters for the 2D gaussian fit to the zero-moment maps of the low and high upper state energy transitions mapped in Fig. \ref{fig:moment_maps} and Figs. \ref{fig:mom_13CH3OH} -- \ref{fig:mom_SO2}.}
\end{table*}

\end{document}